\documentclass[fleqn,usenatbib]{mnras}

\usepackage{newtxtext,newtxmath}

\usepackage[T1]{fontenc}
\usepackage{ae,aecompl}

\usepackage{calc}
\usepackage{longtable}

\usepackage{graphicx}	
\usepackage{amsmath}	
\usepackage{amssymb}	

\newcommand{\etal}{et al. \ }
\newcommand{\kms}{km\,s$^{-1}$}
\newcommand{\pmu}{$\mu$as\,yr$^{-1}$}
\newcommand{\teff}{T$_{\rm eff}$ }
\newcommand{\logg}{log\,$g$ }
\newcommand{\esp}{ESPaDOnS}
\newcommand{\Pristine}{{\it Pristine}}
\newcommand{\msun}{M$_{\odot}$}

\title[CFHT ESPaDOnS Spectra of Pristine Survey Stars]{Pristine IX: CFHT ESPaDOnS Spectroscopic Analysis of 115 Bright Metal-Poor Candidate Stars }

\author[Venn \etal ]
{Kim Venn,$^{1}$\thanks{E-mail: kvenn@uvic.ca}
Collin Kielty,$^{1}$
Federico Sestito,$^{2,3}$
Else Starkenburg,$^{3}$
Nicolas Martin,$^{2,4}$
\and
David Aguado,$^{5}$
Anke Arentsen,$^{3}$
Piercarlo Bonifacio,$^{6}$
Elisabetta Caffau,$^{6}$
Vanessa Hill,$^{7}$
\and
Pascale Jablonka,$^{6,8}$
Carmela Lardo,$^{8}$
Lyudmilla Mashonkina,$^{9}$
Julio Navarro,$^{1}$
\and
Chris Sneden,$^{10}$
Guillaume Thomas,$^{11}$
Kris Youakim,$^{3}$
Jonay I. Gonz\'alez-Hern\'andez,$^{12,13}$
\and
Rub{\'e}n S{\'a}nchez Janssen,$^{14}$
Ray Carlberg,$^{15}$
Khyati Malhan$^{16}$
\\$\;$
\\
$^{1}$Department of Physics and Astronomy, University of Victoria, Victoria, BC, V8W 3P2, Canada \\
$^{2}$Universit\'e de Strasbourg, CNRS, Observatoire astronomique de Strasbourg, UMR 7550, F-67000 Strasbourg, France \\
$^{3}$Leibniz-Institut f\"ur Astrophysik Potsdam (AIP), An der Sternwarte 16, D-14482, Potsdam, Germany \\
$^{4}$Max-Planck-Institut f\"ur Astronomie, K\"onigstuhl 17, D-69117, Heidelberg, Germany \\
$^{5}$Institute of Astronomy, University of Cambridge, Madingley Road, Cambridge CB3 0HA, UK \\
$^{6}$GEPI, Observatoire de Paris, Universit\'e PSL, CNRS, Place Jules Janssen, F-92195 Meudon, France \\
$^{7}$ Universit\'e Cote d'Azur, Observatoire de la Cote d'Azur, CNRS, Laboratoire Lagrange, 
   Bd de l'Observatoire, CS 34229, F-06304 Nice cedex 4, France \\
Observatoire de la Cote d'Azur, BP 4229, F-06304, Nice, France \\ 
$^{8}$Institute of Physics, Laboratoire d'astrophysique, Ecole Polytechnique F\'ed\'erale de 
      Lausanne (EPFL), Observatoire, 1290 Versoix, Switzerland\\
$^{9}$Institute of Astronomy, Russian Academy of Sciences, Pyatnitskaya st. 48, 119017 Moscow, Russia \\
$^{10}$Department of Astronomy, University of Texas at Austin, Austin, TX, 78712, USA \\
$^{11}$National Research Council Herzberg Astronomy \& Astrophysics, 4071 West Saanich Road, Victoria, BC, Canada  \\
$^{12}$Instituto de Astrof\'isica de Canarias, Via L\'actea, 38205, La Laguna, Tenerife, Spain \\
$^{13}$Departamento de Astrof\'isica, Universidad de La Laguna, La Laguna, Tenerife, Spain \\
$^{14}$UK Astronomy Technology Centre, Royal Observatory Edinburgh, Blackford Hill, Edinburgh, EH9 3HJ, UK \\
$^{15}$Department of Astronomy \& Astrophysics, University of Toronto, Toronto, ON, M5S 3H4, Canada \\
$^{16}$The Oskar Klein Centre for Cosmoparticle Physics, Department of Physics, Stockholm University, AlbaNova, SE-10691 Stockholm, Sweden
}
\date{Accepted XXX. Received YYY; in original form ZZZ}

\pubyear{2019}

\begin{document}
\label{firstpage}
\pagerange{\pageref{firstpage}--\pageref{lastpage}}
\maketitle


\begin{abstract}

A chemo-dynamical analysis of 115 metal-poor candidate stars selected from 
the narrow-band \Pristine\ photometric survey is presented based on CFHT 
high-resolution ESPaDOnS spectroscopy.  We have discover 28 new bright 
(V $<$ 15) stars with [Fe/H]$<-2.5$ and 5 with [Fe/H]$<-3.0$ for success 
rates of 40\% (28/70) and 19\% (5/27), respectively. A detailed model 
atmospheres analysis is carried out for the 28 new metal-poor stars. 
Stellar parameters were determined from SDSS photometric colours, 
Gaia DR2 parallaxes, MESA/MIST stellar isochrones, and the initial 
\Pristine\ survey metallicities, following a Bayesian inference method. 
Chemical abundances are determined for 10 elements 
(Na, Mg, Ca, Sc, Ti, Cr, Fe, Ni, Y, Ba). Most stars show chemical 
abundance patterns that are similar to the normal metal-poor stars in the 
Galactic halo; however, we also report the discoveries of a new r-process 
rich star, a new CEMP-s candidate with [Y/Ba]>0, and a metal-poor star 
with very low [Mg/Fe]. The kinematics and orbits for all of the highly 
probable metal-poor candidates are determined by combining our precision 
radial velocities with Gaia DR2 proper motions. Some stars show unusual 
kinematics for their chemistries, including planar orbits, unbound orbits, 
and highly elliptical orbits that plunge deeply into the Galactic bulge 
(R$_{\rm peri}<0.5$ kpc); also, eight stars have orbital energies and actions 
consistent with the Gaia-Enceladus accretion event.  This paper contributes 
to our understanding of the complex chemo-dynamics of the metal-poor Galaxy, 
and increases the number of known bright metal-poor stars available for 
detailed nucleosynthetic studies.

\end{abstract}



\section{Introduction}

Very old stars are witness to the earliest epochs of galaxy formation and evolution. 
Most theoretical models of star formation at early times predict the
formation of high mass stars \citep[e.g.,][]{Nakamura2001,
Abel2002, Bromm2013} that contributed to the reionization of the Universe.
During their short lives, these massive stars initiate the formation
of the chemical elements beyond hydrogen, helium, and lithium, and yet no star 
with such a primordial composition has yet been found.  
Fragmentation of the early star forming regions has also been predicted 
\citep[e.g.,][]{Schneider2003, Clark2011, Greif2015, Hirano2015},  
providing an environment where lower mass ($\sim 1$ M$_\odot$) stars could form,
which would have much longer lifetimes.  These old stars are expected to be
metal-poor, having formed from nearly pristine gas, and could be used to trace the 
chemical elements from the massive (first) stars and their subsequent supernovae 
\citep[e.g.,][]{FN2015, Salvadori2019, Hartwig2018}.  

In recent years, abundance patterns of metal-poor stars have been examined extensively
\citep[e.g.,][]{Keller2014, Nordlander2019, Ishigaki2018}, pointing to the significance 
of low-energy (faint) supernovae, whose ejecta falls back onto their iron-cores, thereby 
mainly expelling light elements.  It is not
clear if these low-energy supernovae were more common at ancient times, or if concurrent massive stars
underwent direct collapse to black holes and ceased nearby star formation, erasing any direct 
evidence of their presence in the next generation of stars. 
Overall, metal-poor stars allow us to examine nucleosynthetic yields from one or a few supernovae
events to constrain the detailed physics of these events, such as neutron star masses, rotation rates, 
mixing efficiencies, explosion energies, etc. \citep{Heger2010, Thielemann2018, 
Wanajo2018, Muller2019, Jones2019}.  
These yields are relevant for understanding the early chemical build-up and
the initial conditions in the early Galaxy.
  
Chemical abundances also show variations between old metal-poor stars in different
environments such as dwarf galaxies, suggesting that the first stages of enrichment were not
uniform.  Stars in the nearby dwarf galaxies typically have lower abundances of $\alpha$-
and odd-Z elements, attributed to their slower star formation histories and/or fewer number of
high mass stars overall \citep{Venn2004, Tolstoy2009, NS2010, McWilliam2013, FN2015, Hayes2018},
while significant variations in heavy r-process elements in some dwarf galaxies, and globular clusters,
are discussed in terms of contributions from individual compact binary merger events, like GW170817
\citep[e.g.,][]{Roederer2011, Roederer2018, Ji2016, Ji2019}. 
In addition, about a third of the [Fe/H] $< -2.5$ stars\footnote{We adopt standard notation, such
that [X/H] = log(X/H)$_*$ $-$ log(X/H)$_\odot$. }
 in the Galactic halo show very high 
enhancements in carbon (the carbon-enhanced metal-poor stars, "CEMP"; \citealt{Yong2013, Aguado2019}
also see \citealt{Kielty2017, Mardini2019}),
discussed as a signature of the earliest chemical enrichment in the Universe.  
However, at least one ultra metal-poor star is not carbon-enhanced \citep[SDSS J102915+172927, ][]{Caffau2012},
and the known metal-poor stars in the Galactic bulge do not show carbon-enhancements 
\citep{Howes2016, Lamb2017}.  
\citet{Norris2013} suggest that there are likely multiple chemical enrichment pathways for old 
metal-poor stars dependent on the star formation environment, and also possibly binary mass transfer
effects (also see discussions by \citealt{Starkenburg2014, Arentsen2019}).

The majority of old, metal-poor stars in the Galatic halo are thought to have been accreted 
from dwarf galaxies at early epochs, based on cosmological hydrodynamical simulations of the Local Group 
\citep{Ibata1994, Helmi1999, Ibata2004, Abadi2010, Starkenburg2017b, ElBadry2018}.  
This is consistent
with the high-velocity, eccentric, orbits determined from the exquisite Gaia DR2 data \citep{gaia18}
and spectroscopic radial velocities  for a majority of the 
ultra metal-poor stars \citep{Sestito2019} and the ultra faint dwarf galaxies \citep{Simon2018}.
Interestingly, many of these orbits are also highly retrograde, 
similar to the diffuse halo merger remnants, Gaia-Enceladus 
\citep{Helmi2018, Haywood2018, Belokurov2018, Myeong2018} 
and Gaia-Sequoia \citep{Myeong2019, Barba2019}.
However, some metal-poor stars have been found to have orbits that place them in the 
Galactic plane \citep{Sestito2019}, even with nearly circular orbits
\citep[e.g., SDSS J102915+172927,][]{Caffau2012}.
These latter observations challenge the cosmological simulations since metal-poor stars
are assumed to be old, and yet the Galactic plane is thought to have formed only $\sim$10 Gyr ago
\citep[e.g.,][]{Casagrande2016, Gianninas2015}.
Alternatively, \citet{Sestito2019} suggest these stars may have be brought into the Galaxy 
from a merger that helped to form the disk.

Progress in this field will require large statistical samples of metal-poor
stars in a variety of environments within the Local Group.
Unfortunately, metal-poor stars are exceedingly rare and difficult to find,
being overwhelmed by the more numerous metal-rich populations in the Galaxy.
Examination of the Besan\c{c}on model of the Galaxy \citep{Robin2003}, 
which is guided by a theoretical framework for the formation and evolution of
the main stellar populations, suggests that a typical halo field has only 
one in $\sim$2000 stars with [Fe/H] $< -3$ between 14 $< V <$ 18 \citep{Youakim2017}.  
Enormous effort has gone into the discovery and study of extremely, ultra, and 
hyper metal-poor stars with [Fe/H] $< -3.0$, [Fe/H] $< -4.0$, and [Fe/H] $< -5.0$,
respectively.
Most of the known metal-poor stars have been found in dedicated surveys, such as
objective prism surveys \citep[the HK survey and Hamburg-ESO survey, ][]{Beers1992, 
Beers2005, Christlieb2002, Christlieb2008, Frebel2006, Schoerck2009},
wide-band photometric surveys \citep{Schlaufman2014}, and blind spectroscopic 
surveys, such as the the Sloan Digital Sky Survey (SDSS) SEGUE and BOSS surveys
\citep{Yanny2009, Eisenstein2011, Dawson2013},
and from the Large Sky Area Multi-Object Fibre Spectoscopic Telescope \citep[LAMOST, ][]{Cui2012}.
According to the SAGA database \citep[see][and references therein]{Suda2017}, there are 
$\sim$500 stars with [Fe/H] $< -3.0$, though fewer than half have detailed chemical
abundances.
Recently, narrow-band photometric surveys have shown higher success rates for finding
metal-poor stars, particularly SkyMapper \citep{Keller2007, DaCosta2019} 
and the \Pristine\ survey \citep{Starkenburg2017, Youakim2017, Aguado2019}.  
\Pristine\ photometry with follow-up Keck II/DEIMOS spectroscopy
has also been used to increase sample sizes and improve 
the chemodynamical studies of faint satellites 
\citep[Draco II and Sgr II,][]{Longeard2018, Longeard2019}.
At the same time, \citet{Simon2018} has shown that Gaia DR2 proper motion cleaning may also be 
a promising way to find new metal-poor members of ultra faint dwarf galaxies. 

The \Pristine\ survey uses a unique narrow-band filter centered on the \ion{Ca}{II} H \& K spectral
lines ("CaHK") mounted on MegaPrime/MegaCam at the 3.6-metre Canada France Hawaii Telescope (CFHT). 
When combined with broad-band SDSS $gri$ photometry \citep{York2000}, 
this CaHK filter has been calibrated to find 
metal-poor candidates with 4200 $<$ T $<$ 6500~K.
The \Pristine\ survey has proven successful at predicting metallicities for
faint objects (18 $> V >$ 15), based on results from medium resolution spectroscopic follow-up
\citep{Youakim2017, Aguado2019}.  For brighter objects, the success of the
\Pristine\ calibration is less certain. \citet{Caffau2017} observed 26 bright (g $< 15$) 
candidates with the FEROS spectrograph at the MPG/ESO 2.2-metre telescope, but found only
5 stars with [Fe/H] $< -2.0$.   It was thought that the selection may be affected by
previously unrecognized saturation effects in the SDSS photometry.  Thus, 
\citet{Bonifacio2019} selected bright candidates using a new \Pristine\ calibration 
with APASS photometry \citep[c.f., APASS DR10][]{Henden2019}; observations of 40 targets 
with the SOPHIE spectrograph at Observatoire de Haute Provence found only 8 stars with
[Fe/H] $< -2.0$, and none with [Fe/H] $< -3.0$. 
Until now, confirmation of the \Pristine\ metallicity predictions below [Fe/H] $= -3.0$
has only been carried out for one star from high resolution spectroscopy, 
Pristine\_221.8781+09.7844 at [Fe/H] = $-4.7$ (1D, LTE) and $V = 16.4$ \citep{Starkenburg2018}.

In this paper, we present the analysis of 115 bright ($V < 15$) metal-poor candidates 
from the \Pristine\ survey, calibrated using the original SDSS $gri$ photometry and observed at the 
CFHT with the high resolution \esp\ spectrograph.
Such high resolution spectra are necessary for detailed chemical abundances,
as well as precision radial velocities for determining the kinematic properties.
The power of combining chemical abundances with kinematic properties of stars is the
backbone of the field of Galactic Archaeology \citep[e.g.,][]{FBH2002, Venn2004, Tolstoy2009, FN2015}.
We confirm the success of the \Pristine\ survey to find metal-poor stars even at bright magnitudes,
determine the chemical abundances for 10 elements, calculate the kinematics of the stars in our
sample, and interpret in the context of variations in nucleosynthetic sites, locations, and time-scales.
The study of metal-poor old stars is unique to our Local Group, since only here can we resolve 
individual stars and study these rare targets that guide our understanding of the physics of 
star formation, supernovae, the early build-up of galaxies, and the epoch of reionization. 

\begin{figure}
\includegraphics[scale=0.4]{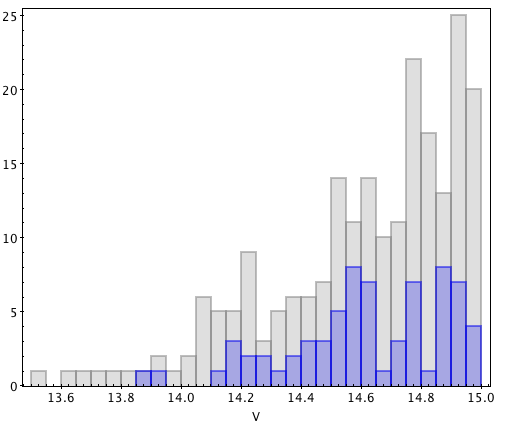}
\caption{Histogram of the V magnitudes of 223 stars with high probabilities to
be metal-poor ([Fe/H]$<-2.50$) from the (g-i) or (g-r) calibrations in the 
\Pristine\ survey original $\sim$1000 sq.deg.$^2$ footprint (grey bars). 
The 70 stars observed with CFHT \esp\ that also meet these criteria 
are overplotted (blue bars).
\label{fig:Vhist}}
\end{figure}
 
\section{Target Selection } \label{sect:selection} 

Targets were selected from the \Pristine\ survey catalogue\footnote{Internal-Catalogue-1802.dat},
which includes 28557 bright (V $<$ 15) stars in the 
original $\sim$1000 sq.deg.$^2$ footprint between
$180 < RA < 256$$^{\circ}$ and $+00 < Dec < +16$$^{\circ}$ 
\citep{Starkenburg2017, Youakim2017}.  

\Pristine\ survey targets were cross-matched with SDSS photometry to obtain $ugri$ broad-band 
magnitudes used for colour temperature determinations and point source identification.
Additional selection criteria were adopted, as described by \citet{Youakim2017},
including the removal of non-star contaminants (based on SDSS and CaHK flags), 
white dwarf contaminants (removing SDSS $u-g > 0.6$, \citealt{Lokhorst2016}), 
variability flags from Pan-STARRS1 photometry \citep{Hernitschek2016},
and the quality of SDSS $gri$-band photometry.   
The SDSS $gri$-band photometry was further used for a colour selection, where 
$0.25 < (g-i)_o < 1.5$ and $0.15 < (g-r)_o < 1.2$ correspond to the temperature
range 4200~K $<$ \teff\ $<$ 6500~K, covering the tip of the red giant branch 
and the cooler main sequence to the main sequence turnoff.

\begin{figure*}
\includegraphics[width=1\textwidth, scale=0.99]{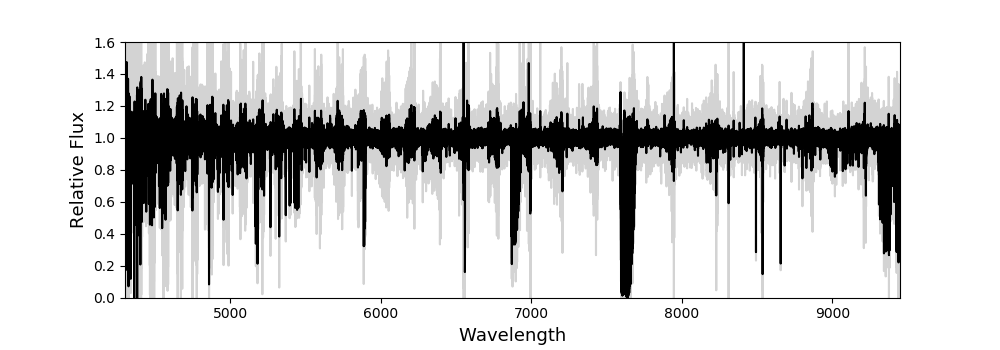}
\caption{Full CFHT \esp\ spectrum for Pristine-235.1449+08.7464 (grey), and smoothed by
a 3-pixel boxcar (black).   A signal-to-noise ratio (SNR) =30 near 520 nm was adopted for 
this exploratory survey, leaving very low SNR and non-smooth continuum on the red side of 
the detector and therefore in the inter-order regions.}
\label{fig:p7full}
\end{figure*}

The 115 stars observed at CFHT with the high resolution (R$\sim$68,000)  \esp\ 
spectrograph \citep{Donati2006}
are listed in Table~\ref{tab:targets} including RA and DEC (from SDSS, in degrees), 
the dereddened SDSS ($ugri$)$_0$ and \Pristine-CaHK$_0$ magnitudes, the $V$ and $I$
magnitudes (converted from SDSS photometry using \citet{Jordi2006} and not dereddened, 
thus in observer units), and the reddening E(B-V) value.
Extinction values are small for most stars, and we assume that all the extinction is
in the foreground, therefore using the \citet{Schlegel1998} extinction maps.
A summary of the CFHT \esp\ observing runs that comprise this program are
16AC031 (23 targets), 16AC096 (17 targets), 16BC008 (25 targets),
17AC002 (30 targets), 18BC018 (25 targets),
which is 120 targets, with 5 repeat targets, thus 115 independent objects.

We note that this program began immediately after the initial \Pristine\ survey 
MegaCam observing runs, and the metallicity calibrations have improved over the 
course of these spectroscopic follow-up observations (2016A to 2018B).
Of the 115 observed stars, 88 remain in the \Pristine\ survey catalogue.
In Table~\ref{tab:targets}, we have 59 stars with $>$80\% probability to 
have [Fe/H]$<-2.25$ using both the SDSS $g-r$ and $g-i$ colour calibrations, 
and with individual metallicity estimates of [Fe/H] $<-2.5$.
Another 10 stars follow these selections using the SDSS $g-r$ colour alone.
\citet{Youakim2017} showed that the SDSS i filter has saturation effects
in some fields for stars in our magnitude range that can affect the
SDSS $g-i$ selection criterion. 
An additional 46 stars were observed with \esp, 
however we now recognize 19 of those to have low probabilities to be metal-poor, 
and 27 are no longer in the \Pristine\ survey catalogue (e.g., due to 
the saturation effects in the SDSS photometry recognized later).  
Ironically, of those latter 27 stars, one star (Pristine\_213.7879+08.4232)
does appear to be metal-poor, e.g. its \ion{Ca}{II} triplet lines are weak 
and narrow.  Possibly the \Pristine\ survey selection function is now slightly 
overly strict; we retained this one metal-poor candidate.
Thus we have observed a total of 70 (59 + 10 + 1) metal-poor candidates selected
from the original $\sim$1000 sq.deg.$^2$ footprint of the \Pristine\ survey. 
In total, there are 223 bright stars that meet all of the selection criteria described
in this section, thus we have observed 31\% (70/223) of these candidates.  
Both of these distributions are shown in Fig.~\ref{fig:Vhist}.

The selection criteria used here differ slightly from \citet{Youakim2017} and
\citet{Aguado2019}, where stars with probability over 25\% in both $g-r$ and 
$g-i$ were selected for their medium resolution spectroscopic program.   These 
lower limits were also adopted by \citet{Caffau2017} and \citet{Bonifacio2019} in 
their target selections, though using APASS photometry in the latter paper.   
We emphasize that our target selections were made without a priori knowledge of 
the spectroscopic metallicities, other than for a small subset of five 
stars\footnote{Five stars had interesting results from our concurrent medium resolution 
spectral campaign, and were selected for observations with \esp\ during our final 2018B run.  
Three were confirmed to be metal poor ([Fe/H]$<-2.5$), but two were not ([Fe/H]$>-2.0$).  
If we recalculate our success rates without these five stars, then 
38\% (25/65) are found with [Fe/H]$<-2.5$ and 16\% (4/25) with [Fe/H]$<-3.0$.}
in our final 2018B observing run.

\section{ \esp\ Observations } \label{sect:observations} 

The CFHT high resolution spectrograph \esp\ was used between 2016A
and 2018B to observe 115 new bright, metal-poor candidates found in 
the original CFHT-MegaCam survey footprint as part of the \Pristine\ survey.
\esp\ was used in the "star+sky" mode, providing a high resolution (R$=68,000$)
spectrum from 400 to 1000 nm,
making it possible to determine precision radial velocities, 
stellar parameters, and chemical abundances.

\begin{figure}
\includegraphics[scale=0.4]{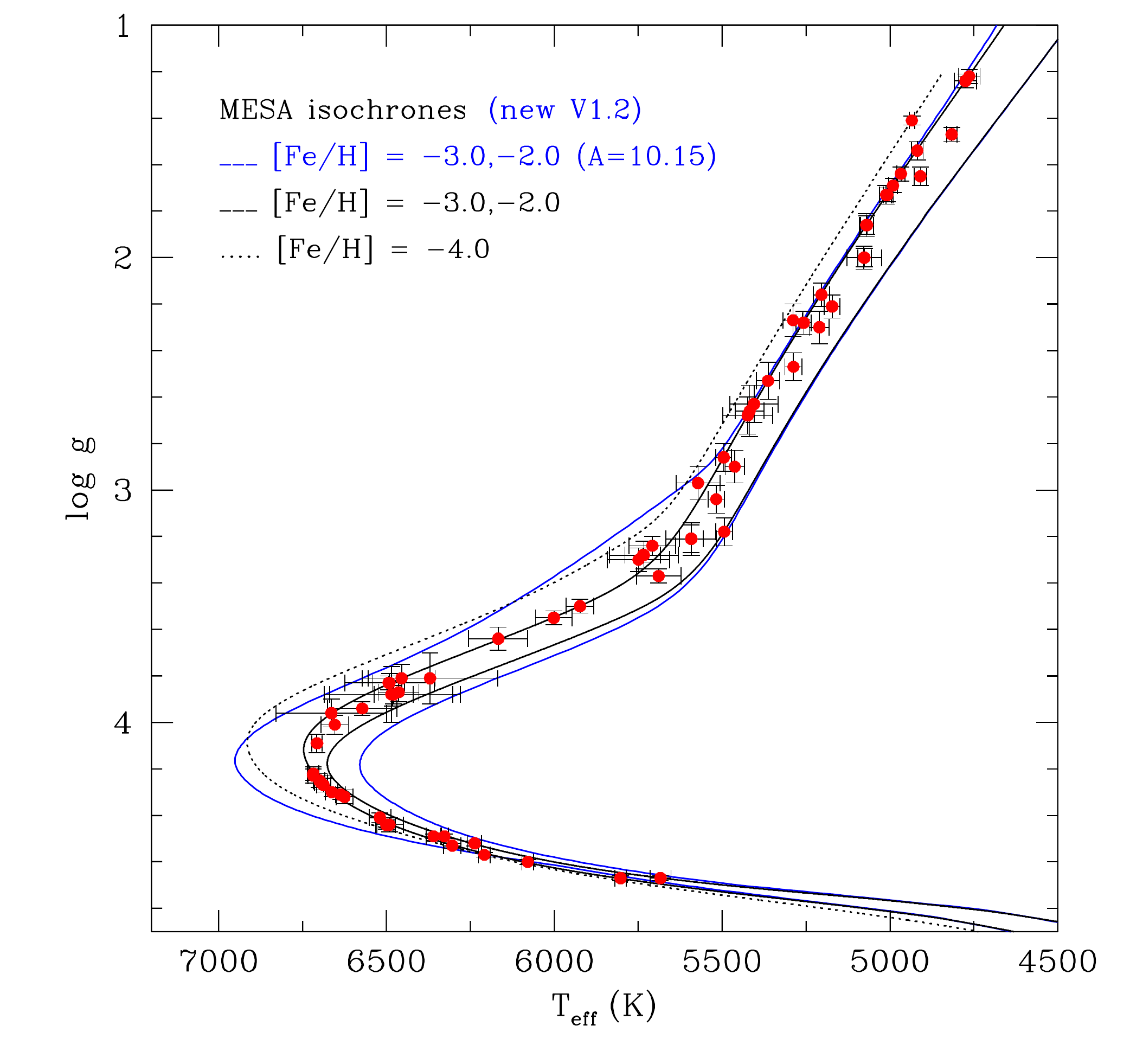}
\caption{\teff\ vs \logg\ for 70 high-probability metal-poor stars 
selected from the \Pristine\ survey.  For illustration purposes, the
isochrones for a single age of 14.1 Gyr are shown (or log(A/yr) $= 10.15$).
The isochrones used for the stellar parameter estimates are from a previous version 
of MESA/MIST (shown in black), compared with isochrones from the newer version 
of MIST (V1.2, shown in blue).  
\label{fig:hrdmist}}
\end{figure}

Each observation was fully reduced using the Libre-Esprit pipeline\footnote{Libre-ESpRIT 
is a self-contained data reduction package developed specifically for reducing the \esp\
echelle spectropolarimetric data  developed by \citet{Donati1997}}.
This included subtraction of a bias and dark frames, flat fielding for pixel to pixel variations, 
and masking bad pixels.  \esp\ records 40 orders, each one of them curved, such that Libre-ESpRIT 
performs a geometric analysis from the calibration exposures before it performs an optimal extraction.  
It also corrects for the tilt of the slit, determines the wavelength calibration from a thorium lamp 
exposure, and applies the heliocentric correction. The "star+sky" mode enables good sky subtraction
during the pipeline reductions. 
The final (combined) spectra were renormalized using an asymmetric k-sigma clipping routine. 

As this is an exploratory program, spectra were collected until signal-to-noise SNR$>$30 
near 520 nm was reached per target;  multiple exposures were coadded for fainter targets to 
reach this SNR.
A full sample spectrum for one metal-poor target is shown in Fig.~\ref{fig:p7full},
where it can be seen the SNR worsens at shorter wavelengths. 
In addition, the red side of the CCD detector in this cross-dispersed echelle spectrograph 
is less illuminated than the centre of each order, causing lower SNR in the interorder regions. 
Overall, this impacts the smoothness of the spectra.   
Spectral lines in the low SNR regions were rejected from this analysis.  
In total, this observing campaign used over 150 hours of CFHT \esp\ time.

\begin{figure}
\includegraphics[scale=0.4]{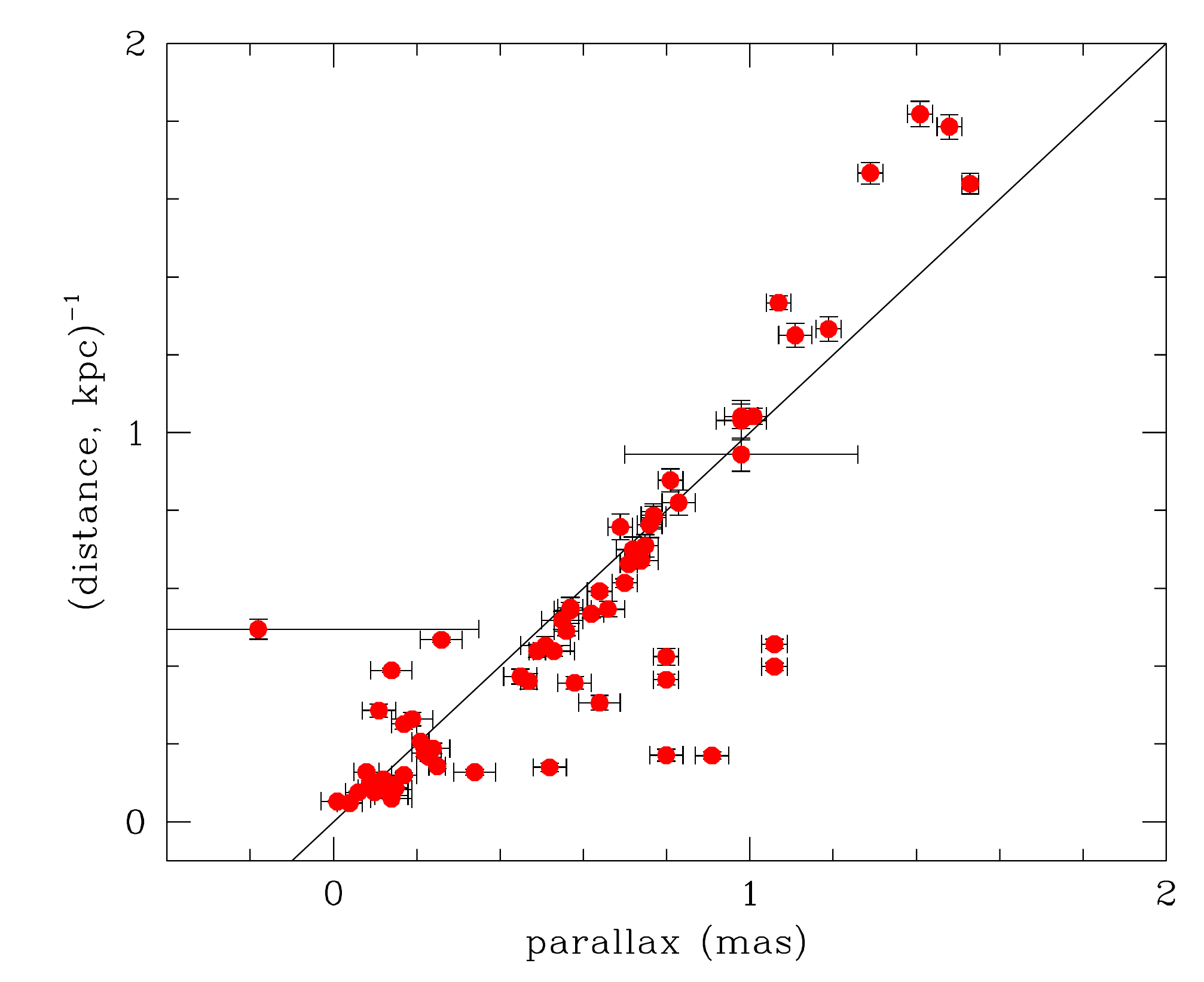}
\caption{A comparison of the Gaia DR2 parallax measurements (with zero point correction,
see text) and 1/(distance, in kpc) from the Bayesian inference method developed by
\citet{Sestito2019} for our 70 metal-poor candidates. 
}
\label{fig:parallax}
\end{figure}

Radial velocities (RV; see Tab.~\ref{tab:newfs}) were determined by fitting several strong
lines per star, and averaging the results from the individual lines together. 
This method was selected rather than a more rigorous use of a cross correlation
technique (e.g., IRAF/{\sl fxcorr}) because of slight wavelength solution 
variations for lines in common between orders and the significant noise in the
interorder regions.  
The typical uncertainty in RV is $\sigma$RV $\le$ 0.5~\kms\ for lines below 6000~\AA.
Variations between the RV solutions were noticed between the CaT lines ($\sim$8500~\AA) 
vs lines in the blue (below 6000~\AA), ranging from 0 to 3~\kms.   
A similar offset was seen in CFHT \esp\ spectra for CEMP stars by 
\citet{Arentsen2019}, who showed that the RV derived from lines below 6000~\AA\
provide better agreement with radial velocity standards.  
Therefore, we did not use any lines above 6000~\AA\ for the RV measurements.
The variations for common lines in overlapping orders was
small (1-2 pixels, or $\le$0.8~\AA\ per line); when averaged over several
lines ($>$10) this intrinsic variation corresponds to $\le$0.5~\kms, the
RV uncertainty that we adopt for all of our spectra.
Multiple observations were spaced over a narrow range in time,
so that no RV variability information is available 
for identifying potential binary systems.

\section{Spectroscopic Analysis}

The analysis of stellar spectra requires a comparison with synthetic spectral
calculations of the radiative transfer through a model atmosphere.   
In this paper, we adopt the ATLAS12 \citep{atlas12} and
MARCS \citep[][further expanded by B. Plez]{marcs}
1-D, hydrostatic, plane parallel models, in local thermodynamic equilibrium.  
These models are represented by an effective temperature (\teff), surface gravity (log g),
mean metallicity (represented as the iron abundance, [Fe/H]).  The model atmospheres are
generated with scaled solar abundances, but increased $\alpha$ element abundances to represent
the majority of metal-poor stars in the Galaxy ([$\alpha$/Fe]=0.0 to +0.4).   
Microturbulence ($\xi$) is assumed to scale with gravity, using the scaling relations 
by \citet{Sitnova2015} and 
\citet{Mashonkina2017} for Galactic metal-poor dwarfs and giants, respectively.

Initial stellar parameters (temperature and metallicity) were determined from photometry.  
A colour temperature was determined from the SDSS $gri$ colours and the semi-empirical 
calibrations from \citet{GH2009}, and metallicity was determined from the SDSS $gri$ photometry
with the \Pristine\ CaH\&K filter, with calibrations described by \citet{Starkenburg2017}.
Our targets range in colour temperature (=T$_{\rm SDSS}$) 
from 4700 to 6700~K, and have \Pristine\ metallicities [Fe/H]$_{\rm Pristine} \lesssim -2.5$; 
see Tab.~\ref{tab:targets}.

\begin{figure*}
\includegraphics[width=1\textwidth, scale=0.99]{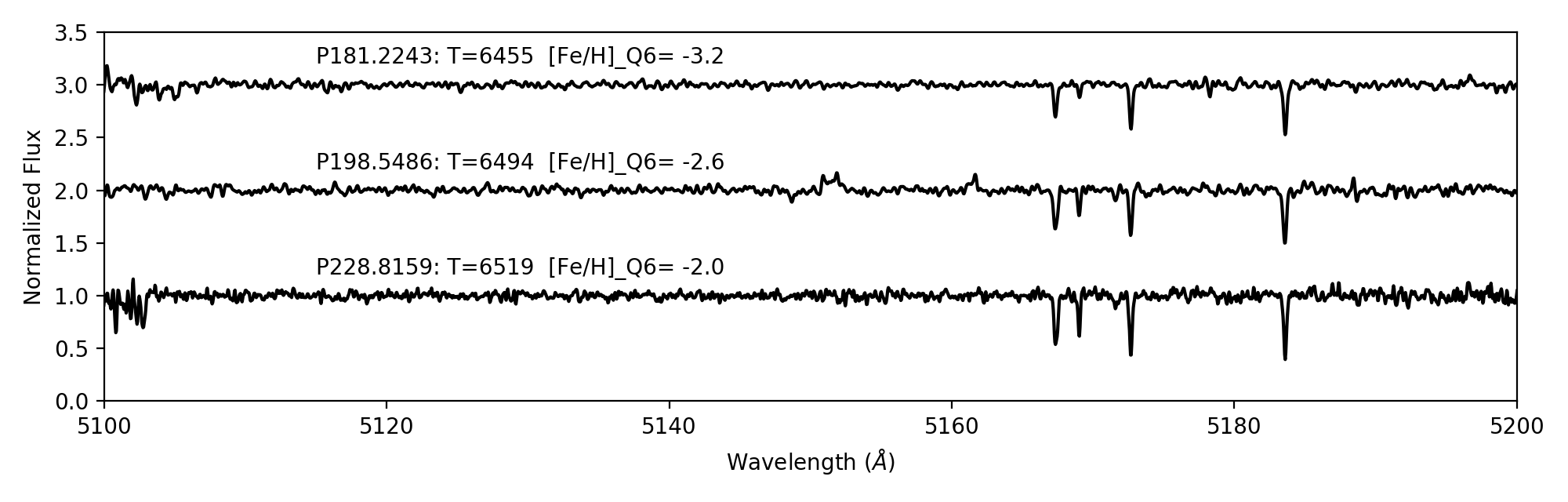}
\includegraphics[width=1\textwidth, scale=0.99]{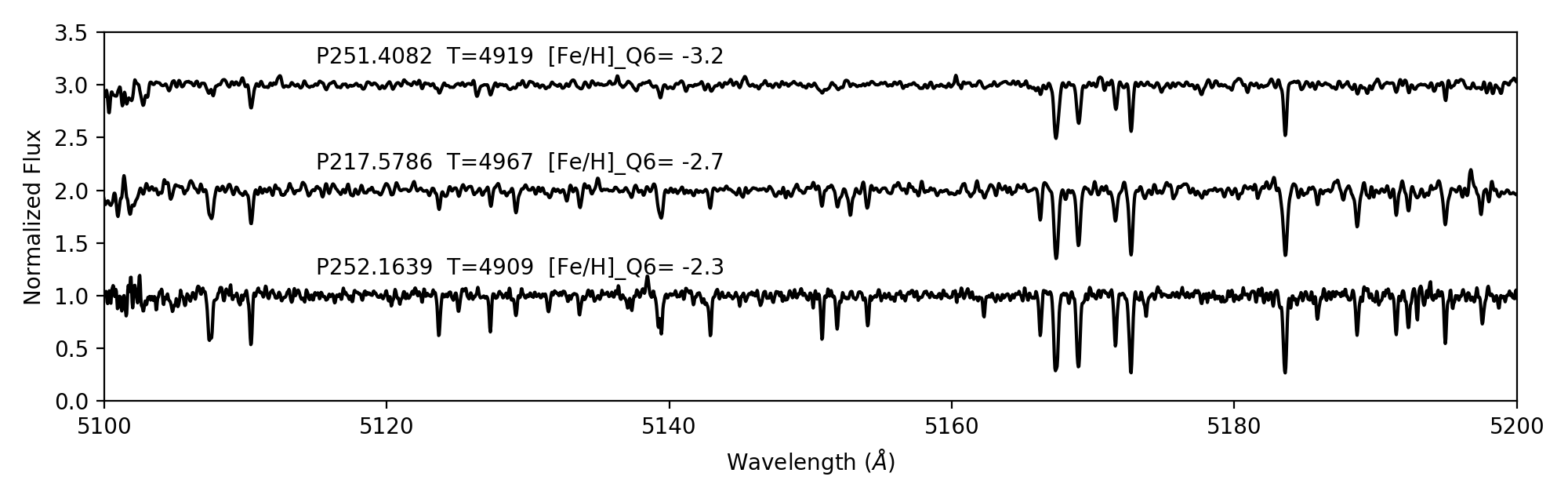}
\caption{Sample CFHT \esp\ spectra for three hot (T$\sim$6500~K)
main sequence turn-off stars and three cool (T$\sim$4900~K) red giants.  
Each spectrum is labelled with
the target name, temperature from the Bayesian inference method,
and spectroscopic metallicity from our "Quick Six" analysis.  }
\label{fig:hotplot}
\end{figure*}

\subsection{Stellar parameters using SDSS and Gaia DR2 data, and MIST isochrones
("Bayesian inference" method) \label{sect:Bayesian}}

Improved stellar temperatures and the gravity estimates were determined 
using a "Bayesian inference" method developed by \citet{Sestito2019}. 
A probability distribution function of the heliocentric distance to each star 
was inferred by combining the SDSS photometric colours and 
Gaia DR2 parallaxes data, with stellar isochrones, and a Milky Way 
stellar density prior.
We apply the zero point offset on the parallax of $-0.029$ mas 
recommended by \citet{Lindegren2018}, but note that the Gaia 
team have discussed the possbilitiy of spatially correlated parallax
errors ranging from 0.1 to 0.01 mas; see discussion by \citealt{Zinn2019}.  
Isochrones are from the MESA/MIST library 
\citep{Paxton2011, Dotter2016, Choi2016}, which 
reach the lowest metallicities ([Fe/H] $\le -4$); see Fig~\ref{fig:hrdmist}.  
A flat age prior
was assumed between ~11 and 14 Gyr (or log(A/yr)$=10.05$ to $10.15$),
and we adopted the \Pristine\ metallicities a priori.

Unique solutions for the stellar parameters were found for 85 of our targets
(out of 89 stars; the 88 stars that remain in the \Pristine\ survey catalogue
after photometric quality cuts, and 
one star that we have retained, see Section~\ref{sect:selection}).
Another four stars have sufficiently large parallax errors that we could not 
distinguish between the dwarf or giant solutions; both are given in 
Tab.~\ref{tab:newfs}.  
It is recognized that determining the distance to a star simply
by inverting the parallax measurement can lead to substantial errors,
especially when the parallax is small (or even negative), and when
there is a relatively large measurement uncertainties (e.g., $\Delta\pi$/$\pi >0.1$);
see \cite{bailer2018}.  
One advantage of this Bayesian inference method is that stars with negative 
parallax results and stars with very large parallax errors can be placed onto 
the isochrones and assumed to be distant. 
In Fig.~\ref{fig:parallax}, 
the Bayesian inferred distances are compared to the Gaia DR2 parallax measurements.

For two stars (Pristine\_200.5298+08.9768 and Pristine\_187.9785+08.7294), 
the Bayesian inferred distance method seemed to fail, placing these stars 
in the outer Galactic halo, even though they have relatively large parallax 
measurements with small uncertainties (0.46 $\pm0.04$ mas and 0.74 $\pm0.04$ mas
in the Gaia DR2 catalogue), and they are metal-rich 
(e.g., visibly strong \ion{Ca}{II} triplet lines).  
Since we had assumed these stars are metal-poor (from their \Pristine\ 
metallicities), then the metal-poor isochrone used to compute their
distances was incorrect, and resulted in a poor distance estimate.
By adjusting their distances to simply 1/parallax (i.e., not using the 
metal-poor stellar isochrones), then both of these stars are located 
closer to the Sun, consistent with the majority of metal-rich stars in the Galaxy.
For our main targets, stars that the \Pristine\ survey identifies as 
metal-poor and that are truly metal-poor, then this will not be a problem,
and we expect this Bayesian inference method will provide 
very precise stellar parameters.

\subsection {Initial ("Quick Six") spectroscopic metallicities \label{sect:quick6}}

Adopting the stellar parameters from the Bayesian inference method described 
above (Section~\ref{sect:Bayesian}), 
then a model atmosphere was generated from both the MARCS and ATLAS grids.
Elemental abundances were computed using a recent version of the 1D LTE 
spectrum analysis code MOOG \citep{Sneden1973, Sobeck2011}.

As an initial spectroscopic metallicity estimate,
a subset of six iron lines were selected that are observable in the good SNR 
regions of the \esp\ spectra;  
4x \ion{Fe}{I} ($\lambda$4957.6, $\lambda$5269.5, $\lambda$5371.5, $\lambda$5397.1) 
and 2x \ion{Fe}{II} ($\lambda$4923.9, $\lambda$5018.4).  
These are well-known and fairly isolated spectral 
lines, with good atomic data\footnote{Atomic data for the \ion{Fe}{I} lines
are from \citet{Blackwell1979} with high precision, or from the
laboratory measurements from \citet{OBrian1991}.  The \ion{Fe}{II} lines
have less certain atomic data from \citet{Raassen1998}, however a NLTE
investigation by \citet{Sitnova2015} showed that these lines have tiny NLTE
corrections {\sl and} yield iron abundances in metal-poor stars within 0.1~dex 
of all other \ion{Fe}{I} and \ion{Fe}{II} lines that they studied.
We also note \citet{Roederer2018b} used 3 \ion{Fe}{I} and 1 \ion{Fe}{II}
of these lines in their detailed iron analysis of six warm metal-poor stars.}  
and line strengths across the parameter range. 
The equivalent widths of
these six lines were measured using IRAF/$splot$\footnote{IRAF is distributed 
by the National Optical Astronomy Observatories, which is operated by the 
Association of Universities for Research in Astronomy, Inc. (AURA) 
under cooperative agreement with the National Science Foundation}, measuring both the area
under the continuum and by fitting a Gaussian profile, comparing the results.
We call the average of these six LTE line abundances 
our "Quick Six" spectroscopic metallicities ([Fe/H]$_{\rm Q6}$),
and these are used as an initial test of the \Pristine\ metallicity estimates.
 
Sample spectra are shown for six targets; three hot (T$\sim$6500~K),  
main sequence turn-off stars and three cool (T$\sim$4900~K) red
giants in Fig.~\ref{fig:hotplot}.   These spectra are labelled with their target name,
temperature (from the Bayesian inference method, see Section~\ref{sect:Bayesian}), and 
metallicity [Fe/H]$_{\rm Q6}$ from this "Quick Six" analysis.

Departures from LTE are known to overionize the \ion{Fe}{I} atoms due to the impact
of the stellar radiation field, particularly in hotter stars and metal-poor giants.  
These non-LTE (NLTE) effects can be significant in our stellar parameter
range, such that NLTE corrections typically reduce the line scatter 
and improve the \ion{Fe}{I}=\ion{Fe}{II} ionization balance
\citep{Amarsi2016, Sitnova2015, Mashonkina2019}. 
NLTE effects are explored in this "Quick Six" analysis, by
comparing the results from \citet{Mashonkina2017, Mashonkina2019} and 
the INSPECT table\footnote{INSPECT non-LTE corrections available at http://inspect-stars.com.} 
\citep{Amarsi2016, Lind2012}.
INSPECT provides data for one of the selected lines, \ion{Fe}{I} $\lambda$5269, 
where the NLTE correction is  $\Delta$(\ion{Fe}{I}) $\le 0.15$, 
over our parameter space, where \ion{Fe}{I}(NLTE) = \ion{Fe}{I}(LTE) + $\Delta$(\ion{Fe}{I}).   
Based on a similar treatment for inelastic collisions (of \ion{Fe}{I} with \ion{H}{I}),
\citet{Mashonkina2017} predict similar NLTE corrections for the other three 
\ion{Fe}{I} lines ($\lambda$4957, $\lambda$5372 and $\lambda$5397).
The largest NLTE corrections ($\Delta$(\ion{Fe}{I}) $\sim 0.3$) 
are for stars on the subgiant branch,
while main sequence stars have $\sim$zero corrections.  
Recently, \citet{Mashonkina2019} have examined the impact of quantum 
mechanical rate coefficients on the inelastic collisions, and find that
the NLTE corrections could be even larger (more positive) in the atmospheres 
of warm metal-poor stars, but smaller (even negative) in cool metal-poor
stars and with a wide variation depending on the specific spectral line. 
This suggests that the NLTE calculations for \ion{Fe}{I} need further 
study; however, given that these corrections in the literature are smaller 
than or equal to our measurement errors, then we do not apply the NLTE 
corrections in this "Quick Six" analysis.

\begin{figure}
\includegraphics[scale=0.43]{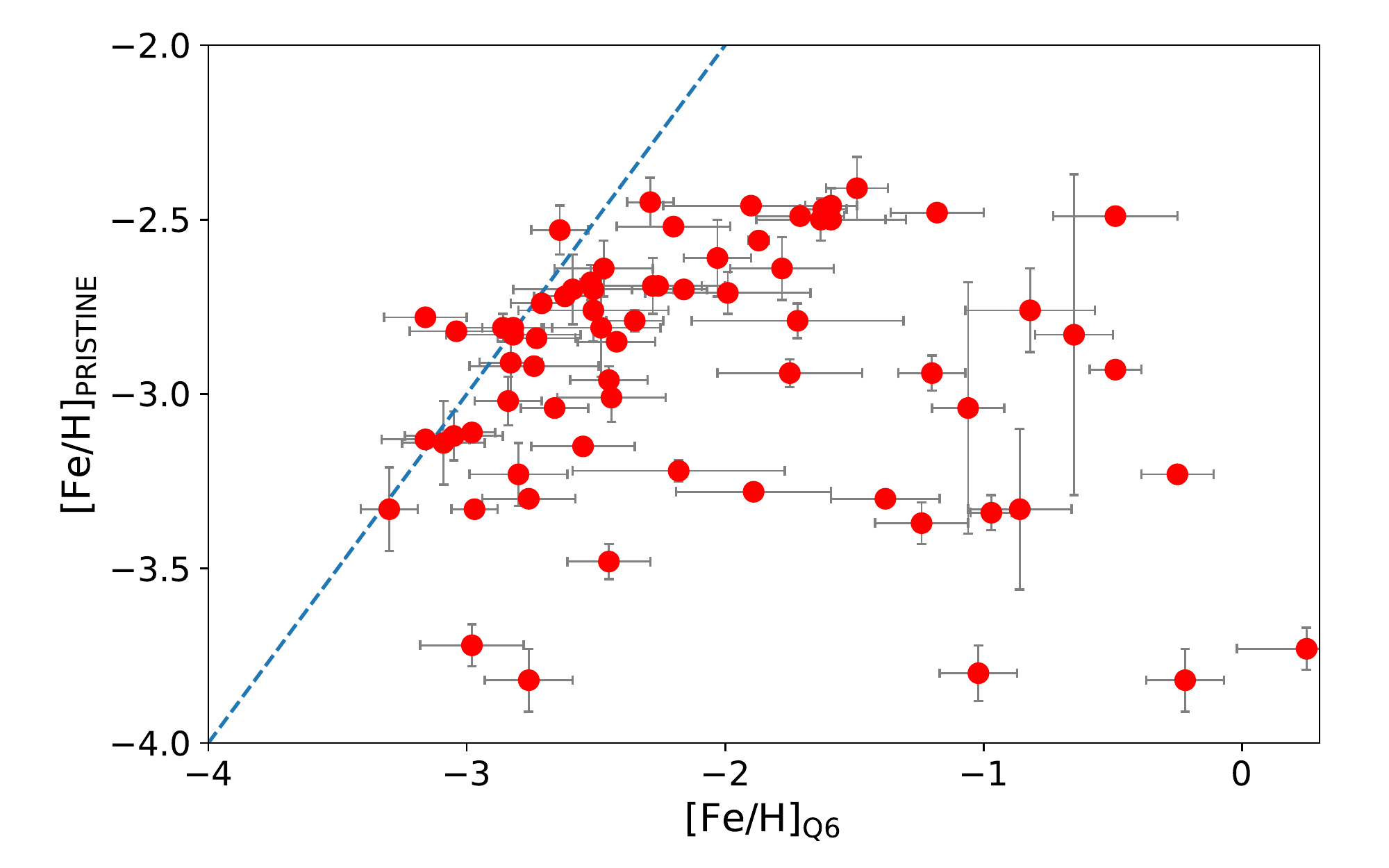}
\caption{Comparisons of the "Quick Six" [Fe/H]$_{\rm Q6}$ spectral abundances
compared with the \Pristine\ [Fe/H]$_{\rm Pristine}$ photometric predictions.
Clearly some of the \Pristine\ metal-poor candidates are not
metal-poor stars.
}
\label{fig:ironPQ}
\end{figure}

The \ion{Fe}{I} and \ion{Fe}{II} individual line abundances are averaged together
to find [Fe/H]$_{\rm Q6}$ and the standard deviation $\sigma$[Fe/H]$_{\rm Q6}$. 
Each of these results and the total number of lines used ($\le$6) 
are shown in Table~\ref{tab:newfs}. 
From this analysis, we find that several of the \Pristine\ metal-poor candidates 
are not metal-poor.  A comparison of the [Fe/H]$_{\rm Q6}$ iron abundances to the
[Fe/H]$_{\rm Pristine}$ predictions are shown in Fig.~\ref{fig:ironPQ}.  
These results are similar to the medium resolution spectral analyses
\citep{Youakim2017, Aguado2019}, and discussed further in Section~\ref{sect:ferre}.

\subsection{Comparing stellar temperatures  }

\begin{figure}
\hspace*{-0.2cm}
\includegraphics[scale=0.43]{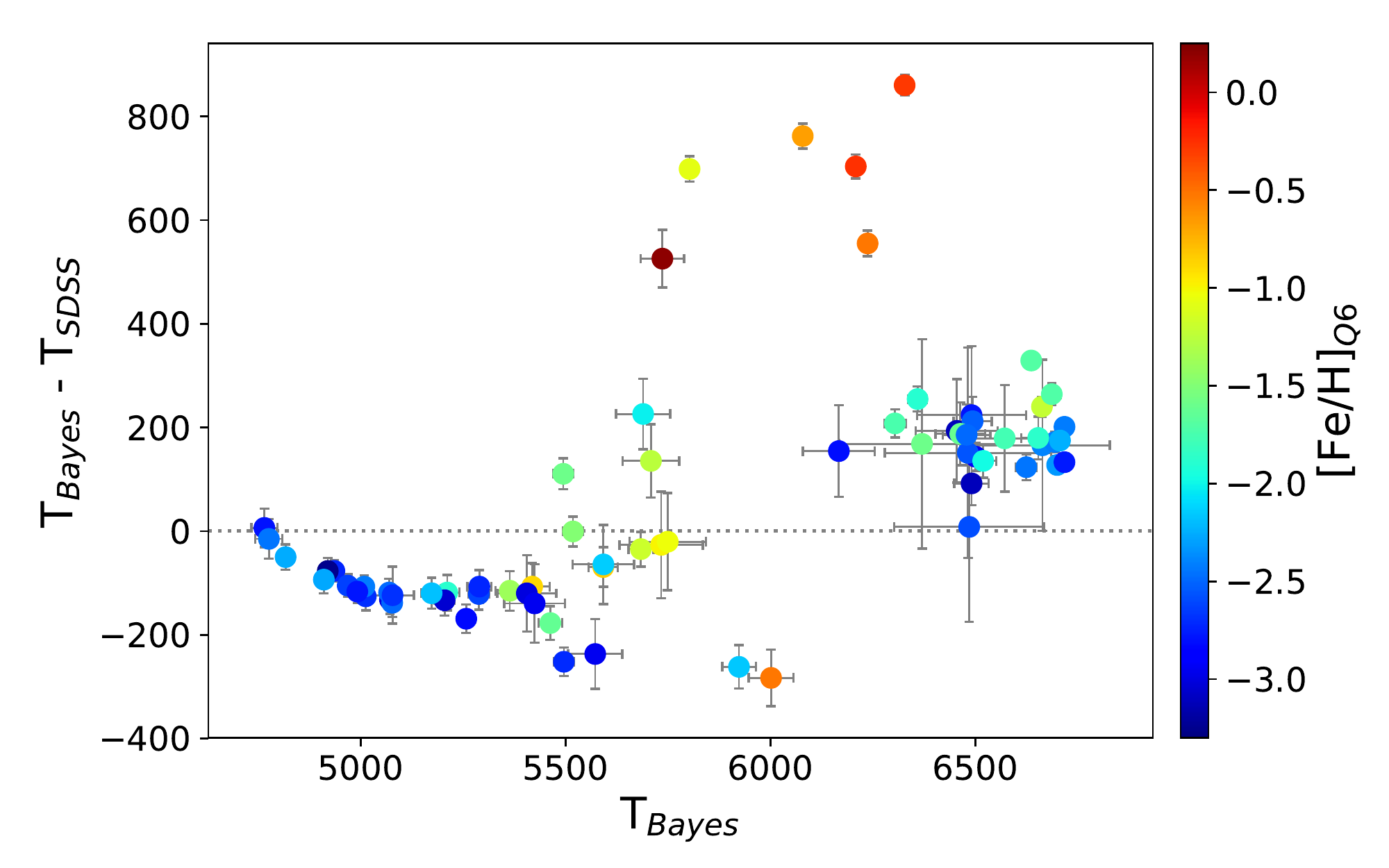}
\caption{Comparisons of the \Pristine\ survey colour temperature (T$_{\rm SDSS}$) and the effective
temperature determined from Bayesian inference method (T$_{\rm Bayes}$) for our 70 metal-poor candidates.
The data points are coloured by their metallicities from our spectroscopic [Fe/H]$_{\rm Q6}$ analysis.
As both temperature estimates adopte the \Pristine\ photometric metallicity estimates 
[Fe/H]$_{\rm Pristine}$ a priori, then clearly the metal-rich stars are not well calibrated.
}
\label{fig:Temp}
\end{figure}

A comparison of stellar temperatures from the Bayesian inference method 
(Section~\ref{sect:Bayesian}) to the SDSS colour temperature (T$_{\rm SDSS}$) 
is shown in Fig.~\ref{fig:Temp}.
T$_{\rm SDSS}$ were the initial temperature estimates calculated using the InfraRed 
Flux Method\footnote{IRFM, see https://www.sdss. org/dr12/spectro/sspp\_irfm/.}, 
assuming [Fe/H] $=-2.5$, and based on SDSS (g-i) photometry. 
An average of the dwarf and giant solutions was used.  
For 10 stars, their ($g-i$) colours are unreliable because of saturation 
flags, and we adopt the relation based on the ($g-r$) colours from 
\citet{Ivezic2008}.  With this relation, a 200~K offset was applied to move 
from [Fe/H] $= -0.5$ to [Fe/H] $= -2$.  
Thus, we expect these values of T$_{\rm SDSS}$ to be an oversimplification, 
and are not surprised by the comparisons in 
Fig.~\ref{fig:Temp}, which are colour-coded by the "Quick Six" 
metallicities [Fe/H]$_{\rm Q6}$.  

Ignoring the metal-rich stars, then there is still a systematic offset 
between these methods for the metal-poor stars: the 
T$_{\rm SDSS}$ colour temperatures are too hot by $\sim$150~K for stars 
between T = $4700 - 5700$~K, but they are too cool by $\sim$200~K for stars 
with T $> 6000$~K.
This offset is similar to the uncertainties in the Bayesian inference 
method temperatures (T$_{\rm Bayes}$) for most stars, where 
$\sigma$T$_{\rm Bayes}$ ranges from $\sim$10 to 200~K (Table~\ref{tab:newfs}).
The very small colour temperature errors dT$_{\rm SDSS} \lesssim$10~K in 
Table~\ref{tab:targets} are based on the difference between the dwarf/giant solutions, 
and are not realistic uncertainties.

\subsection{Comparing gravity and \ion{Fe}{I}=\ion{Fe}{II} }

Ionization balance has traditionally been used as an indicator of surface gravity in 
a classical model atmospheres analyses.   Therefore, we compare the \logg values from the 
Bayesian inference method (Section~\ref{sect:Bayesian}) to the difference 
in the [\ion{Fe}{I}] and [\ion{Fe}{II}] abundances, in Fig.~\ref{fig:GravFe}.  
This figure is colour-coded by the "Quick Six" spectroscopic metallicities ([Fe/H]$_{\rm Q6}$).   
For the metal-poor stars, the majority of our stars show \ion{Fe}{I}=\ion{Fe}{II} to within 
2$\sigma$ of the measurement errors, with a mean offset of [\ion{Fe}{I}] $-$ [\ion{Fe}{II}]
$= +0.2$.  The measurement errors are calculated as the line weighted average of 
\ion{Fe}{I} and \ion{Fe}{II}.   

For stars with poor agreement between iron ionization states, the cause cannot be due
to neglected NLTE effects which appear to increase the \ion{Fe}{I} abundance even further
(see in Section~\ref{sect:quick6}).  The offset is primarily seen in the cooler stars 
in our sample that are on the red giant branch (with lower gravities).  For these stars,
the NLTE corrections are expected to be small ($\Delta$(\ion{Fe}{I})$\lesssim$0.15).  
For stars closer to the main sequence turn off, the NLTE corrections can be larger;
however, the offset between the \ion{Fe}{I} and \ion{Fe}{II} abundances seems smaller
for those stars in our results. 
Therefore, the source of ionization equilibrium offsets is not yet clear.
 
For the metal-rich stars, we expect the surface gravities to be unreliable since 
the photometric \Pristine\ metallicities [Fe/H]$_{\rm Pristine}$ were assumed 
a priori in the Bayesian inference method.  
We do not investigate the metal-rich stars beyond our "Quick Six" analysis.

\subsection{Comparisons with medium resolution spectroscopic analyses (FERRE) \label{sect:ferre}}

A simultaneous \Pristine\ survey program has been carried out for fainter stars (15 $<$ V $<$ 17) 
with medium resolution (R $\sim 1800$) spectroscopy at the 2.4-m Isaac Newton Telescope (INT), 
4.2-m William Herschel Telescope, and 3.6-m New Technology Telescope \citep{Aguado2019}.   
These spectra have been observed with uniform spectral wavelength coverage, 360-550 nm, 
and analysed using the ASSET synthetic spectral grid \citep{Koesterke2008}.  
Both the observed and the synethetic 
spectra have been continuum normalized with the same functions, and the 
$\chi^2$ minimization algorithm FERRE \citep{AllendePrieto2006} is applied to
derive the stellar parameters (temperature, gravity, metallicity, carbonacity). 

\begin{figure}
\includegraphics[scale=0.42]{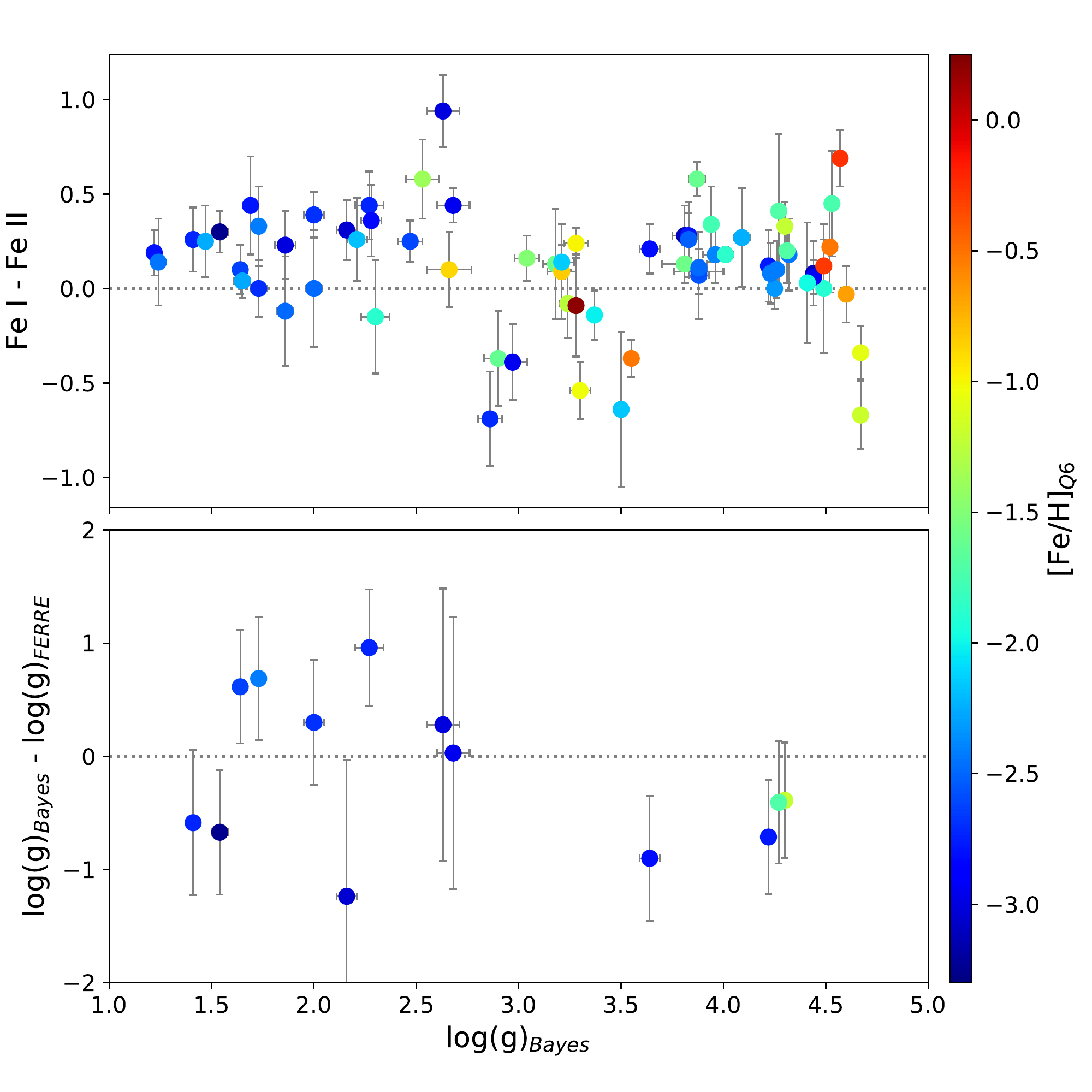}
\caption{ Comparisons of the surface gravities and iron ionization balance 
estimates for our 70 metal-poor candidates from the \Pristine\ survey (top panel), 
and comparisons of our surface gravities versus those from the FERRE analysis of medium
resolution spectra \citep{Aguado2019} for 13 stars in common.   
The uncertainties in the gravities from FERRE can be quite large for the metal-poor stars 
due to a lack of suitable spectral signatures.
The data points are coloured by their metallicities from 
our spectroscopic [Fe/H]$_{\rm Q6}$ analysis.   
}
\label{fig:GravFe}
\end{figure}

\begin{figure}
\includegraphics[scale=0.42]{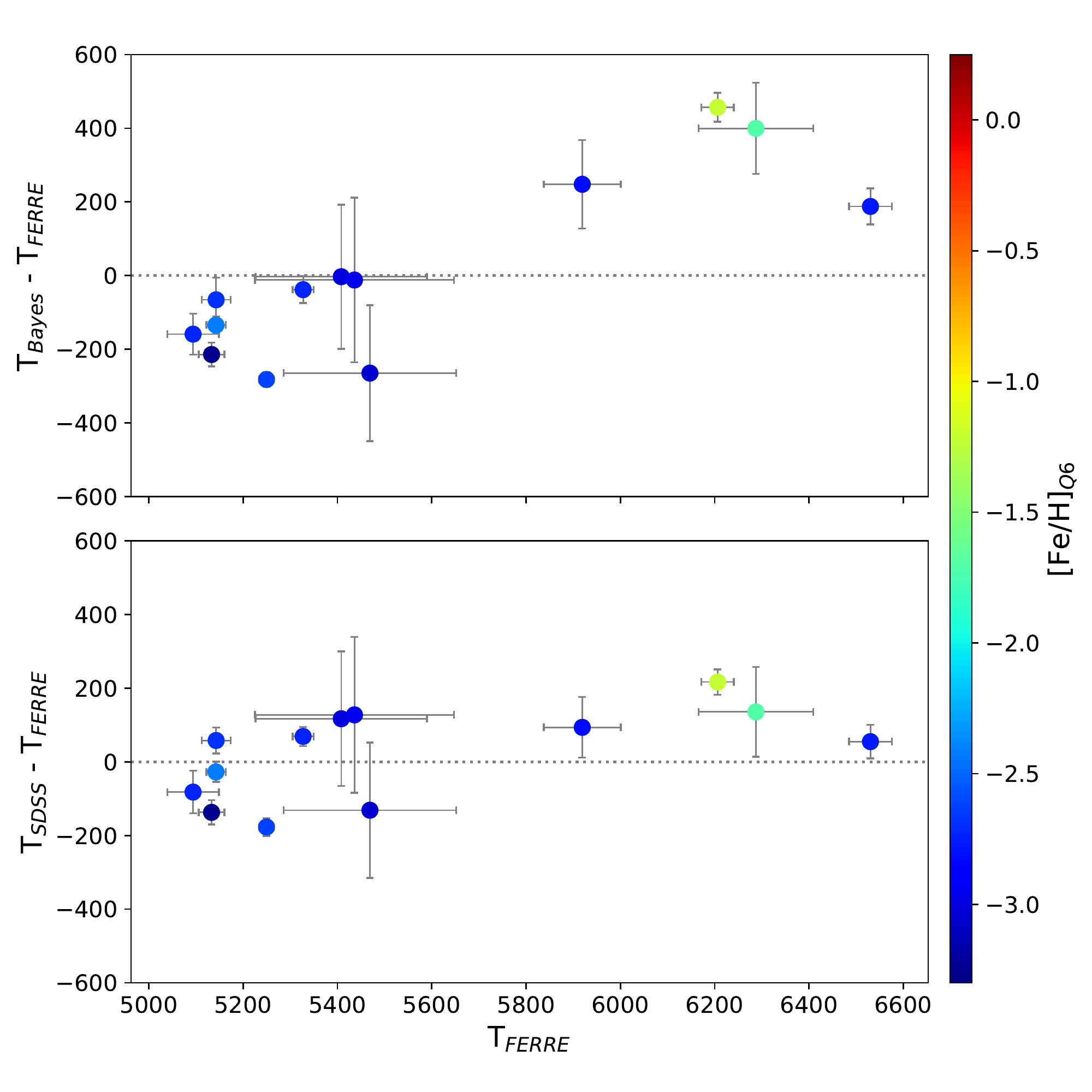}
\caption{Temperature comparisons for 13 stars in common between the Bayesian inference 
analysis of our CFHT \esp\ spectra and the FERRE analysis of medium resolution spectra 
\citep[top panel, ][]{Aguado2019}.   The temperature offsets are slightly smaller when compared
with the $\Pristine$ colour temperatures (T$_{\rm SDSS}$, bottom panel).
 } 
\label{fig:Tferre}
\end{figure}

\begin{figure}
\includegraphics[scale=0.42]{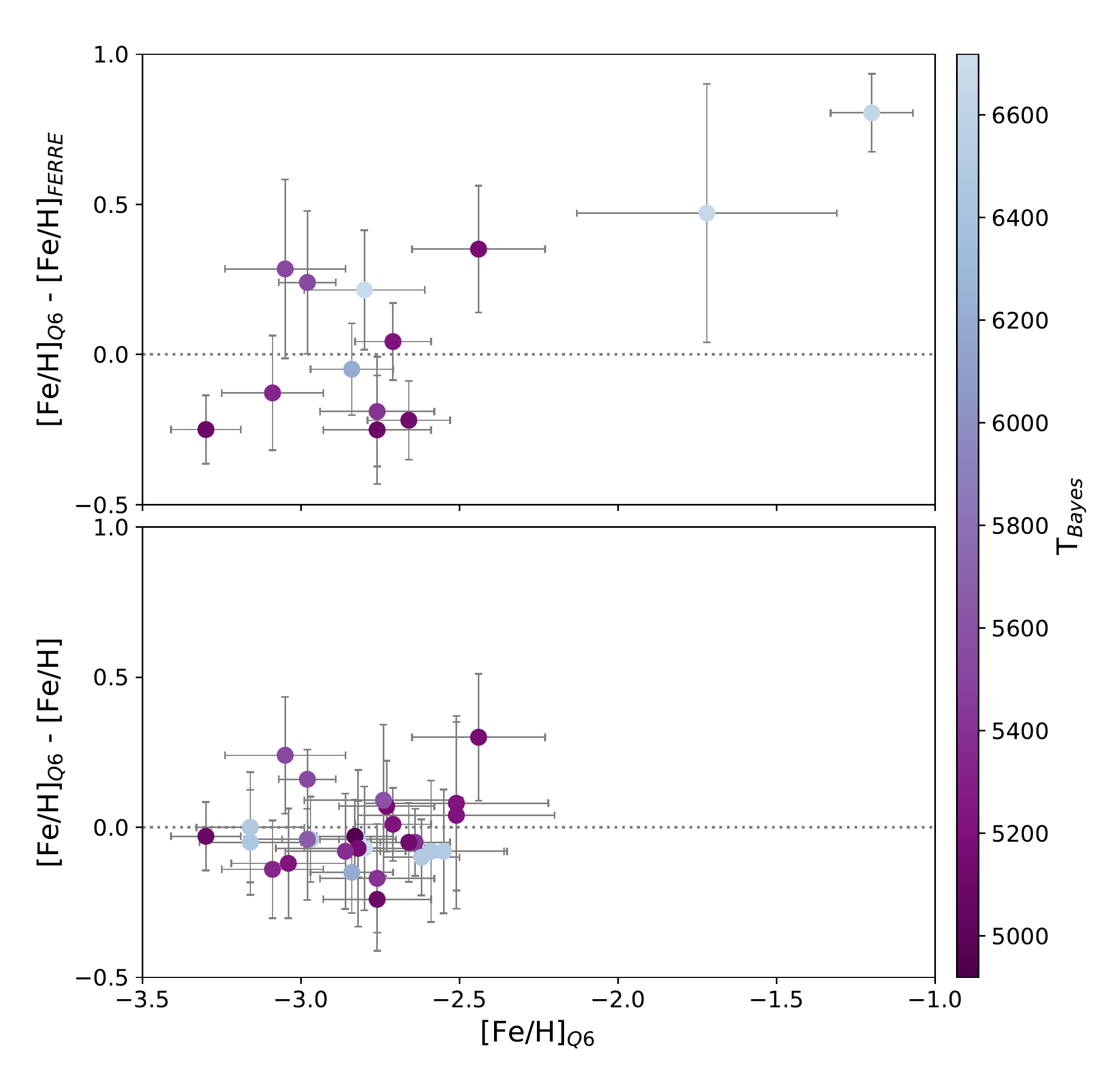}
\caption{Metallicity comparisons for 13 stars in common between the [Fe/H]$_{\rm Q6}$ 
analysis of our CFHT \esp\ spectra and the FERRE analysis of medium resolution spectra 
\citep[top panel, ][]{Aguado2019}. 
[Fe/H]$_{\rm Q6}$ values are also compared to the improved [Fe/H] values for our 28 
very metal-poor stars, which include more lines of both \ion{Fe}{I} and \ion{Fe}{II}.  
The errors in the bottom panel are dominated by the "Quick Six" $\sigma$[Fe/H]$_{\rm Q6}$.
analysis.
 } 
\label{fig:Feferre}
\end{figure}

The most recent analysis of the medium resolution spectroscopic data 
includes 946 stars \citep{Aguado2019}, where 13 of those stars are also 
in our sample of 70 high probability metal-poor stars 
(recall, that only 5 were observed at the INT first, and did not 
affect our target selections). 
In Figs.~\ref{fig:GravFe}, \ref{fig:Tferre}, and \ref{fig:Feferre}, 
the surface gravities, temperatures, and metallicities 
are compared between the two analyses for the 13 stars in common.   
The large differences in gravity are from the {\it systematic} 
errors in the medium resolution FERRE analysis.  While the FERRE analysis
struggles with precision gravities, both methods are still able to
break the dwarf-giant degeneracies sufficiently.

There is a clear relationship between the temperatures such that those 
determined from isochrones in the Bayesian inference method are cooler by 
$\sim$200\,K near 5000~K and hotter by $\sim$500\,K near 6700~K compared
to the FERRE temperatures.  
These offsets are slightly smaller when compared with the 
SDSS colour temperatures T$_{\rm SDSS}$.
These temperature differences correlate with small-to-moderate metallicity offsets 
($\Delta$[Fe/H]$\le0.3$) for stars cooler than 6000~K, whereas two of the hotter stars 
show larger metallicity offsets, $\Delta$[Fe/H] $\sim 0.5$.  In summary, 
this analysis adopts the stellar parameters from the Bayesian inference method,
and finds that the hot stars are hotter and less metal-poor than
the results from the medium resolution FERRE analysis.

\section{New stars with [Fe/H]$\le-2.5$  \label{sect:newstars} }

We have identified 28 new metal-poor stars, with spectroscopic metallicity [Fe/H]$_{\rm Q6} \le -2.5$, 
and where both [\ion{Fe}{I}/H] and [\ion{Fe}{II}/H] are below $-2.5$ dex (with the exception of 
Pristine\_198.5486+11.4123, with [\ion{Fe}{I}/H]$ = -2.42$, which we retain because of its interesting 
orbit, discussed below).   In this section, a more complete LTE, 1D model atmosphere analysis is
carried out for a larger set of spectral lines and chemical elements. 

As a comparison star, a spectrum of HD~122563 from the CFHT archive was analysed 
using the same methods as for the \Pristine\ survey targets.   
Its metallicity is adopted from the literature, i.e., [Fe/H] $= -2.7 \pm 0.1$ 
\citep[see][and references therein]{Collet2018}, 
and our methods using its SDSS colours and Gaia DR2 parallax measurements
yield stellar parameters that are in good agreement with the literature:
\teff = $4625 \pm50$~K, \logg = $1.6 \pm0.1$.  Microturbulence ($\xi$) was set to 2.0
\kms\ using the relationship with gravity from \citet{Mashonkina2017}.

\begin{figure}
\includegraphics[scale=0.4]{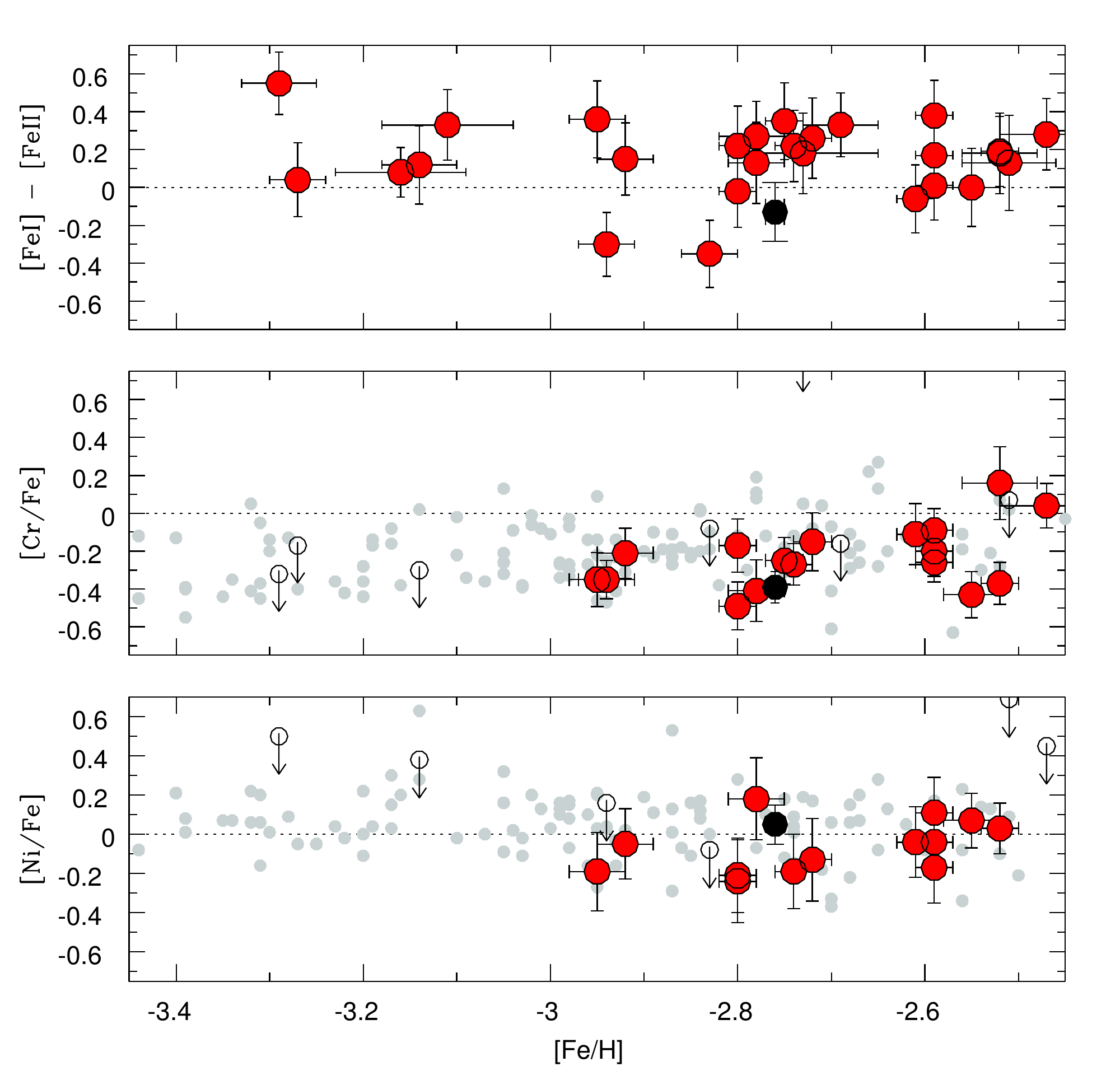}
\caption{Iron-group (Fe, Cr, Ni) abundances and upper limits in our 28 new 
very metal-poor stars ([Fe/H] $< -2.5$, red points).  
Analysis results of the CFHT \esp\ spectrum for the  standard star 
HD~122563 are included (black point).
Errorbars are the measurement errors and systematic errors combined in quadrature. 
Galactic comparison stars are included from the homogeneous analysis by 
\citet[small grey points]{Yong2013}. 
}
\label{fig:iron}
\end{figure}

For all 28 \Pristine\ survey stars and HD122563, we identify and measure as many clean spectral lines as possible
for a detailed abundance analysis, including more lines of \ion{Fe}{I} and \ion{Fe}{II} for higher 
precision iron abundances (than from the [Fe/H]$_{\rm Q6}$ analysis).   
Starting with the spectral line list from \citet{Norris2017}, spectral features were identified 
and measured using DAOSpec \citep{Stetson2008}, and frequently checked by measuring 
the area under the continuum using IRAF/$splot$.
Atomic data were updated when appropriate by comparing to the {\it linemake} 
atomic and molecular line database\footnote{ {\it linemake} 
contains laboratory atomic data (transition probabilities, hyperfine and
isotopic substructures) published by the Wisconsin Atomic Physics and
the Old Dominion Molecular Physics groups.  These lists and accompanying
line list assembly software have been developed by C. Sneden 
and are curated by V. Placco at https://github.com/vmplacco/linemake.}.
Abundance results from the model atmospheres analysis are compared to the solar 
(photospheric) abundances from \citet{Asplund2009}.

\subsection{Iron-group elements \label{sect:iron} }

The 28 new very metal-poor stars were initially identified from their [Fe/H]$_{\rm Q6}$ 
abundances in Table~\ref{tab:newfs}.  

The iron abundances have been recalculated from 2-86 lines of \ion{Fe}{I}, 2-6 lines of \ion{Fe}{II};
see Table~\ref{tab:iron}.  A 3$\sigma$ minimum equivalent width was used to calculate an upper 
limit for \ion{Fe}{II} for one star.  The line-to-line 
scatter in the \ion{Fe}{I} abundances range from $\sigma$(\ion{Fe}{I}) = 0.12 to 0.24, even when only 
a small number of lines were measured.   This is noteworthy because when other elements have $<4$ lines, 
we adopt the larger of $\sigma$(X) or $\sigma$(\ion{Fe}{I})/$\sqrt(N_X)$ 
as a better representation of their line scatter.

\begin{figure}
\includegraphics[scale=0.4]{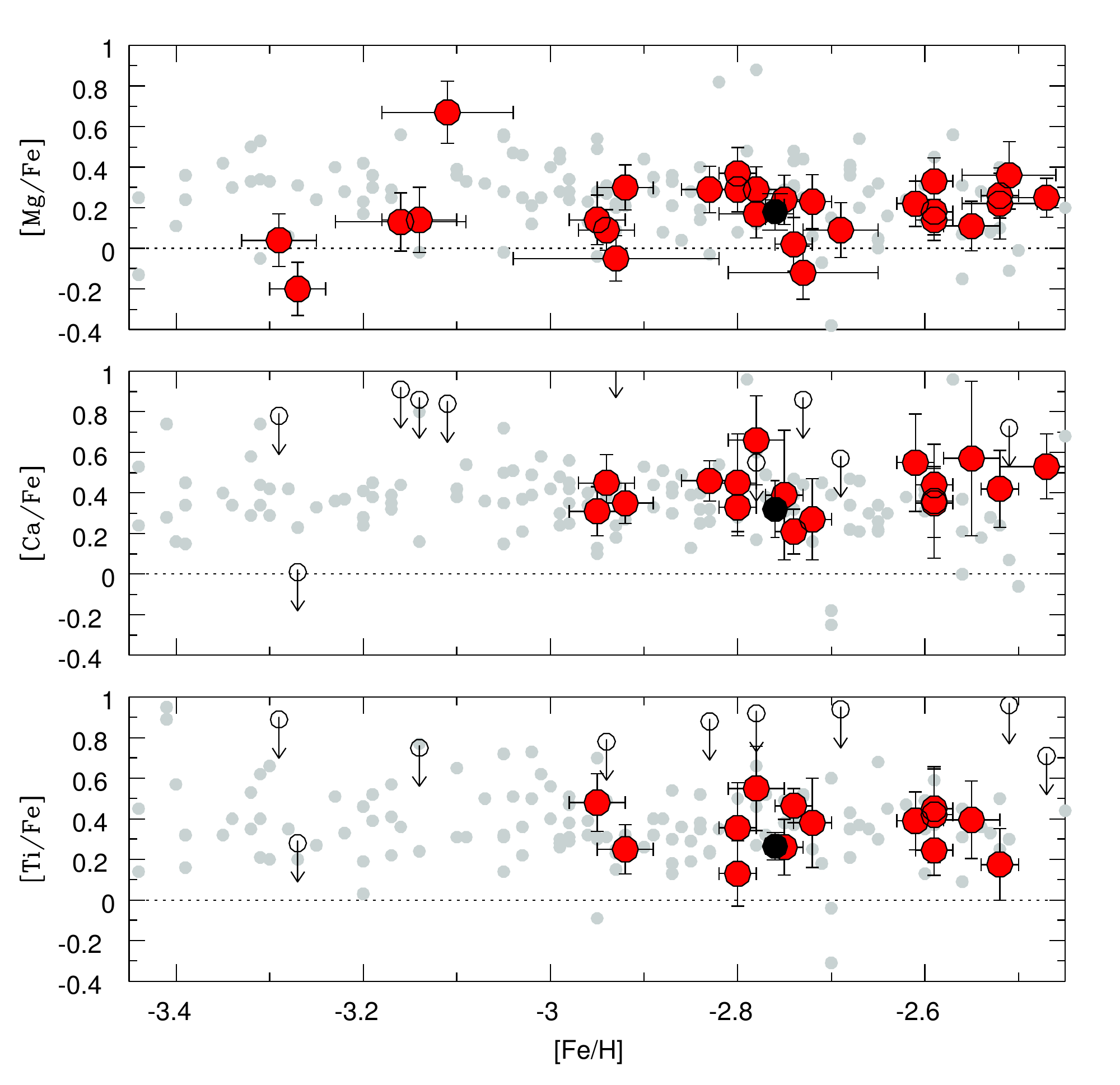}
\caption{ Mg, Ca, and Ti abundances and upper limits in the 28 new metal-poor stars 
([Fe/H] $< -2.5$); symbols the same as in Fig.~\ref{fig:iron}.
}
\label{fig:MgCaTi}
\end{figure}

These extended iron line measurements and abundances are not used to redetermine the spectroscopic 
stellar parameters for three reasons: (1) low sensitivity to the precise metallicity in the
Bayesian inference method for the confirmed metal-poor stars, (2) insufficient number of lines
of \ion{Fe}{II} (and often \ion{Fe}{I}) for a fully independent analysis, and (3) the SNR of our 
CFHT \esp\ spectra ($\le30$) is such that individual measurements of weak lines remain somewhat uncertain.  
The total iron abundance [Fe/H] is calculated as a weighted mean of \ion{Fe}{I} and \ion{Fe}{II},
and the total error $\delta$[Fe/H] as $\sigma$([Fe/H])/(N\ion{Fe}{I}+N\ion{Fe}{II})$^{1/2}$.
These iron abundances are shown in Fig.~\ref{fig:iron} (top panel), where the errorbars include
the systematic errors from the stellar parameter uncertainties added in quadrature 
(see Section~\ref{sect:uncs},
though the systematic errors tend to be much smaller).

There is good to fair agreement between \ion{Fe}{I} and \ion{Fe}{II}, such that 
[\ion{Fe}{I}] $-$ [\ion{Fe}{II}] ranges from $\sim -0.2$ to $+0.2$.  There is a median offset 
$\sim +0.2$ for the sample, which is {\it not} due to NLTE corrections
(see the discussion in Section~\ref{sect:quick6}).  This may be due to the lack of \ion{Fe}{II} lines
in our metal-poor stars spectra for robust measurements, but another possibility is a systematic 
gravity uncertainty $\Delta\logg \sim 0.5$.   High resolution spectra at bluer wavelengths
(4000~\AA) would provide more lines of \ion{Fe}{II} to test this in the future.
We also examine the slopes in the \ion{Fe}{I} line abundances vs excitation potential ($\chi$, in eV)
to test our temperature estimates.  A meaningful slope could be measured when 
N(\ion{Fe}{I})$>15$ and $\Delta\chi>3$ eV, and all slopes were found to be 
relatively flat, $< 0.1$ dex/eV.
This gives us more confidence in the fidelity of the temperatures T$_{\rm Bayes}$, and thereby the
Bayesian inference method for calculating stellar parameters and uncertainties.

The other iron-group elements (Cr and Ni, listed in Table~\ref{tab:iron}) 
are in good agreement with [Fe/H], and/or other Galactic halo
stars at similar metallicities; see Fig.~\ref{fig:iron}.  
Cr is determined from 1-3 lines of \ion{Cr}{I} 
(5206.0, 5208.4 , and 5409.8~\AA);
only the spectrum of Pristine\_245.8356+13.8777 had sufficient SNR at blue wavelengths
that the lines at 4254.3, 4274.8, and 4289.7~\AA\ could also be observed. 
[Cr/Fe] is subsolar in metal-poor stars, suggested as a NLTE effect \citep{Bergemann2010}. 
Ni is determined from 1-2 lines of \ion{Ni}{I} (5035.4, 5476.9~\AA). 
Three additional lines were available in the high SNR spectrum of HD~122563
(5080.5, 6643.6, and 6767.8~\AA).  The [Ni/Fe] results  
are within $1\sigma$ of the solar ratio, similar to other Galactic halo stars.

\begin{figure}
\includegraphics[scale=0.75]{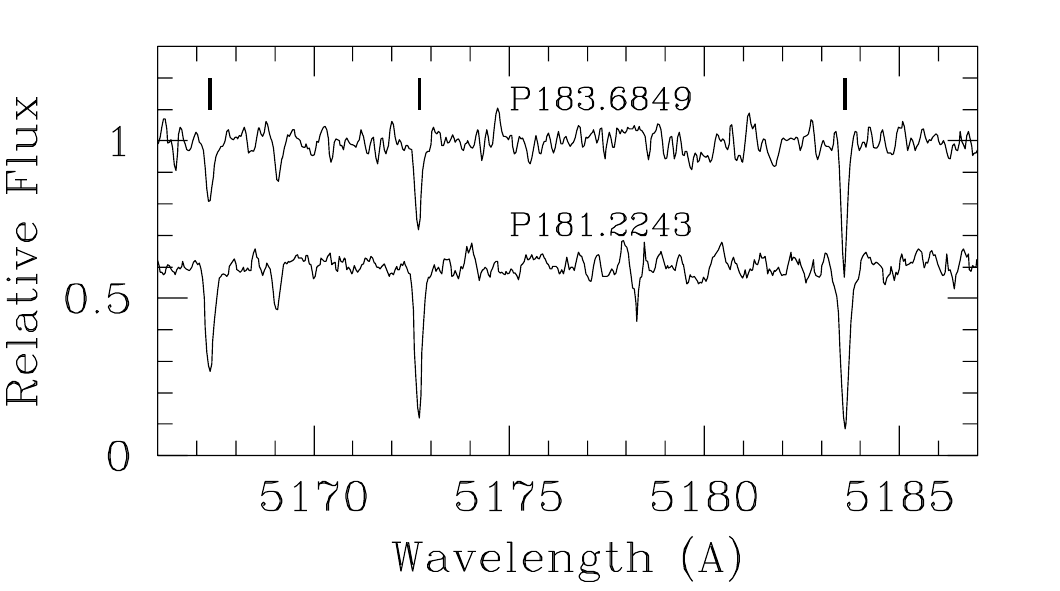}
\caption{ The spectrum of the \ion{Mg}{I} b lines in the Mg strong star, P181.2243.  
This star is compared to Pristine\_183.6849+04.8619, which has very similar stellar parameters (T$\sim$6450,
log(g)$\sim$4, [Fe/H]$\sim$-3.2), but is Mg normal.  
}
\label{fig:mgb}
\end{figure}

The $\alpha$-element abundances (Mg and Ca) in the  28 new very metal-poor 
stars are listed in Table~\ref{tab:alpha}.  Upper limits are determined for some stars by 
computing 3$\sigma$ minimum equivalent widths.  The $\alpha$-elements form through 
hydrostatic H- and He-core burning stages, though some Ca can also form later during SN Ia
events.  Because of these different nucleosynthetic sites, the [Mg/Ca] ratio
need not scale together at all metallicities, as seen in some dwarf galaxies such as  
the Carina and Sextans dwarf galaxies 
\citep[e.g., ][]{Norris2017, Jablonka2015, Venn2012},
also the unusual star cluster NGC~2419 \citep{Cohen2012}.
We also include our discussion of Ti in this section even though it does not
form with the $\alpha$-elements.  The dominant isotope $^{48}$Ti forms 
primarily through Si-burning in massive stars \citep[e.g.,][]{Woosley2002}, 
and yet it seems to scale with other $\alpha$-elements in metal-poor stars 
in the Galaxy.  

Mg is determined from 2-3 lines (5172.7, 5183.6, 5528.4~\AA), and a fourth line (4703.0~\AA)
was measurable in one star (Pristine\_245.8356+13.8777).
In Fig.~\ref{fig:MgCaTi}, a larger scatter can be seen in the [Mg/Fe] results,  
though this is similar to the Galactic comparison stars.
One star shows sub-solar [Mg/Fe] by more than $1\sigma$ (Pristine\_251.4082+12.3657). 
Another star has high [Mg/Fe]$\sim+0.6$, validated from all three \ion{Mg}{I} lines
(Pristine\_181.2243+07.4160), also shown in Fig~\ref{fig:mgb}.

\subsection{ $\alpha$-elements}
The calcium abundances are determined from 1-9 lines of \ion{Ca}{I}.
The [Ca/Fe] abundances are in good agreement with each other, and with the Galactic comparison stars,
as seen in Fig.~\ref{fig:MgCaTi}.
The same star with low [Mg/Fe] (Pristine\_251.4082+12.3657) also has a very low [Ca/Fe] upper limit.
This star is discussed further in Section 6.2.

Titanium has been determined from 1-9 lines of \ion{Ti}{I} and 2-11 lines of \ion{Ti}{II}.  When both
are unavailable, upper limits are determined from two \ion{Ti}{II} lines (which provide stronger 
constraints than the \ion{Ti}{I} features).   
In Fig.~\ref{fig:MgCaTi}, the unweighted average results of [\ion{Ti}{I}/Fe] and [\ion{Ti}{II}/Fe] are shown.
 
NLTE corrections have {\it not} been incorporated for Mg, Ca, or Ti because they tend to be small to
negligible ($\Delta \le 0.1$ dex) 
according to the INSPECT database (for \ion{Mg}{I}) and \citet[][ for \ion{Ca}{I}]{Mashonkina2017b}.
For \ion{Ti}{I}, three lines (4981.7, 4991.1, 4999.5~\AA) are available
in the INSPECT database, which suggests large corrections $\Delta\sim+0.5$ dex.
However NLTE corrections for the same lines from \citet{Sitnova2016}, using a model atom that includes
important high excitation levels of \ion{Ti}{I}, are significantly smaller, $\Delta\sim+0.2$ dex.
NLTE corrections should be included, but most of our stars have \ion{Ti}{I} $\sim$ \ion{Ti}{II} 
to within 1$\sigma$ (our measurement errors) in LTE.  
Therefore, for this analysis, where the maximum SNR per star is $\le30$, we do not include
the small NLTE corrections, and note that the good agreement with the Galactic comparison stars
and Ti ionization balance furthers our 
confidence in the stellar parameters from the Bayesian inference method.

No oxygen abundances or upper limits were determined since the [OI] 6300 and 6363~\AA\ lines 
are weak and in a region that is poorly cleaned of telluric contaminants.

\begin{figure}
\includegraphics[scale=0.4]{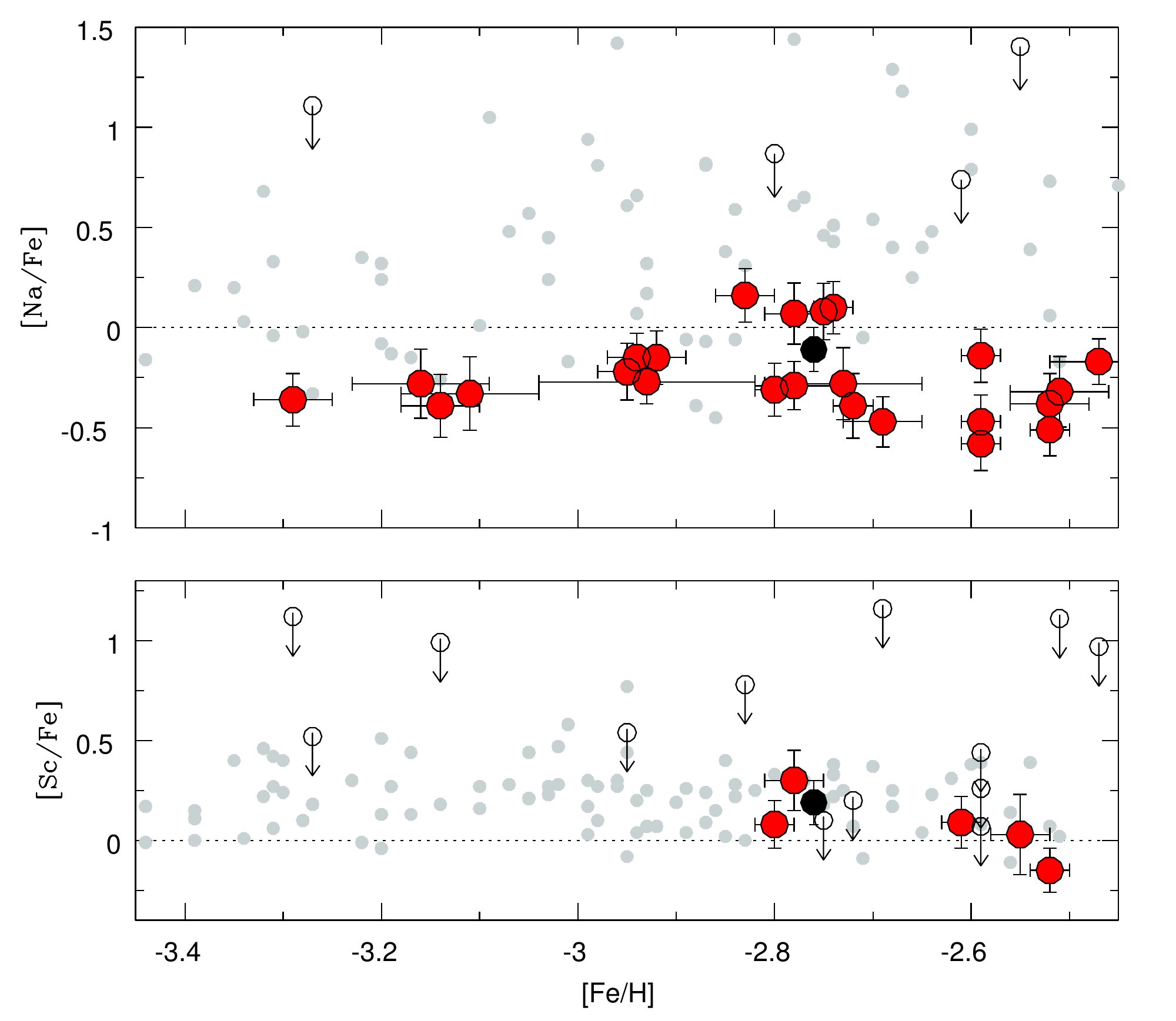}
\caption{ Na and Sc abundances and upper limits in the 28 new metal-poor stars 
([Fe/H] $< -2.5$); symbols the same as in Fig.~\ref{fig:iron}.
NLTE corrections for Na have been applied from INSPECT \citep{Amarsi2016}. 
}
\label{fig:NaSc}
\end{figure}

\subsection{Odd elements}

Odd elements, Na and Sc, are listed in Table~\ref{tab:alpha}. These have different nucleosynthetic 
sources and are not related to one another.   We also include a comment on Li upper limits at the
end of this section.

In metal-poor stars, sodium typically forms with the $\alpha$-elements during core collapse SN. 
On the other hand, scandium forms in the iron core of a massive star with a yield
that strongly depends on the proton-to-neutron ratio (Y$_{\rm e}$), and it is very sensitive 
to neutrino processes \citep[e.g.,][]{Woosley2002, Curtis2019}. 

Sodium abundances are initially from the LTE analysis of the 
\ion{Na}{I} D lines (5889.9, 5895.9~\AA), which unfortunately can be strong,
therefore sensitive to microturbulence in a 1D LTE analysis, and also
contaminated by interstellar Na.  
Furthermore, since they originate from the \ion{Na}{I} ground state, 
they are subject to NLTE effects.  NLTE corrections are similar between
the INSPECT database and \citet{Mashonkina2017c}; 
[Na/H]$_{\rm NLTE} = $[Na/H]$_{\rm LTE} + \Delta$Na, where $\Delta$Na =
$-0.1$ to $-0.6$ dex.  The \ion{Na}{I} subordinate line (5688.2~\AA) could only 
be used for upper limit estimates at the SNR of our spectra.

Despite the large NLTE corrections, four stars were found with initially 
very high \ion{Na}{I} abundances 
(Pristine\_251.4082+12.3657, Pristine\_193.8390+11.4150, 
Pristine\_217.5786+14.0379, and Pristine\_250.6963+08.3743, 
in order of decreasing metallicity).   
These four stars also have the lowest radial velocities in 
our sample (-5, +4, -16, and -4 \kms, in order of decreasing metallicity), and  
we suggest they are contaminated by the interstellar Na lines. 
To test this, their \ion{Na}{I} D line shapes were compared with other spectral lines
in the same stars and found to be slightly broader (occassionally, the line core is even split); 
their \ion{Na}{I} D line shapes are also broader than similar stars with
higher radial velocities (where the interstellar lines are often seen offset from the stellar lines).
Thus, in Fig.~\ref{fig:NaSc}, the highest Na abundances are noted as upper limits only
since they are most likely blended, and for the other stars the NLTE corrected Na
abundances are shown. 
 
\ion{Sc}{II} has been measured from 1-3 lines (5031.0, 5526.8, 5657.9~\AA) in five
metal-poor \Pristine\ stars, and the comparison star HD~122563, and upper limits
were determined in the others.
With an odd number of nucleons, this species undergoes strong hyperfine splitting,
which affects line formation through de-saturation.  The HFS corrections were found to be 
small ($< 0.1$).  Upper limits have also been determined for
\ion{Sc}{II} in most of the other new metal-poor stars.    
Upper limits were examined for \ion{Mn}{I} as well, but did not provide 
interesting constraints. 

Lastly, we mention Li in this section. 
Estimates from the \ion{Li}{I} 6707~\AA\ line provide upper limits that do 
not provide meaningful constraints, i.e., the upper limits are above the standard 
Big Bang nucleosynthesis value of A(Li) = 2.7 \citep[e.g., from WMAP,][]{Spergel2003}.   
Only two stars (Pristine\_229.1219+00.9089 and Pristine\_237.8246+10.1426) 
have 3$\sigma$ equivalent width (35 m\AA) upper limits of A(Li) $\le2.2$, 
which is similar to most metal-poor stars that lie on (or below) the Spite Plateau
\citep[e.g., see][]{Aguado2019Li,Bonifacio2018}.

\begin{figure}
\includegraphics[scale=0.4]{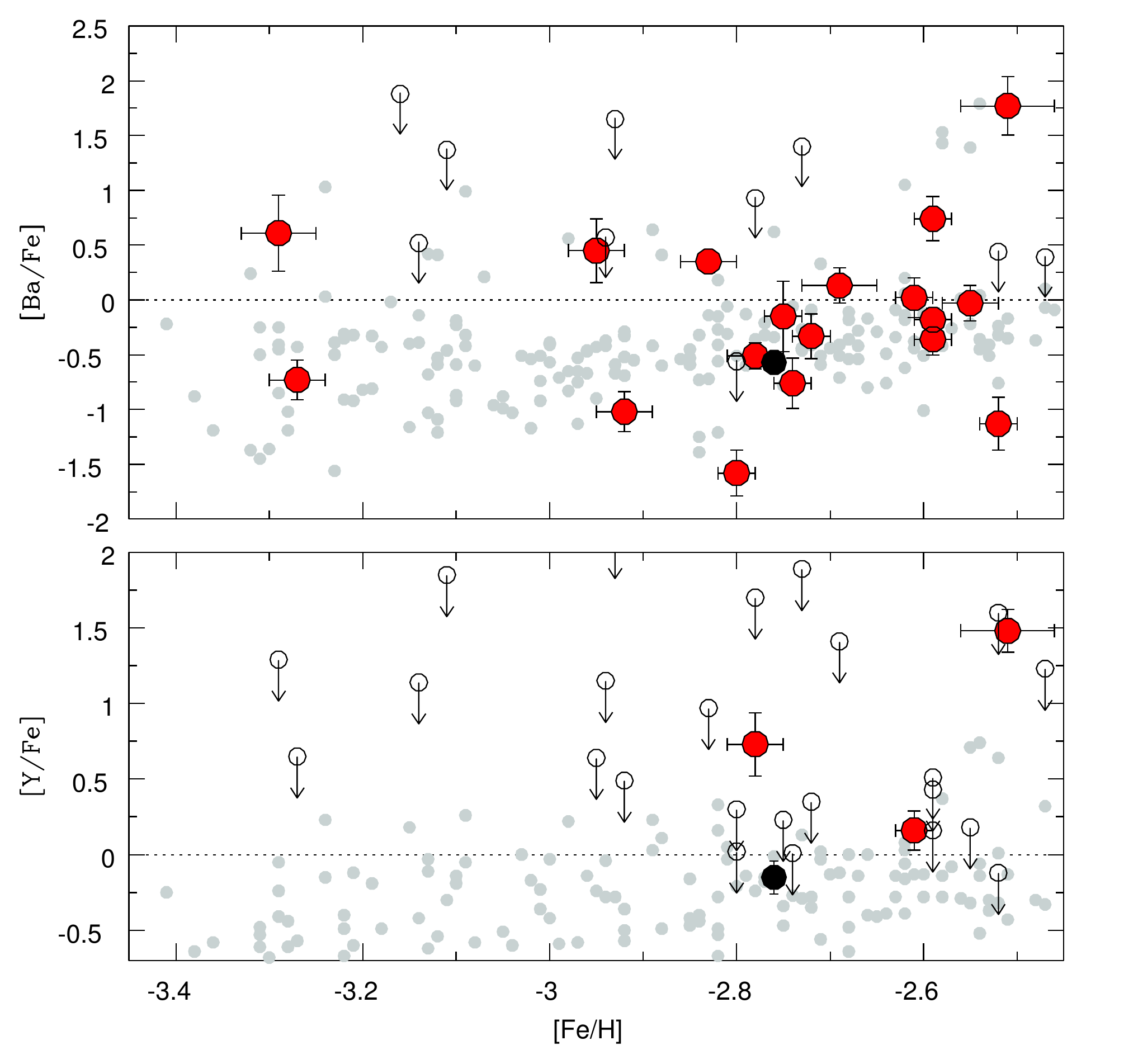}
\caption{ Ba and Y abundances and upper limits in the 28 new metal-poor stars 
([Fe/H] $< -2.5$); symbols the same as in Fig.~\ref{fig:iron} with the
exception of the Galactic comparison stars from \citet[small grey points]{Roederer2014}. 
We identify one star near [Fe/H]$=-2.5$ (Pristine\_214.5556+07.4669) as an r-process rich 
star, significantly enriched in both Y and Ba.
}
\label{fig:BaY}
\end{figure}

\subsection{Heavy elements \label{sect:heavy} }

Abundances for the neutron-capture elements Y and Ba 
in the 28 new very metal-poor stars are listed in Table~\ref{tab:iron}.
Up to four lines of \ion{Ba}{II} (4554.0, 5853.7, 6141.7, and 6496.9~\AA) and two lines of \ion{Y}{II} 
(4883.7, 4990.1~\AA) could be measured.   Unfortunately no lines or useful upper limits for Eu are
available in our CFHT spectra.
When no lines were observable, we determined upper limits from
3$\sigma$ minimum equivalent width estimates.
Hyperfine splitting and the isotopic splitting has been included in the Ba analysis.
Most stars have [Ba/Fe] in good agreement with the Galactic comparison stars.

All six lines were measured in only one star near [Fe/H]$=-2.5$ (Pristine\_214.5555+07.4670). 
This star is enriched in both Y and Ba, and we identify it as an r-process rich star.
Without Eu, it cannot be further classified as r-I or r-II \citep{Christlieb2004, Sakari2018b}. 
Studies of r-process rich stars have found a nearly identical main r-process pattern 
(from barium, A=56, to hafnium, A=72) 
in all types of stars, in all environments, and with variations only between the 
lightest and heaviest elements \citep[see ][ and references therein]{Roederer2010, Hill2017, Sakari2018}.
No other elements stand out in this star; however, as one of the hotter turn-off stars in our 
sample, there are not many other features or elements to analyse at the SNR of our spectra. 

Two more stars show [Ba/Fe]$\gtrsim$+0.5 (Pristine\_237.8246 +10.1426, Pristine\_210.0166+14.6289).  
These lie above the typical [Ba/Fe] values found in the Galactic halo metal-poor stars 
by \citet{Roederer2014}, and their results are securely derived from 2 - 4 \ion{Ba}{II} line measurements.   
However, no \ion{Y}{II} lines were observed in either 
(and the \ion{Y}{II} upper limits do not provide useful constraints).
The two may be moderately r-process enriched stars.

Possibly of greater interest are the two most Ba-poor stars  
(Pristine\_181.4395+01.6294 and Pristine\_193.8390+11.4150). 
Low Ba is very unusual at their metallicities when compared with the other Galactic halo stars,
as seen in Fig.~\ref{fig:BaY}.  This composition is similar to stars in the Segue 1 and
Hercules ultra faint dwarf (UFD) galaxies \citep{Frebel2014, Koch2013}.
In Segue 1, the Ba-poor stars were discussed as representative of 
inhomogeneous enrichment by a single (or few) supernova events, 
and therefore possibly related to first stars.   
Higher SNR data for these two stars is warranted in order to test this hypothesis.

Finally, one star (Pristine\_245.8356+13.8777) shows a high \ion{Y}{II} abundance, 
but normal \ion{Ba}{II} abundance.   
A similar star was recently studied by \citet[][J0222-0313]{Caffau2019},
where the authors show it is a CEMP-s star, having undergone mass transfer 
in a binary system with an Asymptotic Giant Branch (AGB) star.
However, they also suggest that the AGB star in this system may have undergone
a proton ingestion event just before the mass transfer that produced an enhancement 
in only the first s-process peak elements.

\begin{figure}
\includegraphics[scale=0.45]{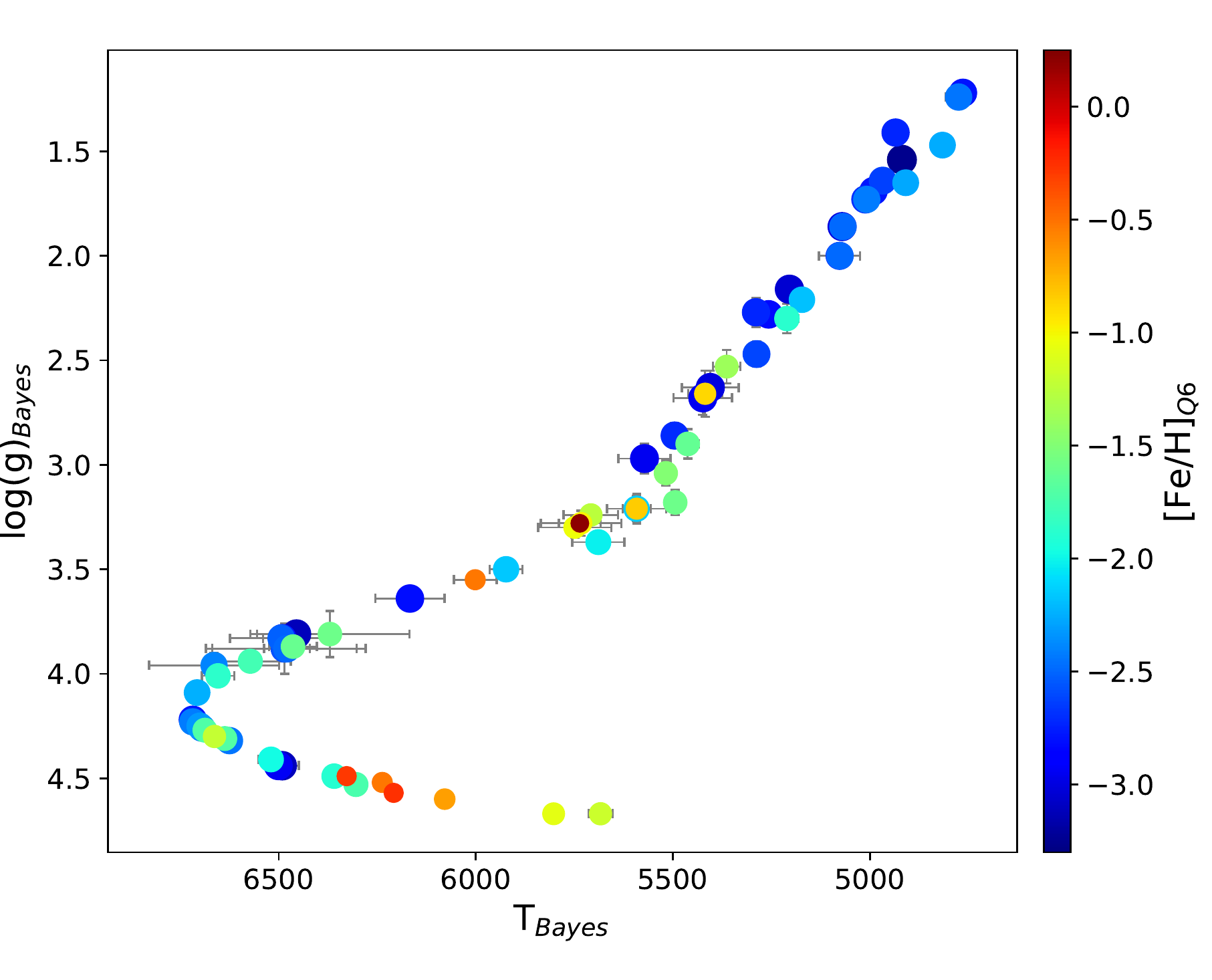}
\caption{The HRD for the 70 metal-poor candidates in the \Pristine\ survey, colour-coded 
by their ("Quick Six") metallicities [Fe/H]$_{Q6}$ as determined from our high resolution 
CFHT \esp\ spectrum and Bayesian inference analysis.   Stars that are not very metal-poor, 
with [Fe/H]$>-2.0$, are located over all stellar parameters.
}
\label{fig:HRD}
\end{figure}

\subsection{Abundance uncertainties \label{sect:uncs} }

Total uncertainties in the chemical abundances are a combination of the measurement 
uncertainties and systematic errors in the stellar parameters, added in quadrature.   
For the measurement errors, when fewer than 4 lines are available for an element X, 
then we adopt the larger of $\sigma$(X) or $\sigma$(\ion{Fe}{I})/sqrt(N$_X$).
Since \ion{Fe}{I} lines are measured across the entire spectrum and over a range of equivalent
widths and excitation potentials, then this assumes that $\sigma$(\ion{Fe}{I}) captures the
minimum measurement quality of our spectra. 
For the systematic errors, due to uncertainties in the stellar parameters, 
we determine the impact of the $1\sigma$ changes in temperature, gravity, and metallicity
listed in Table~\ref{tab:newfs}.

A sample of the systematic uncertainties for three stars that cover the parameter space 
of this sample are shown in Table~\ref{tab:errsample}.
It can be seen that temperature tends to
be the dominant systematic error in the analysis of most elements.  
While we could further investigate the impact of the final metallicities [Fe/H] 
and uncertainties $\sigma$([Fe/H]) through iterations in the Bayesian inference method 
on the model atmospheres parameters, we did not;  
the only stars that we follow up in detail are those that did prove to be very metal-poor,
therefore the impact of adjusting for the final metallicities on the other stellar parameters
is very small.

\begin{figure}
\hspace{-0.2 cm}
\includegraphics[scale=0.45]{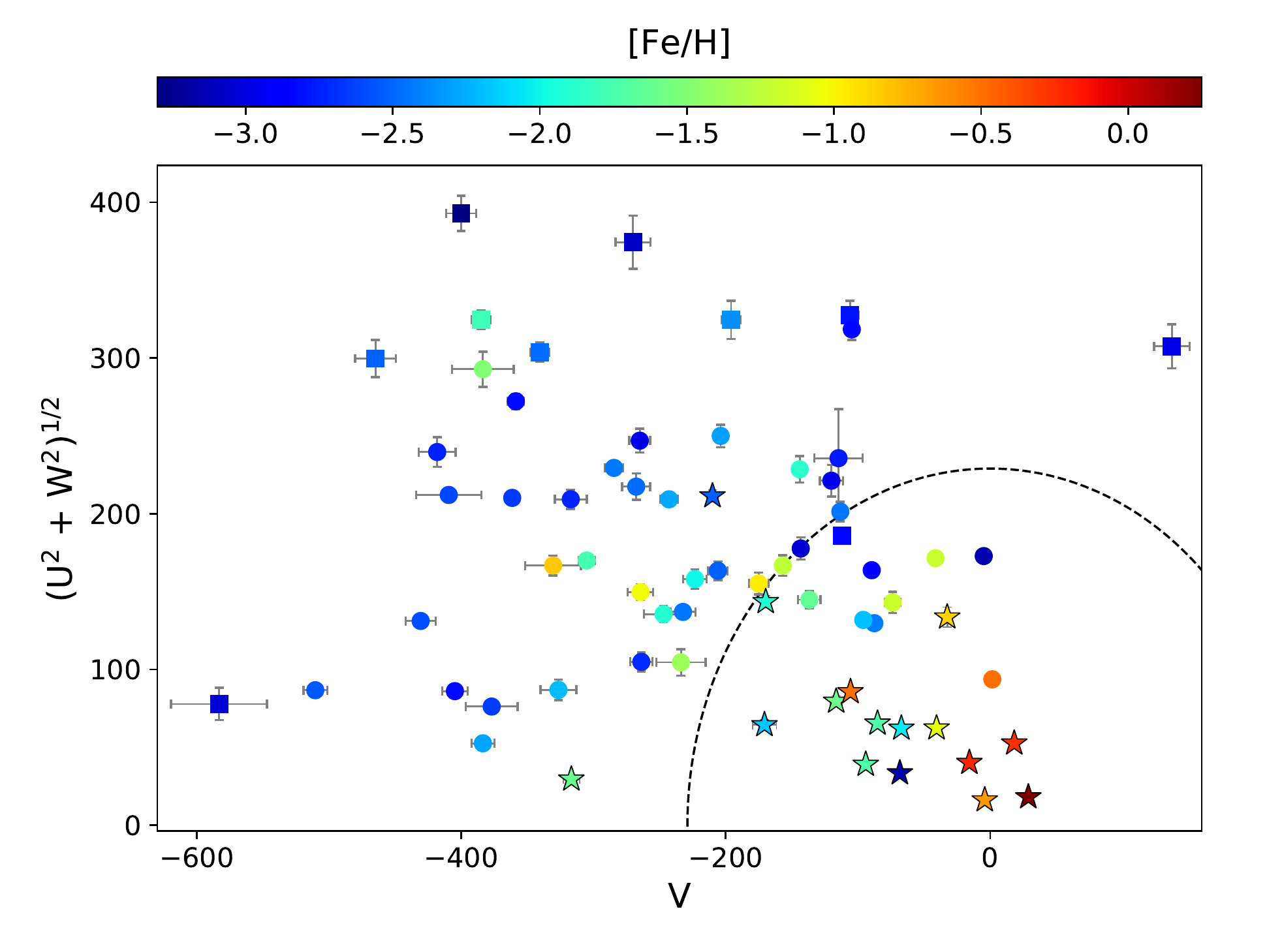}
\caption{Toomre diagram for the 70 highly probable metal-poor stars in our \Pristine\ survey
sample.   Dashed line represents stars potentially with disk dynamics (V$_{\rm circ}$ = 239 \kms).
Symbols the same as in Fig.~\ref{fig:apo}.
}
\label{fig:toomre}
\end{figure}

\section{Discussion }

A total of 70 (out of 115) bright, metal-poor candidates have been observed with the CFHT \esp\ 
spectrograph from the original footprint ($\sim$1000 sq deg) of the \Pristine\ survey.
These targets were selected to have a high probabilily for [Fe/H]$_{\rm Pristine}<-2.5$, 
when the \Pristine\ CaHK filter was calibrated with the SDSS g-i and g-r colours (60 stars), or 
only the SDSS g-r colour alone (10 stars).
We carry out a model atmospheres analysis by adopting stellar parameters
determined from a Bayesian inference method that uses the SDSS colours, Gaia DR2 parallaxes, 
and MESA/MIST isochrones, assuming the initial \Pristine\ survey metallicities.   
Out of these 70 selected stars, we have found 28 to indeed have low metallicities,
[Fe/H]$\le-2.5$ (40\%).   
The \Pristine\ survey had also predicted 27 stars would have [Fe/H]$\le -3.0$, 
and 5 were found (19\%).  Of the 42 remaining stars ($-2.5 <$ [Fe/H]$_{Q6} < +0.25$), there are 
no obvious relationships with any other stellar parameters (e.g., see Fig.~\ref{fig:HRD}),
although we notice that all of the candidates on the upper red giant branch were 
successfully selected and confirmed to be metal-poor stars.  
 
The selections made in this paper differ from those used by \cite{Youakim2017} and \cite{Aguado2019},
see Section~\ref{sect:selection}, being far more strict in the metal-poor probability cuts.  
Furthermore, about 1/3 of the targets in this program were observed before the selection criteria were finalized.  
Nevertheless, our success rates are very similar to the results from the medium resolution surveys.
We do not reproduce the (lower) success rates for bright stars seen in earlier \Pristine\ survey 
papers \citep{Caffau2017, Bonifacio2019}, partially due to our improved (more strict) selection 
criteria, partially due to differences between the SDSS and APASS photometry, and possibly due 
to the larger number of stars in this sample.

In the remainder of this Discussion, we examine the kinematic and orbital properties of the 70 
metal-poor candidates in this paper, and correlate those with their chemical abundances.
We caution that these calculations and our interpretations are highly dependent on the 
accuracy of the adopted Milky Way potential (described in the next section).
For example, our orbit integrations do not account for effects like the Galactic bar, 
which can significantly influence halo star orbits \citep[e.g.,][]{PriceWhelan2016, 
Hattori2016, Pearson2017}.

\subsection{Kinematics and Orbits \label{sect:orbits}}

Galactocentric velocities (U, V, W) are calculated for each star
from their Galactic Cartesian coordinates (X,Y,Z) following the methods of
\citet{Bird2019}.
The distance between the Sun and the Galactic centre is taken to be 
8.0 kpc, the Local Standard of Rest circular velocity is 
V$_{\rm circ}$ = 239 \kms, and the peculiar motion of the Sun is 
(U$_0$ = 11.10 \kms; V$_0 +$ V$_{\rm circ}$ = 251.24 \kms; 
W$_0$ = 7.25 \kms, as described in \citet{Schonrich2010}.
The sign of U$_0$ is changed so that U is positive towards the Galactic anticentre.
Errors in these velocities are propagated from the uncertainties 
in proper motion, radial velocities, and distance by calculating
the mean dispersions from 1000 Monte-Carlo realisations, and 
selecting from a Gaussian distribution in each of the original quantities.

\begin{figure}
\includegraphics[scale=0.47]{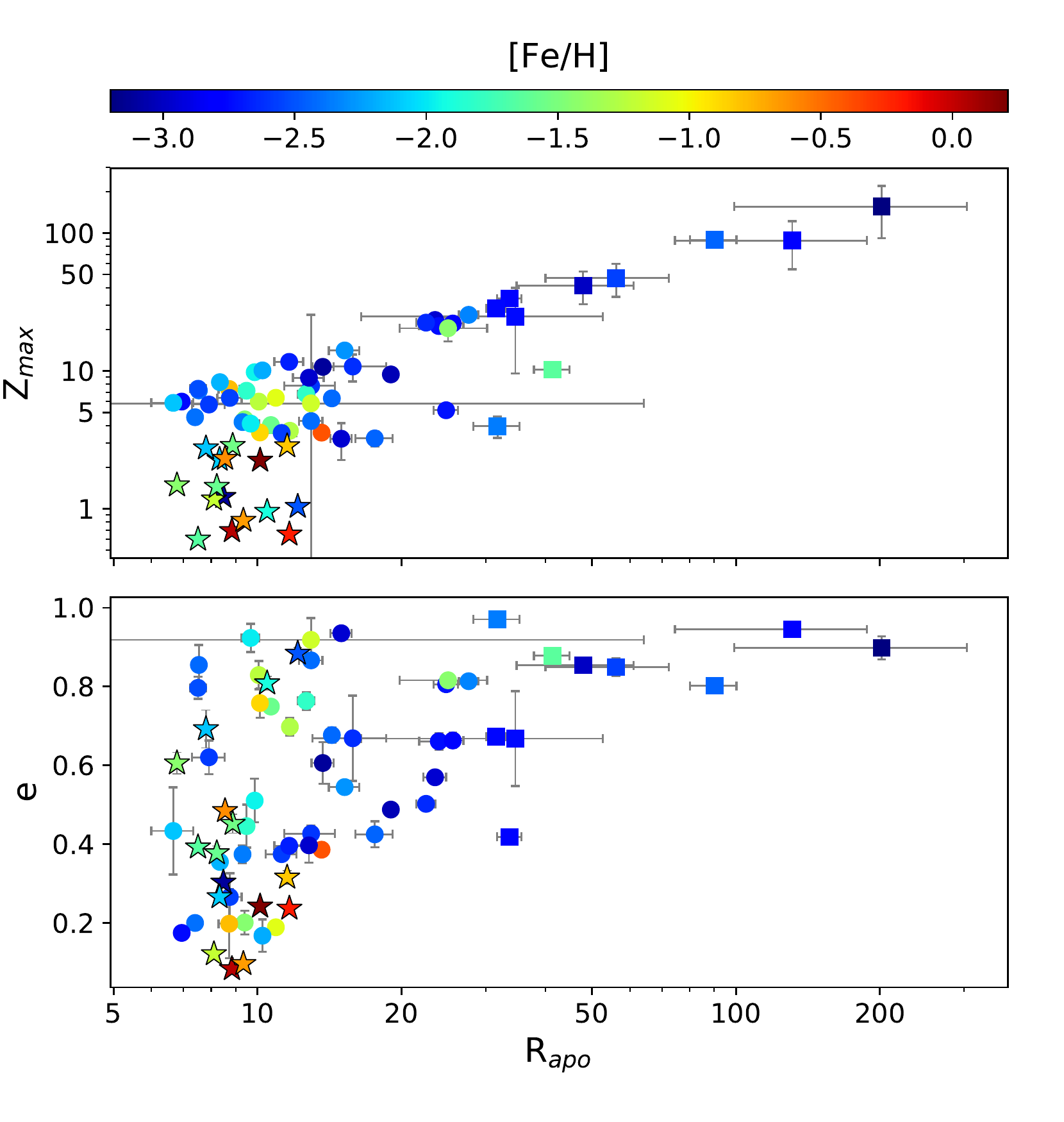}
\caption{ Perpendicular distance from the Galactic plane (Z$_{\rm max}$) 
and eccentricity (e) of the orbits vs apocentric distance (R$_{\rm apo}$)
for the 70 high probability metal-poor stars in this paper. 
For targets with R$_{\rm apo} < 15$ and Z$_{\rm max} < 3$ kpc we adopt
"star" symbols, for R$_{\rm apo} < 30$ kpc we adopt circle symbols,
and when R$_{\rm apo} > 30$ kpc we adopt square symbols.
All targets are colour-coded by their [Fe/H]$_{Q6}$ metallicities.
}
\label{fig:apo}
\end{figure}

With the distances from the Bayesian inference analysis\footnote{ 
For three stars, we reverted back to distances from their 1/parallax values
based on unrealistical outer halo distances and other orbital properties. 
Two of these stars were discussed at the end of Section~\ref{sect:Bayesian},
and a third star is discussed in Appendix A.}, precision
radial velocities from our high resolution spectra, and proper motions 
from the Gaia DR2 database, then the orbital parameters for the sample are
calculated using the Galpy package \citep{Bovy2015}.  
The {\it MWPotential14} is adopted, a Milky Way gravitational potential composed 
of a power-law, exponentially cutoff bulge, Miyamoto Nagai Potential disk, 
and \citet{NFW1997} dark matter halo.  A more massive halo is chosen following 
\citet{Sestito2019}, with a mass of of 1.2 x 10$^{12}$ \msun\
which is more compatible with the value from \citet{BH2016}.

The UVW velocities for the 70 highly probable metal-poor stars 
in this sample are given in Table~\ref{tab:orbits}.
The Toomre diagram for these objects are shown in Fig.~\ref{fig:toomre},
colour-coded by the [Fe/H]$_{\rm Q6}$ metallicities.
Most of the metal-poor stars in our sample have halo-like velocities, 
as expected for their metallicities.  
One very metal-poor star (Pristine\_183.6849+04.8619, discussed below)
appears to have disk-like dynamics.

\subsection {Orbit Analyses}

To investigate the relationships between the chemical and kinematic properties of the stars 
in our sample, we examine their maximum excursions.  This includes the apocentric 
and pericentric distances (R$_{\rm apo}$ and R$_{\rm peri}$), perpendicular
distance from the Galactic plane (Z$_{\rm max}$), and eccentricity (e) of the derived orbits;
see Table~\ref{tab:orbits}.

\begin{figure}
\includegraphics[scale=0.47]{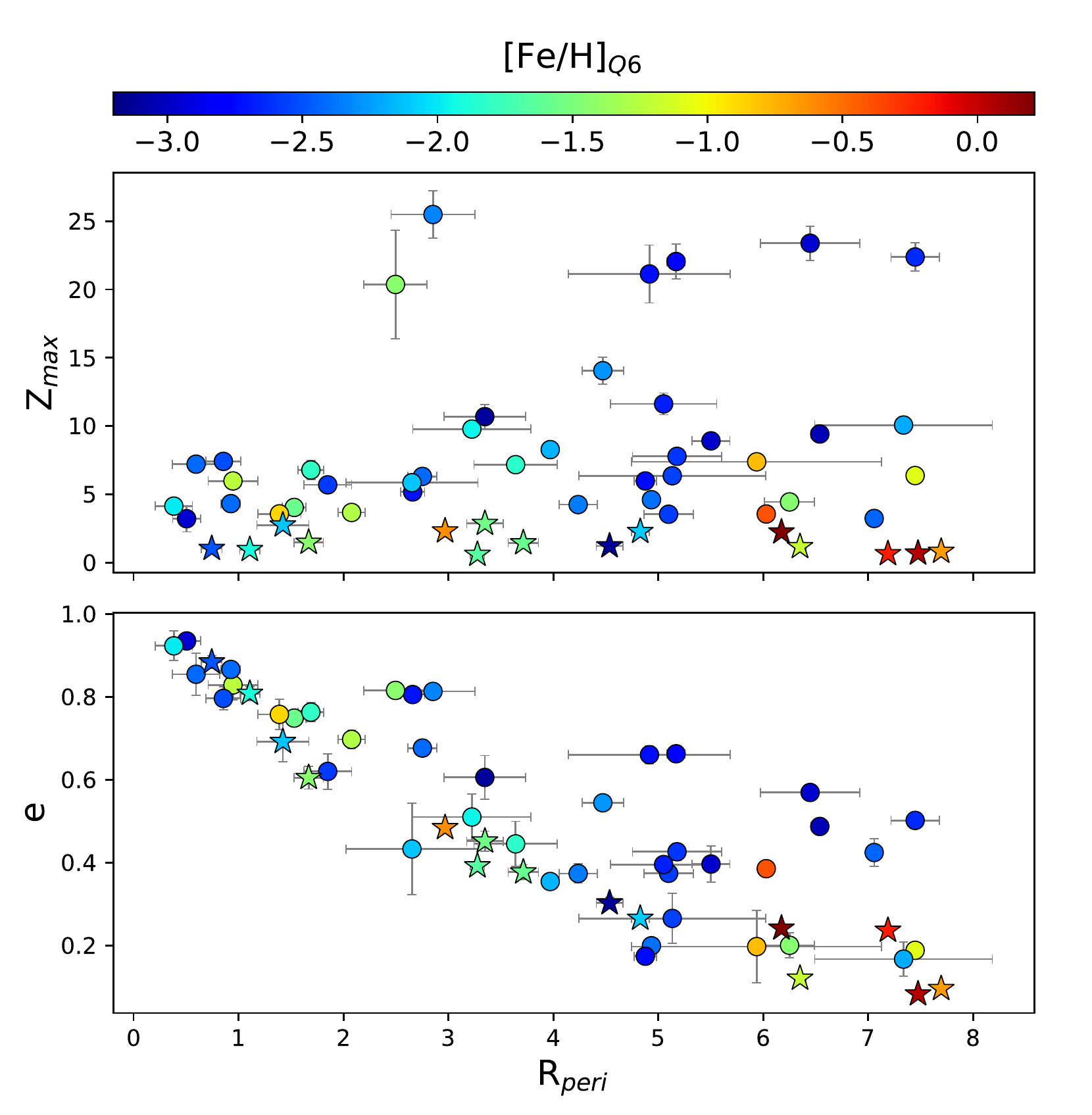}
\caption{ Z$_{\rm max}$ and eccentricity of the orbits vs pericentric distance R$_{\rm peri}$
for the stars within R$_{\rm apo} <$ 30 kpc.
Symbols the same as in Fig.~\ref{fig:apo}.
The very metal-poor star confined within Z$_{\rm max} = 1$ kpc (Pristine\_183.6849+04.8619)
near R$_{\rm peri}=$4.5 kpc can be seen more clearly in this plot than Fig.~\ref{fig:apo}. 
}
\label{fig:peri}
\end{figure}

In Fig.~\ref{fig:apo}, stars with R$_{\rm apo} < 15$ kpc and Z$_{\rm max} < 3$ kpc are
considered to be confined to the Galactic plane (16 stars), while stars with 
R$_{\rm apo} > 30$ kpc are considered to be members of the outer halo (10 stars).  
The outer halo star Pristine\_251.4082+12.3657 has the largest $R_{\rm apo}$ 
distance in our sample, with a highly eccentric orbit, and it is one of the most 
metal poor stars ([Fe/H]=$-3.3$), with low abundances of [Mg/Fe] and [Ca/Fe] 
(see Fig.~\ref{fig:MgCaTi}), and also low [Ba/Fe].  
This chemical signature is typical of stars in or accreted from the nearby dwarf 
galaxies. 
Alternatively, it may have been accreted from an {\it ultra} faint dwarf
galaxy, since its chemistry is also similar to the unique
stars CS~29498-043 and CS~29249-037 \citep{Aoki2002b, Depagne2002}, both near 
[Fe/H]$=-4$.  These stars have been proposed to be second-generation stars, 
that formed from gas enriched by a massive Population III first star, exploding 
as a fall-back supernova \citep[see also][]{Frebel2019}, and as such they would 
have formed in a now accreted {\it ultra} faint dwarf galaxy.  

\begin{figure}
\includegraphics[scale=0.32]{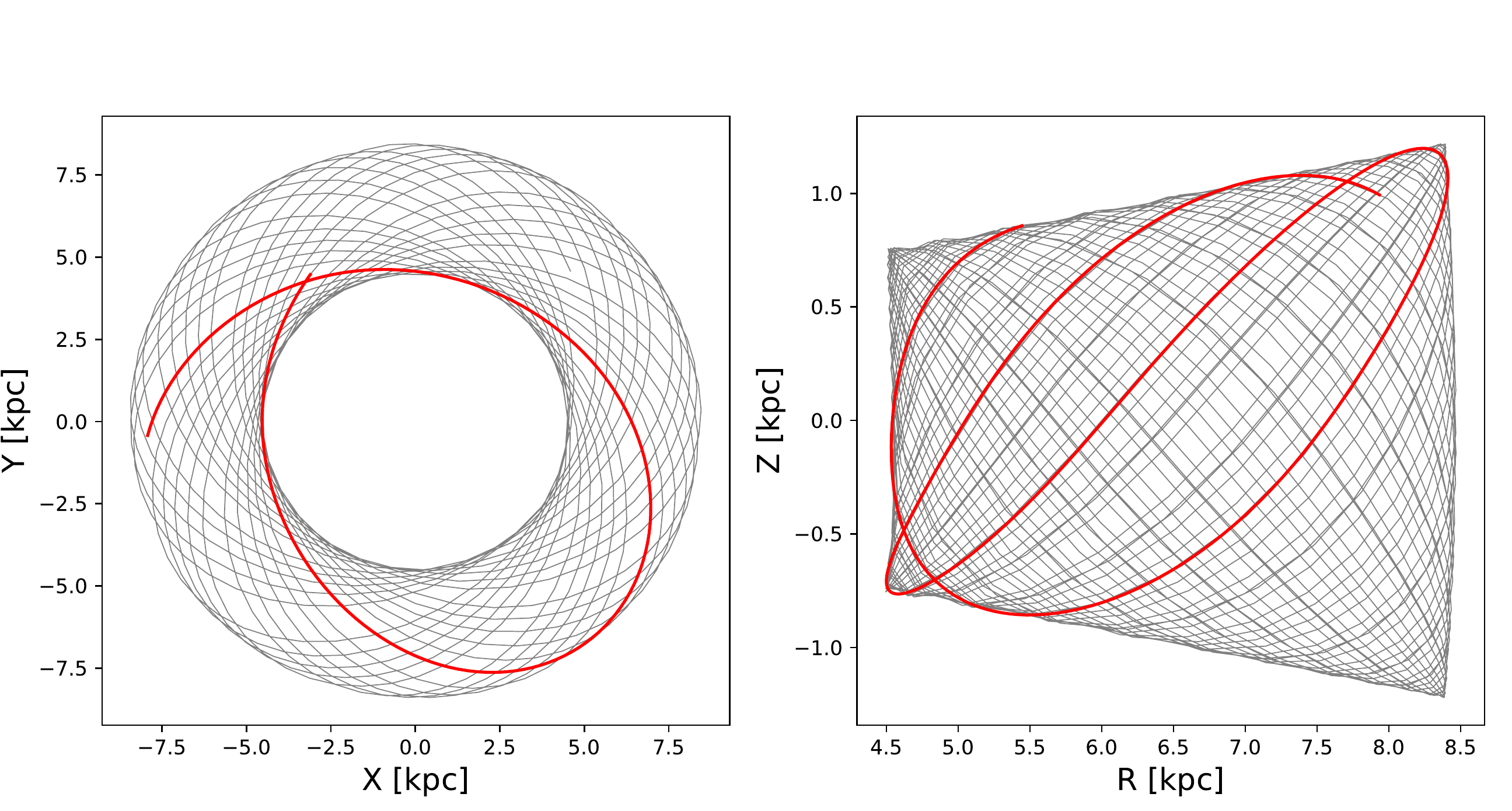}
\caption{The orbit for the very metal-poor star Pristine\_183.6849+04.8619,
from our adopted Galactic potential. 
The orbital properties are R$_{\rm apo}=8.5$ kpc, Z$_{\rm max}=1.2$ kpc, 
and eccentricity e$=0.3$.
A sample single orbit is shown in red. }
\label{fig:P183orbit} 
\end{figure}

In Fig.~\ref{fig:peri}, only stars with R$_{\rm apo}<30$ kpc are shown.  Clearly,
most of the stars confined to the Galactic plane (Z$_{\rm max}<3$ kpc) are 
the relatively metal-rich (interloping) stars in our sample.    However, one 
of the most metal-poor stars (Pristine\_183.6849+04.8619, [Fe/H] $=-3.1$) 
is also confined to the Galactic plane with a nearly circular orbit (e=0.3).
This was also seen in the Toomre diagram (Fig.~\ref{fig:toomre}).
A detailed view of the orbit of this star is shown in Fig.~\ref{fig:P183orbit}.  
Most of the spectral lines in this star are weak and so we were unable to 
determine many elemental abundances, only [Mg/Fe]=+0.13 ($\pm$0.14) 
and [Na/Fe]=$-0.18$ ($\pm0.17$), which are both quite low for a typical halo 
metal-poor star. 
Ultra metal-poor stars ([Fe/H]$<-4$) have been found on similar quasi-circular 
and planar orbits by \citet{Sestito2019}, and interpreted as stars that may
have been brought in during the early merger phase of 
the building blocks of the proto-MW that eventually formed the disk.

Several (8) stars in our sample have orbits that take them 
deep into the Galactic bulge (R$_{\rm peri} < 1$ kpc).  All of these stars are on 
highly radial orbits (e $>0.8$), and two are very metal-poor;
Pristine\_250.6963+08.3743 at [Fe/H]$=-2.55 \pm0.03$, and Pristine\_201.8710+07.1810 
at [Fe/H]$=-2.93 \pm0.11$.   
While the former star shows typical halo abundances in [(Mg,Ca,Ti)/Fe] $= +0.4$ ($\pm$0.4),
the latter is clearly challenged in $\alpha$-elements, [(Na,Mg)/Fe] $= -0.1$ ($\pm$0.2).
It is difficult to discern whether these stars formed in the bulge and have been 
flung out or if they have been accreted from the halo (or a dwarf galaxy) and 
moved inwards.
As metal-poor stars in the bulge are thought to be older in absolute age 
\citep{Tumlinson2010, Howes2016, Starkenburg2017b, ElBadry2018, Frebel2019}, 
then these could be extremely valuable objects for studies of the earliest stages 
of star formation in the Galaxy.

\subsection{Action Parameters \label{sect:action}  }

\begin{figure}
\includegraphics[scale=0.44]{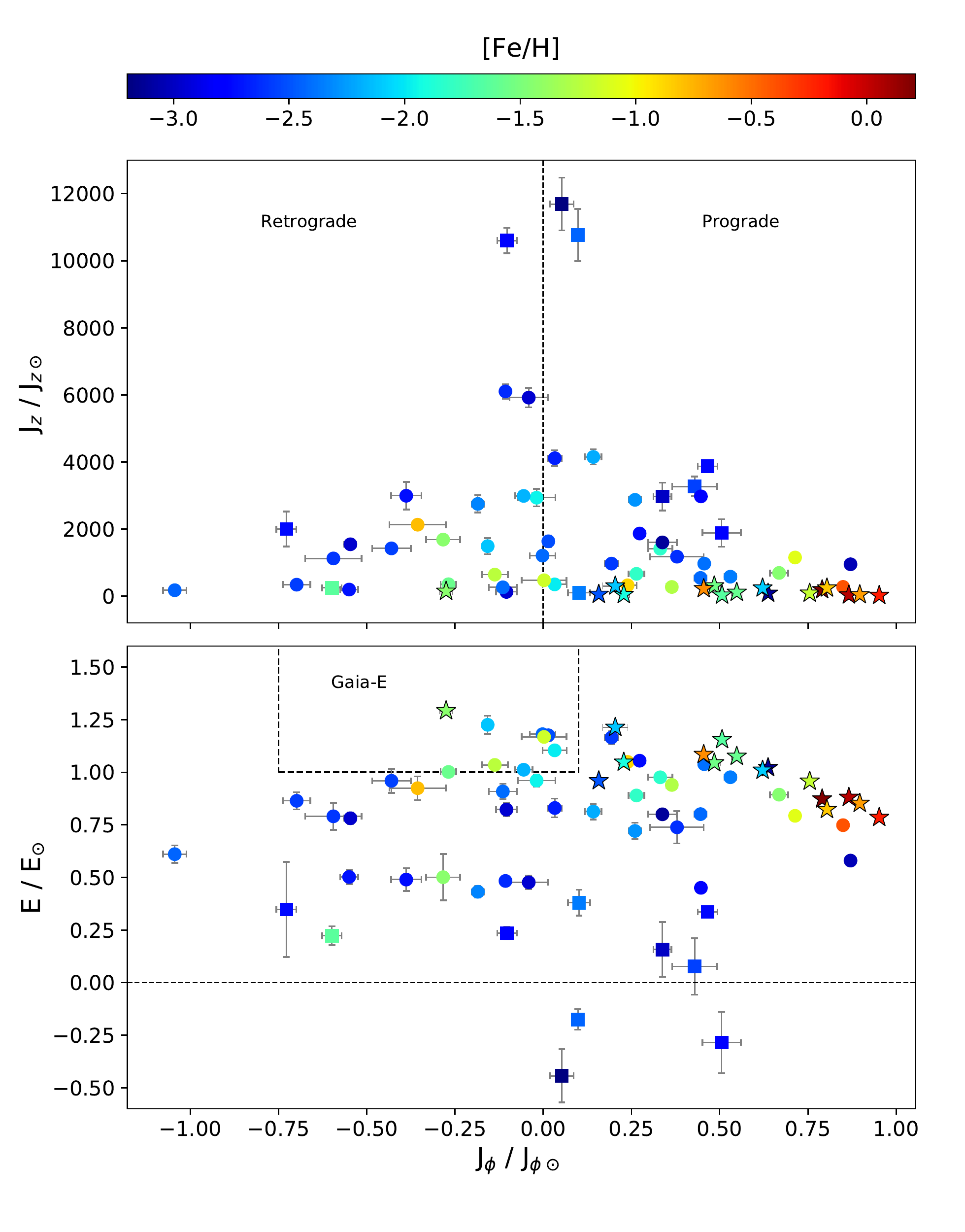}
\caption{The orbit energies and rotational actions 
for the 70 high probability metal-poor stars in this paper. 
The rotational action J$_\phi$ (= L$_z$) is compared with the vertical action 
J$_z$ space (top panel) and the orbit energy (bottom panel), all normalized by
the solar values.  
Prograde and retrograde regions are identified in the top panel.
Symbols are as in Fig.~\ref{fig:apo}.
The region associated with Gaia-Enceladus is marked, 
above E/E$_\odot >1$ and $-0.75<$J$_\phi$/J$_\phi\odot<$0.1 
\citep{Helmi2018, Haywood2018, Belokurov2018, Myeong2018, Myeong2019}. 
}
\label{fig:action}
\end{figure}

The orbital energy (E) and action parameters (vertical J$_z$, azimuthal J$_\phi$)
were determined during the Galpy orbit calculations (discussed above); these
are shown in Fig.~\ref{fig:action} and provided in Table~\ref{tab:orbits}.
Values are scaled by the solar values, where 
J$_{\phi\odot}$ = 2009.92 \kms\ kpc, J$_{z\odot}$ = 0.35 \kms\ kpc and E$_\odot$ = -64943.61 (\kms)$^2$.
It is worth noting that stars with J$_\phi$/J$_{\phi\odot}$ = 1 rotate like the Sun around the 
Galactic Centre.

Amongst the very metal-poor stars, we note they are roughly evenly distributed between
retrograde and prograde orbits, i.e., between $-1 <$ J$_\phi$/J$_{\phi\odot} < 1$.
The most retrograde metal-poor star with a bound orbit (near J$_\phi$/J$_{\phi\odot} = -1$) is
Pristine\_198.5486+11.4123.  This star has Z$_{\rm max} = 3.2$ kpc, placing it very close
to the Galactic plane.   Therefore, this star is travelling at nearly the speed of the Sun 
but in the opposite direction, close to the Galactic plane.    This orbit is certainly unusual
and suggests it may have been accreted from a dwarf galaxy; however, its chemistry is like 
that of a normal metal-poor star, [Fe/H]$=-2.5$, [Mg/Fe]=+0.3, and [Ba/Fe]$<$+0.4.
 
The very metal-poor star Pristine\_251.4082+12.3657, identified as having the 
largest $R_{\rm apo}$ value in this sample is also found to have a large 
Jz/Jz$_\odot$ value and an unbound orbit (E/E$_\odot<0$).  In total, three stars 
in Fig.~\ref{fig:action} appear to have unbound orbits, although we caution 
that our uncertainties in their orbits are quite small when based on the 
very small distance errors from the Bayesian inference method.   
Examination of their parallax errors show that their orbits could be bound, 
consistent with E/E$_\odot\sim0$.
In Appendix A, we examine five more stars that appear to be dynamically unbound 
when their Bayesian inferred distances are used to determine their orbits.   
Two of those stars were discussed in Section~\ref{sect:Bayesian}, and it was 
shown that the orbital properties for these two metal-rich stars were significantly 
improved when 1/parallax was adopted for their distances.  The same was found for 
a third star Pristine\_213.7879+08.4232, even though this star has been confirmed
to be metal-poor.  The parallax errors for these three stars are all very small,
and therefore we have adopted the 1/parallax distance for the orbital analysis
of these three stars.
Finally, we removed two stars from this kinematic analysis,
Pristine\_181.4395+01.6294 and Pristine\_182.5364+00.9431.
Both stars have R$_{\rm apo}>$500 kpc and e$\sim$1 resulting in
extreme and unbound orbits for any distance that we adopt.

One of the most exciting discoveries from the Gaia DR2 dataset has been the 
identification of the Gaia-Enceladus dwarf galaxy (or galaxies) 
dissolved into the Milky Way halo.
The region where stars may be associated with Gaia-Enceladus is shown in 
Fig.~\ref{fig:action}, i.e., E/E$_\odot>1$ and $-0.75<$J$_\phi$/J$_{\phi\odot} <0.1$
\citep{Helmi2018, Haywood2018, Belokurov2018, Myeong2018, Myeong2019}. 
This includes eight stars in our sample that range from 
$-2.5 <$ [Fe/H]$_{\rm Q6} < -1.0$, with a mean metallicity 
of $<$[Fe/H]$>$=$-2.0 \pm0.5$; see Table~\ref{orbits}.
Only one of these stars is sufficiently metal-poor to have made it into our
detailed analysis sample, Pristine\_250.6963+08.3743 ([Fe/H]=$-2.55 \pm0.03$).
This star has high $\alpha$-element abundances [(Ca,Ti)/Fe]$\sim+0.4$, but lower
magnesium such that [Mg/(Ca,Ti)]=$-0.3$, which is has been seen in 
some dwarf galaxies (e.g., Tri II, \citealt{Venn2017}).   
However, unlike most stars in dwarf
galaxies, this star appears to have 
solar-like [Ba/Fe]$\sim 0$ and [Sc/Fe]$\sim 0$. 
It is unclear if this star is a true member of the original Gaia-Enceladus 
accretion event, but if so it would be amongst the most metal-poor stars yet 
found in that system (though also see \citealt{Monty2019}).
As a final test, we examine the action-energy space of the newly discovered
Gaia-Sequoia accretion event \citep{Myeong2018, Myeong2019}, i.e., 
E/E$_\odot >1$ and J$_\phi$/J$_{\phi\odot} <-1.5$, but find no targets in 
that parameter space.

\section{Conclusions and Future Work}

The results from our follow-up spectroscopy of 115 bright metal-poor candidates 
selected from the \Pristine\ survey has been presented based on CFHT \esp\ spectra.
We have discovered 28 new very metal-poor stars with [Fe/H]$<-2.5$ and five stars 
with [Fe/H]$<-3.0$, which imply success rates of 40\% (28/70) and 19\% (5/27), 
respectively.  These rates are higher than previous surveys,
though in line with the \Pristine\ medium resolution programs.
A detailed model atmospheres analysis for the 28 new very metal-poor stars,
has provided stellar parameters and chemical abundances for 
10 elements (Na, Mg, Ca, Sc, Ti, Cr, Fe, Ni, Y, Ba) 
and Li upper limits.
Most stars show chemical abundance patterns that are similar to
the normal metal-poor stars in the Galactic halo; however, we
also report the discoveries of a new r-process rich star (Pristine\_214.5556+07.4670),   
a new CEMP-s candidate with [Y/Ba]$>0$ (Pristine\_245.8356+13.8777), 
and a [Mg/Fe] challenged star (Pristine\_251.4082+12.3657) which has 
an abundance pattern typical of stars
in dwarf galaxies or, alternatively, gas enriched by a
massive Population III first star exploding as a fall-back supernova.
Two stars are also interesting because they are quite Ba-poor 
(Pristine\_181.4395+01.6294 and Pristine\_193.8390+11.4150),
and resemble stars in the Segue 1 and Hercules UFDs, which have been
interpreted as evidence for inhomogeneous enrichment by a single 
(or few) supernovae events, and therefore possibly related to first stars.   

The kinematics and orbits for all 70 of the metal-poor candidates have
been determined using Gaia DR2 data, our radial velocities, and adopting
the {\it MWPotential14} in the Galpy package (with a slightly more massive halo).
The majority of the confirmed metal-poor stars are members of the Galactic halo, 
although some stars show unusual kinematics for their chemistry.
We report the discovery of a very metal-poor ([Fe/H] $= -3.2 \pm0.1$) star 
(Pristine\_183.6849+04.8619) 
with a slightly eccentric ($e = 0.3$) prograde orbit confined to
the Galactic plane (Z$_{\rm max} < 1.2$ kpc). 
We also find a metal-poor ([Fe/H] $= -2.5 \pm0.1$) star  (Pristine\_198.5486+11.4123) 
on a highly retrograde orbit (V$ = -510$ \kms, J$_\phi$/J$_{\phi\odot} = -1.0$) that
remains close to the Galactic plane (Z$_{\rm max} < 3.2$ kpc).
These two stars do not fit standard models for the formation of the Galactic plane,
pointing towards more complex origins.
An additional eight stars were found to have orbital energy and actions consistent
with the Gaia-Enceladus accretion event, including one very metal-poor star 
(Pristine\_ 250.6963+08.3743) with [Fe/H]=$-2.5$ and chemical abundances that
are common for stars in dwarf galaxies.
Finally, eight stars have highly radial orbits that take them deep into the
Galactic bulge (R$_{\rm peri} < 1$ kpc), including two very metal-poor
stars (Pristine\_250.6963+08.3743 at [Fe/H]$=-2.55 \pm0.03$, 
and Pristine\_201.8710+07.1810 at [Fe/H]$=-2.93 \pm0.11$, the latter
star is also low in $\alpha$-elements).  If these stars formed
in the bulge, they could be extremely valuable for studies of the earliest
conditions for star formation in the Galaxy.

Currently, we are running a Gemini/GRACES Large and Long Program to follow-up 
with high SNR ($>100$) spectra for our best metal-poor candidates ([Fe/H] $< -3.5$) 
and with V $< 17$ selected from medium resolution spectroscopy.
We also plan to observe a selection of these stars with the upcoming 
Gemini GHOST spectrograph \citep{Chene2014, Sheinis2017}, 
which is anticipated to have excellent throughput at blue-UV wavelengths,
providing far more iron-group lines for stellar parameter assessments and many more spectral lines of 
heavy neutron-capture (and light) elements.

In the near future, massively multiplexed high resolution spectroscopic surveys 
(R $> 20,000$) will be initiated, including the European WEAVE survey at the Isaac Newton Telescopes 
\citep{Dalton2012}, the 4MOST survey at ESO \citep{4MOST2019}, and the 
SDSS-V survey comprising fields in both the 
northern and southern hemispheres \citep{Kollmeier2017}.   These will provide the truly
large statistical samples needed for studies of the metal-poor Galaxy.

\section*{Acknowledgements}

This work is based on observations obtained at the Canada-France-Hawaii
Telescope (CFHT) which is operated by the National Research
Council of Canada, the Institut National des Sciences de l'Univers
of the Centre National de la Recherche Scientique of France,
and the University of Hawaii.
This work has made use of data from the European Space Agency (ESA) mission
{\it Gaia} (\url{https://www.cosmos.esa.int/gaia}), processed by the {\it Gaia}
Data Processing and Analysis Consortium (DPAC,
\url{https://www.cosmos.esa.int/web/gaia/dpac/consortium}). Funding for the DPAC
has been provided by national institutions, in particular the institutions
participating in the {\it Gaia} Multilateral Agreement.
This research has made use of use of the SIMBAD database, operated at CDS, Strasbourg, 
France \citep{Wenger2000}.
The authors wish to thank David Yong and Ian Roederer for help 
discussions on the detailed abundances of the new metal-poor stars.

This work was funded in part through the National Science and 
Engineering Research Council Discovery Grants program and the 
CREATE training program on New Technologies for Canadian Observatories.
The authors thank the International Space Science Institute, Berne,
Switzerland, for providing financial support and meeting facilities
to the international team \Pristine. FS and NFM gratefully
acknowledge funding from CNRS/INSU through the Programme
National Galaxies et Cosmologie and through the CNRS grant
PICS07708. FS thanks the Initiative dExcellence IdEx from the
University of Strasbourg and the Programme Doctoral International
PDI for funding his PhD. This work has been published under the
framework of the IdEx Unistra and benefits from funding from
the state managed by the French National Research Agency (ANR)
as part of the investments for the future program.
JIGH acknowledges financial support from the Spanish Ministry of Science, 
Innovation and Universities (MICIU) under the 2013 Ram{\'o}n y Cajal 
program MICIU RYC-2013-14875, and also from the Spanish ministry project 
MICIU AYA2017-86389-P.

The authors wish to recognize and acknowledge the
very significant cultural role and reverence that the summit
of Maunakea has always had within the indigenous Hawaiian
community. We are most fortunate to have the opportunity
to conduct observations from this mountain.

\section{Appendix A}

\begin{figure}
\includegraphics[scale=0.45]{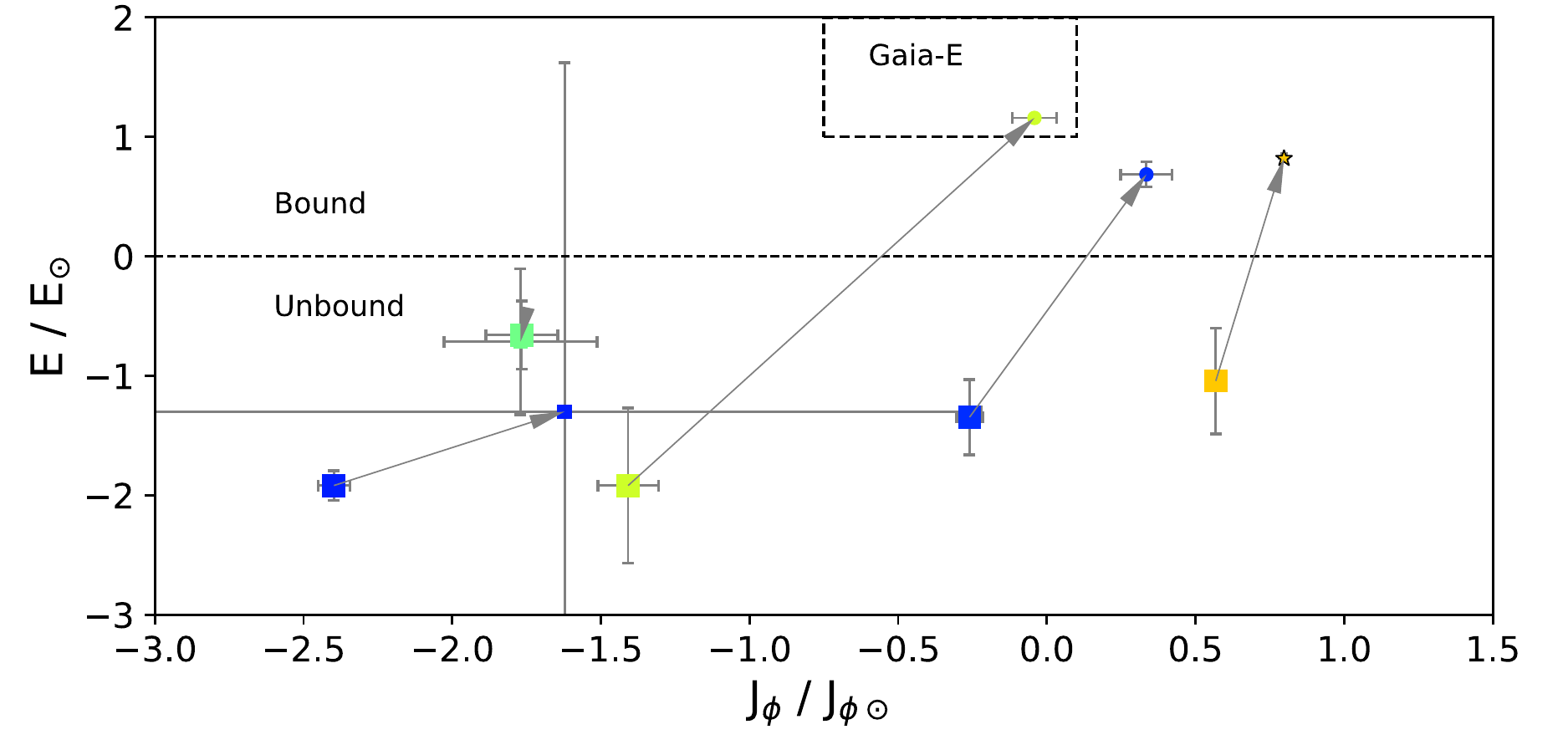}
\caption{
Comparison of the orbital parameters for five stars with unbound
orbits from the dynamical analysis in Section~\ref{sect:orbits}.   
Large symbols are the parameters using the distances from the Bayesian
inference method, and small symbols are those from adopting 1/parallax
from the Gaia DR2 database.
Points coloured by [Fe/H]$_{\rm Q6}$ using the same scheme same 
as in Fig.~\ref{fig:HRD}.
\label{fig:append}
}
\end{figure}

Five stars were identified for a more careful dynamical analysis 
in Section~\ref{sect:orbits}.  
When their distances were determined from the Bayesian inference method,
these stars had highly energetic and unbound orbits (E/E$_\odot < -0.5$),
with R$_{\rm apo} > 500$ kpc.  
These five stars are shown in Fig.~\ref{fig:append}.

As two of these stars 
(Pristine\_200.5298+08.9768 and Pristine\_187.9785+08.7294) 
were found to have higher metallicities than
had been adopted for the MIST stellar isochrone in the Bayesian inference
method, then we assumed their Bayesian inferred distances were unreliable.
Adopting their distances as 1/parallax from the Gaia DR2 database, 
then we found sensible orbits and dynamical parameters for both stars.  
Furthermore, the parallax errors were small for both stars.

We found that the orbit solution for a third star (Pristine\_213.7879+08.4232) 
was also significantly improved by rejecting the Bayesian inferred distance
in favour of the 1/parallax value.   Again, the parallax error is small, and
the resulting orbital properties are less peculiar.   It is not clear why
the Bayesian method did not work for this star, however we note that this was
a target that we observed very early on and it is no longer in the \Pristine\ 
survey catalogue. 
Investigating this star a bit further, we notice that the 1/parallax distance
is closer than the Bayesian inferred distance, suggesting that the surface
gravity for this star may be slightly higher (\logg = 2.3, may be closer to
\logg$\sim$3). In that case, we find small corrections to the abundances,
such that $\Delta$log(\ion{Fe}{II})$\sim$+0.3, bringing \ion{Fe}{II} into
much better agreement with \ion{Fe}{I}.  The impact on [Fe/H] for this star
is negligible though since the iron abundance is dominated by the more 
numerous spectral lines of \ion{Fe}{I}. Minor adjustments to the other elements
would have no signficant effect on the chemical analysis and interpretation
of this star.

Finally, when examining the impact of the distances for the two stars 
Pristine\_182.5364 ([Fe/H]=$-1.6$)  and Pristine\_181.4395 ([Fe/H]=$-2.8$), 
we find they always result in highly retrograde and unbound orbits. 
The orbit for the more metal-poor of these two stars is highly uncertain
when determined from its parallax ($\Delta\pi$/$\pi =0.45$).
Interestingly though, this star is also one of the [Ba/Fe]-poor stars
discussed in Section~\ref{sect:heavy} as possibly accreted from an
ultra faint dwarf galaxy. 
As a sanity check, we also calculated the orbits for all of the other 
metal-poor candidates in this analysis, but found only small
offsets in their orbit and action parameters.



\bibliographystyle{mnras}
\bibliography{Venn.bib}




\begin{table*}
\caption{Metal-Poor Targets (115) from the original {\it Pristine} survey footprint.   
59 stars were selected that have a {\it Pristine} metallicity [Fe/H]$_P < -2.5$ and 
SDSS {\it (g-i)} and {\it (g-r)} calibrations with probabilities for both 
[Fe/H]$_P{\rm gi}$ and [Fe/H]$_P{\rm gr} < -2.25$ greater than 80\% 
(dFe$_P = 1\sigma$).
For 10 stars, only the SDSS {\it (g-r)} calibration was available, as noted 
(where we flagged SDSS $i$ saturation)..
Targets no longer in the {\it Pristine} catalogue, or with [Fe/H]$_P > -2.5$, 
are noted.
The CaHK and SDSS {\it ugri} magnitudes are dereddened using the E(B-V) values
from the \citep{Schlegel1998} maps.
The V and I magnitudes are not dereddened (i.e., observer units).  
The SDSS colour temperature (T$_{\rm SDSS}$) averages the dwarf 
and giant solutions (where dT $= 1\sigma$).
The CFHT semester program labels are in the comments.
}
\label{tab:targets}
\begin{tabular}{cccccccccccclcc} 
\hline
RA$_{\rm SDSS}$ & DEC$_{\rm SDSS}$ & V & I & E(B-V) & CaHK$_0$ & u$_0$ & g$_0$ & r$_0$ & i$_0$ & T$_{\rm SDSS}$ & dT & [Fe]$_P$  & dFe$_P$ & COMM  \\
(deg)           & (deg)            &   &   &        &          &       &       &       &       &  (K)           & (K)  &         &     &       \\
\hline 
\multicolumn{10}{l}{High probability for [Fe/H]$_P \le -2.5$ in the {\it Pristine} catalogue:}   \\
180.2206 & 09.5683 & 14.92 & 13.99 & 0.021 & 15.822 & 16.439 & 15.172 & 14.625 & 14.392 & 5202.6 & 17.6 & -2.82 & 0.02 & 16BC,17AC\\
181.2243 & 07.4160 & 14.95 & 14.33 & 0.015 & 15.361 & 15.949 & 15.044 & 14.804 & 14.707 & 6261.5 & 5.8  & -2.78 & 0.01 & 17AC002 \\
181.3464 & 11.6698 & 14.18 & 13.42 & 0.033 & 14.782 & 15.293 & 14.392 & 13.841 & 13.757 & 5504.8 & 15.6 & -3.82 & 0.09 & 16AC031 \\
181.4395 & 01.6294 & 14.67 & 13.66 & 0.020 & 15.562 & 16.345 & 14.961 & 14.357 & 14.078 & 5011.6 & 18.5 & -3.82 & 0.09 & 16BC008 \\
181.6953 & 13.8075 & 14.52 & 13.73 & 0.031 & 15.135 & 15.701 & 14.692 & 14.228 & 14.087 & 5572.5 & 15.0 & -3.37 & 0.06 & 18BC018 \\
182.5364 & 00.9431 & 14.97 & 14.33 & 0.029 & 15.395 & 15.975 & 15.036 & 14.771 & 14.678 & 6201.8 & 6.8  & -2.46 & 0.05 & 17AC002 \\
183.6849 & 04.8619 & 14.96 & 14.37 & 0.018 & 15.285 & 15.823 & 15.025 & 14.822 & 14.735 & 6398.9 & 3.2  & -3.13 & 0.02 & 18BC018 \\
185.4110 & 07.4777 & 14.72 & 14.07 & 0.022 & 15.168 & 15.657 & 14.826 & 14.523 & 14.430 & 6096.2 & 8.6  & -2.94 & 0.04 & 17AC002 \\
187.9785 & 08.7294 & 14.93 & 14.18 & 0.018 & 15.640 & 16.179 & 15.178 & 14.661 & 14.552 & 5525.0 & 15.5 & -3.33 & 0.23 & 16BC008 \\
189.9449 & 11.5534 & 14.39 & 13.74 & 0.035 & 14.729 & 15.321 & 14.416 & 14.188 & 14.081 & 6267.2 & 5.7  & -2.81 & 0.02 & 18BC018 \\
190.6313 & 08.5137 & 14.92 & 14.16 & 0.021 & 15.601 & 16.402 & 15.101 & 14.673 & 14.534 & 5660.9 & 14.2 & -2.76 & 0.12 & 17AC002 \\
193.1159 & 08.0557 & 14.56 & 13.89 & 0.026 & 15.030 & 15.489 & 14.646 & 14.361 & 14.253 & 6104.5 & 8.5  & -2.46 & 0.01 & 17AC002 \\
193.5533 & 11.5037 & 14.18 & 13.37 & 0.033 & 15.040 & 15.407 & 14.440 & 13.818 & 13.717 & 5316.7 & 17.0 & -2.83 & 0.46 & 16AC031 \\
193.8390 & 11.4150 & 14.75 & 13.61 & 0.030 & 15.962 & 16.671 & 15.055 & 14.364 & 14.017 & 4757.8 & 18.9 & -2.91 & 0.12 & 18BC018 \\
196.3755 & 08.5138 & 14.65 & 13.70 & 0.029 & 15.577 & 16.175 & 14.899 & 14.326 & 14.085 & 5137.4 & 17.9 & -2.84 & 0.01 & 17AC002 \\
196.4117 & 14.3176 & 14.57 & 13.67 & 0.028 & 15.534 & 15.935 & 14.870 & 14.193 & 14.038 & 5103.8 & 18.1 & -3.04 & 0.36 & 16AC096 \\
196.5453 & 12.1211 & 14.71 & 14.16 & 0.026 & 15.020 & 15.626 & 14.739 & 14.562 & 14.506 & 6572.4 & 0.5  & -2.79 & 0.03 & 18BC018 \\
198.5486 & 11.4123 & 14.27 & 13.65 & 0.023 & 14.619 & 15.223 & 14.341 & 14.093 & 14.011 & 6281.6 & 5.4  & -3.15 & 0.01 & 16AC031 \\
200.0999 & 13.7228 & 14.94 & 13.84 & 0.025 & 16.185 & 16.864 & 15.267 & 14.570 & 14.250 & 4790.0 & 19.0 & -2.81 & 0.14 & 18BC018 \\
200.5298 & 08.9768 & 15.15 & 14.32 & 0.027 & 15.821 & 16.575 & 15.342 & 14.870 & 14.695 & 5478.3 & 15.9 & -3.30 & 0.01 & 16BC008 \\
201.1158 & 15.4381 & 14.31 & 13.50 & 0.020 & 15.115 & 15.893 & 14.538 & 14.039 & 13.884 & 5463.0 & 16.0 & -2.61 & 0.11 & 18BC018 \\
201.3732 & 08.4513 & 14.65 & 13.92 & 0.018 & 15.192 & 16.047 & 14.834 & 14.432 & 14.305 & 5759.5 & 13.1 & -3.34 & 0.05 & 16BC,18BC\\
201.8711 & 07.1810 & 14.76 & 14.15 & 0.030 & 15.045 & 15.592 & 14.799 & 14.562 & 14.495 & 6357.4 & 3.9  & -3.33 & 0.01 & 18BC018 \\
202.3435 & 13.2291 & 14.24 & 13.48 & 0.021 & 14.706 & 15.263 & 14.386 & 14.027 & 13.864 & 5769.4 & 13.1 & -3.80 & 0.08 & 16BC,18BC\\
203.2831 & 13.6326 & 14.85 & 13.88 & 0.024 & 15.765 & 16.417 & 15.103 & 14.536 & 14.277 & 5115.0 & 18.1 & -3.01 & 0.07 & 16BC008 \\
204.9008 & 10.5513 & 14.88 & 14.31 & 0.028 & 15.119 & 15.745 & 14.886 & 14.717 & 14.657 & 6584.9 & 0.8  & -3.23 & 0.09 & 17AC002 \\
205.1342 & 13.8234 & 14.61 & 13.84 & 0.023 & 15.313 & 15.955 & 14.788 & 14.368 & 14.212 & 5639.6 & 14.4 & -2.50 & 0.06 & 17AC002 \\
205.8131 & 15.3832 & 14.64 & 14.08 & 0.032 & 14.863 & 15.482 & 14.659 & 14.458 & 14.408 & 6516.8 & 0.7  & -3.48 & 0.05 & 18BC018 \\
206.3487 & 09.3099 & 14.29 & 13.71 & 0.027 & 14.657 & 15.204 & 14.347 & 14.110 & 14.055 & 6393.0 & 3.3  & -2.64 & 0.09 & 18BC018 \\
208.0798 & 04.4266 & 14.54 & 13.79 & 0.028 & 14.988 & 23.848 & 14.664 & 14.308 & 14.158 & 5809.0 & 12.6 & -3.72 & 0.06 & 18BC018 \\
209.9364 & 15.9251 & 14.85 & 14.28 & 0.020 & 15.175 & 15.781 & 14.894 & 14.706 & 14.637 & 6498.4 & 1.1  & -2.85 & 0.02 & 18BC018 \\
210.0166 & 14.6289 & 14.57 & 13.70 & 0.018 & 15.390 & 15.908 & 14.779 & 14.320 & 14.100 & 5408.9 & 16.4 & -2.53 & 0.07 & 16AC031 \\
210.8632 & 08.1797 & 14.50 & 13.72 & 0.023 & 15.161 & 15.653 & 14.667 & 14.260 & 14.098 & 5656.2 & 14.3 & -2.70 & 0.02 & 18BC018 \\
213.2814 & 14.8983 & 14.55 & 13.94 & 0.018 & 14.935 & 15.819 & 14.637 & 14.394 & 14.308 & 6284.5 & 5.3  & -2.93 & 0.01 & 17AC002 \\
214.5556 & 07.4670 & 14.64 & 14.04 & 0.030 & 15.007 & 15.584 & 14.695 & 14.442 & 14.382 & 6331.0 & 4.5  & -2.70 & 0.10 & 18BC018 \\
217.3861 & 15.1650 & 14.61 & 14.03 & 0.025 & 14.937 & 15.546 & 14.660 & 14.436 & 14.378 & 6422.8 & 2.7  & -2.94 & 0.05 & 18BC018 \\
217.5786 & 14.0379 & 14.49 & 13.51 & 0.029 & 15.433 & 16.086 & 14.750 & 14.155 & 13.901 & 5072.5 & 18.2 & -3.04 & 0.02 & 16BC008 \\
217.6443 & 15.9633 & 14.85 & 14.26 & 0.031 & 15.171 & 15.756 & 14.879 & 14.666 & 14.597 & 6422.8 & 2.7  & -2.79 & 0.05 & 18BC018 \\
218.4622 & 10.3683 & 14.90 & 14.26 & 0.023 & 15.280 & 16.105 & 14.986 & 14.716 & 14.622 & 6184.9 & 7.2  & -3.22 & 0.03 & 17AC002 \\
228.4607 & 08.3553 & 14.85 & 14.28 & 0.030 & 15.179 & 15.709 & 14.876 & 14.672 & 14.620 & 6501.5 & 1.1  & -2.64 & 0.08 & 17AC002 \\
228.6557 & 09.0914 & 14.79 & 14.24 & 0.031 & 15.086 & 15.664 & 14.818 & 14.613 & 14.569 & 6523.0 & 0.6  & -2.96 & 0.04 & 17AC002 \\
228.8159 & 00.2222 & 14.76 & 14.13 & 0.053 & 15.023 & 15.581 & 14.719 & 14.497 & 14.424 & 6384.0 & 3.5  & -2.71 & 0.06 & 17AC002 \\
233.5730 & 06.4702 & 14.55 & 13.56 & 0.044 & 15.445 & 16.062 & 14.750 & 14.166 & 13.920 & 5107.5 & 18.1 & -2.83 & 0.01 & 17AC002 \\
235.1449 & 08.7463 & 14.58 & 13.87 & 0.042 & 14.980 & 15.533 & 14.627 & 14.327 & 14.200 & 6012.6 & 9.9  & -3.02 & 0.07 & 16BC008 \\
236.1068 & 10.5311 & 14.54 & 13.45 & 0.050 & 15.689 & 16.383 & 14.775 & 14.097 & 13.806 & 4866.0 & 18.8 & -2.69 & 0.08 & 16AC031 \\
237.8246 & 10.1426 & 15.08 & 14.21 & 0.052 & 15.656 & 16.223 & 15.169 & 14.740 & 14.543 & 5525.0 & 15.4 & -3.12 & 0.07 & 18BC018 \\
240.4216 & 09.6761 & 14.77 & 13.85 & 0.040 & 15.473 & 16.084 & 14.924 & 14.438 & 14.211 & 5337.4 & 16.9 & -3.14 & 0.12 & 16AC,16BC\\
245.5747 & 06.8844 & 14.95 & 14.09 & 0.064 & 15.478 & 16.038 & 15.000 & 14.568 & 14.391 & 5563.4 & 15.1 & -3.11 & 0.02 & 16BC,18BC\\
245.8356 & 13.8777 & 13.86 & 12.97 & 0.045 & 14.547 & 15.069 & 13.987 & 13.522 & 13.316 & 5426.0 & 16.2 & -2.81 & 0.04 & 16AC031 \\
246.5144 & 05.9826 & 14.47 & 13.51 & 0.068 & 15.110 & 16.021 & 14.591 & 14.020 & 13.815 & 5210.4 & 17.5 & -3.73 & 0.06 & 18BC018 \\
246.8588 & 12.3193 & 14.61 & 13.67 & 0.057 & 15.393 & 15.843 & 14.730 & 14.207 & 13.995 & 5292.2 & 17.2 & -2.52 & 0.02 & 16AC096 \\
248.4959 & 15.0776 & 14.57 & 13.58 & 0.058 & 15.365 & 16.025 & 14.698 & 14.168 & 13.910 & 5187.1 & 17.7 & -2.76 & 0.09 & 16AC096 \\
250.6963 & 08.3743 & 14.52 & 13.52 & 0.085 & 15.233 & 15.836 & 14.567 & 14.025 & 13.793 & 5214.4 & 17.6 & -2.70 & 0.02 & 16AC096 \\
251.4082 & 12.3657 & 14.85 & 13.80 & 0.055 & 15.727 & 16.394 & 15.034 & 14.423 & 14.142 & 4995.7 & 18.5 & -3.33 & 0.12 & 17AC002 \\
252.1639 & 15.0648 & 14.43 & 13.38 & 0.069 & 15.417 & 16.018 & 14.578 & 13.938 & 13.690 & 5002.7 & 18.5 & -2.45 & 0.07 & 16AC096 \\
253.8582 & 15.7240 & 14.88 & 13.87 & 0.090 & 15.579 & 16.157 & 14.915 & 14.367 & 14.134 & 5200.6 & 17.6 & -2.74 & 0.02 & 18BC018 \\
254.5469 & 10.9129 & 14.40 & 13.49 & 0.077 & 14.990 & 15.447 & 14.403 & 13.978 & 13.774 & 5518.3 & 15.5 & -2.41 & 0.09 & 16AC031 \\
255.2671 & 14.9711 & 14.41 & 13.72 & 0.081 & 14.611 & 15.170 & 14.292 & 14.048 & 13.966 & 6293.2 & 5.3  & -2.68 & 0.05 & 16AC031 \\
255.5555 & 10.8612 & 14.76 & 13.94 & 0.074 & 15.183 & 15.850 & 14.746 & 14.370 & 14.215 & 5747.4 & 13.3 & -2.92 & 0.02 & 16AC096 \\
\hline
\end{tabular} 
\end{table*} 

\begin{table*}
\contcaption{Metal-Poor Candidates (115) from the Original {\it Pristine} survey footprint.} 
\begin{tabular}{cccccccccccclcc} 
\hline
RA$_{\rm SDSS}$ & DEC$_{\rm SDSS}$ & V & I & E(B-V) & CaHK$_0$ & u$_0$ & g$_0$ & r$_0$ & i$_0$ & T$_{\rm SDSS}$ & dT & [Fe]$_P$  & dFe$_P$ & COMM  \\
(deg)           & (deg)            &   &   &        &          &       &       &       &       &  (K)           & (K)  &          &     &       \\
\hline 
\\
\multicolumn{12}{l}{Used the SDSS-gr solution only for [Fe/H]$_P \le -2.5$ in the {\it Pristine} catalogue:}    \\
180.3789 & 00.9470 & 14.15 & 13.43 & 0.019 & 14.900 & 15.332 & 14.341 & 13.910 & 13.798 & 5718.2 & 13.6 & -2.48 &  ...  & 17AC002 \\
181.3699 & 11.7636 & 14.19 & 13.38 & 0.037 & 15.164 & 15.461 & 14.411 & 13.838 & 13.720 & 5383.4 & 16.6 & -2.50 &  ...  & 16AC031 \\
188.1264 & 08.7740 & 14.22 & 13.42 & 0.018 & 15.162 & 15.593 & 14.515 & 13.917 & 13.798 & 5329.0 & 16.9 & -3.28 &  ...  & 17AC002 \\
200.7620 & 09.4375 & 14.51 & 13.75 & 0.025 & 15.222 & 16.010 & 14.677 & 14.255 & 14.119 & 5682.2 & 13.9 & -2.49 &  ...  & 17AC002 \\
229.1219 & 00.9089 & 14.76 & 14.15 & 0.047 & 15.040 & 15.607 & 14.725 & 14.525 & 14.461 & 6477.1 & 1.6  & -2.72 &  ...  & 17AC002 \\
235.7578 & 09.0000 & 14.92 & 14.34 & 0.041 & 15.257 & 15.819 & 14.916 & 14.704 & 14.651 & 6474.1 & 1.7  & -2.56 &  ...  & 17AC002 \\
237.9600 & 15.4022 & 14.55 & 13.96 & 0.048 & 14.827 & 15.329 & 14.510 & 14.314 & 14.264 & 6532.2 & 0.4  & -2.69 &  ...  & 16AC031 \\
245.4387 & 08.9954 & 14.77 & 14.12 & 0.056 & 15.109 & 15.613 & 14.747 & 14.491 & 14.415 & 6275.9 & 5.5  & -2.47 &  ...  & 17AC002 \\
248.4394 & 07.9229 & 14.41 & 13.74 & 0.074 & 14.678 & 15.182 & 14.320 & 14.066 & 13.999 & 6307.7 & 5.0  & -2.49 &  ...  & 16AC031 \\
254.5207 & 15.4969 & 13.93 & 13.16 & 0.086 & 14.718 & 15.096 & 14.030 & 13.396 & 13.378 & 5467.4 & 15.9 & -3.23 &  ...  & 16AC031 \\
 \\
\multicolumn{10}{l}{Low probability for [Fe/H]$_P \le -2.5$ in the current {\it Pristine} catalogue:}   \\
182.8512 & 14.1595 & 14.19 & 13.14 & 0.042 & 15.258 & 15.959 & 14.396 & 13.823 & 13.521 & 5025.7 & 18.4 & -2.3 &  ...  & 16AC031 \\
196.6015 & 15.6768 & 14.44 & 13.67 & 0.029 & 15.355 & 15.780 & 14.602 & 14.167 & 14.028 & 5644.4 & 14.4 & -1.2 &  ...  & 16AC096 \\
197.9861 & 12.3577 & 14.98 & 14.30 & 0.029 & 15.586 & 16.030 & 15.056 & 14.765 & 14.657 & 6088.0 & 8.7  & -1.4 &  ...  & 18BC018 \\
209.2123 & 01.5275 & 14.45 & 13.72 & 0.034 & 15.013 & 15.504 & 14.546 & 14.202 & 14.071 & 5887.2 & 11.6 & -2.3 &  ...  & 17AC002 \\
209.7175 & 10.8612 & 14.55 & 13.92 & 0.025 & 14.996 & 15.460 & 14.627 & 14.371 & 14.277 & 6224.4 & 6.5  & -2.4 &  ...  & 16AC031 \\
210.7504 & 12.7744 & 14.50 & 13.34 & 0.027 & 16.054 & 16.702 & 14.869 & 14.089 & 13.749 & 4639.0 & 19.0 & -2.0 &  ...  & 16AC096 \\
211.2766 & 10.3280 & 14.71 & 14.05 & 0.019 & 15.223 & 15.807 & 14.823 & 14.525 & 14.424 & 6088.0 & 8.7  & -2.4 &  ...  & 17AC002 \\
212.5834 & 10.5365 & 14.42 & 13.74 & 0.023 & 14.970 & 15.395 & 14.540 & 14.217 & 14.108 & 5999.3 & 10.1 & -2.4 &  ...  & 17AC002 \\
215.6121 & 15.0163 & 14.20 & 13.32 & 0.024 & 15.082 & 15.560 & 14.403 & 13.926 & 13.713 & 5385.5 & 16.6 & -2.2 &  ...  & 16AC031 \\
216.1245 & 10.2135 & 14.49 & 13.65 & 0.027 & 15.257 & 15.773 & 14.666 & 14.219 & 14.029 & 5500.5 & 15.7 & -2.4 &  ...  & 17AC002 \\
223.5273 & 11.1353 & 14.28 & 13.18 & 0.032 & 15.681 & 16.306 & 14.596 & 13.884 & 13.577 & 4786.8 & 18.9 & -2.1 &  ...  & 16AC096 \\
227.2885 & 01.3598 & 14.45 & 13.70 & 0.052 & 15.328 & 15.720 & 14.513 & 14.127 & 14.011 & 5818.9 & 12.4 & -0.5 &  ...  & 16BC008 \\
229.0409 & 10.3020 & 14.45 & 13.39 & 0.039 & 15.709 & 16.330 & 14.701 & 14.054 & 13.768 & 4925.4 & 18.8 & -1.9 &  ...  & 16AC031 \\
232.6956 & 08.3392 & 14.97 & 14.18 & 0.044 & 15.561 & 16.051 & 15.052 & 14.681 & 14.513 & 5727.9 & 13.5 & -2.4 &  ...  & 17AC002 \\
233.9312 & 09.5596 & 14.84 & 14.06 & 0.037 & 15.466 & 15.931 & 14.949 & 14.564 & 14.410 & 5727.9 & 13.5 & -2.3 &  ...  & 17AC002 \\
234.4398 & 12.6286 & 14.46 & 13.61 & 0.039 & 15.604 & 15.979 & 14.635 & 14.132 & 13.961 & 5419.6 & 16.3 & -0.8 &  ...  & 16AC096 \\
235.9710 & 09.1864 & 14.61 & 13.99 & 0.041 & 14.974 & 15.475 & 14.611 & 14.380 & 14.307 & 6357.4 & 3.9  & -2.4 &  ...  & 17AC002 \\
236.0763 & 04.0496 & 14.33 & 13.54 & 0.069 & 15.264 & 15.653 & 14.361 & 13.936 & 13.820 & 5723.0 & 13.5 & -0.4 &  ...  & 16AC096 \\
241.1186 & 09.4156 & 14.48 & 13.80 & 0.046 & 14.896 & 15.395 & 14.499 & 14.221 & 14.117 & 6134.8 & 8.1  & -2.2 &  ...  & 17AC002 \\
 \\
\multicolumn{10}{l}{No longer in the {\it Pristine} catalogue:}  \\
180.7918 & 03.4085 & 13.18 & 12.12 & 0.022 &  ...  & 15.512 & 13.753 & 12.652 & 12.501 & 4494.0 &  ...  & -2.8 &  ...  & 16BC008 \\
182.1679 & 03.4771 & 13.61 & 12.80 & 0.017 &  ...  & 15.397 & 14.199 & 13.099 & 13.138 & 4735.0 &  ...  & -3.8 &  ...  & 16BC008 \\
182.5089 & 03.7165 & 13.27 & 12.74 & 0.016 &  ...  & 14.590 & 13.610 & 12.945 & 13.067 & 5736.0 &  ...  & -4.0 &  ...  & 16BC008 \\
187.8508 & 13.4559 & 13.50 & 13.16 & 0.022 &  ...  & 14.599 & 13.920 & 13.077 & 13.421 &  ...   &  ...  & -2.5 &  ...  & 16AC031 \\
190.5804 & 12.8577 & 13.91 & 13.39 & 0.026 &  ...  & 15.173 & 14.326 & 13.473 & 13.675 &  ...   &  ...  & -2.5 &  ...  & 16AC096 \\
192.2112 & 15.9262 & 14.42 & 13.47 & 0.022 &  ...  & 16.268 & 14.836 & 14.001 & 13.850 &  ...   &  ...  & -2.3 &  ...  & 16AC031 \\
192.8531 & 15.8198 & 13.83 & 13.53 & 0.020 &  ...  & 15.131 & 14.270 & 13.405 & 13.786 &  ...   &  ...  & -2.3 &  ...  & 16AC031 \\
199.9279 & 08.3815 & 13.69 & 13.15 & 0.020 &  ...  & 15.451 & 14.101 & 13.293 & 13.453 & 5514.0 &  ...  & -3.9 &  ...  & 16BC008 \\
207.9961 & 01.1795 & 13.84 & 13.25 & 0.026 &  ...  & 15.156 & 14.302 & 13.373 & 13.545 & 5258.0 &  ...  & -3.9 &  ...  & 16BC008 \\
211.7174 & 15.5515 & 14.52 & 13.48 & 0.014 &  ...  & 16.490 & 14.887 & 14.178 & 13.906 &  ...   &  ...  & -2.5 &  ...  & 16AC096 \\
213.7879 & 08.4232 & 14.81 & 13.90 & 0.026 &  ...  & 16.015 & 14.995 & 14.546 & 14.299 & 5396.0 & 13.4  & -3.3 &  ...  & 16BC008 \\
215.6783 & 07.6929 & 14.09 & 13.61 & 0.025 &  ...  & 14.536 & 13.890 & 13.638 & 15.348 & 5527.0 &  ...  & -3.5 &  ...  & 16BC008 \\
218.4256 & 07.5213 & 14.19 & 13.76 & 0.020 &  ...  & 15.539 & 14.675 & 13.743 & 14.034 &  ...   &  ...  & -2.6 &  ...  & 16AC096 \\
229.8912 & 00.1105 & 14.10 & 13.43 & 0.055 &  ...  & 15.457 & 14.438 & 13.569 & 13.681 & 5296.0 &  ...  & -3.3 &  ...  & 16BC008 \\
230.4662 & 06.5251 & 14.17 & 13.50 & 0.034 &  ...  & 15.392 & 14.534 & 13.732 & 13.803 & 5348.0 &  ...  & -3.5 &  ...  & 16BC008 \\
231.0319 & 06.4866 & 14.26 & 13.67 & 0.037 &  ...  & 15.556 & 14.627 & 13.794 & 13.948 & 5468.0 &  ...  & -3.5 &  ...  & 16BC008 \\
232.8039 & 06.1178 & 14.76 & 14.04 & 0.042 &  ...  & 16.594 & 15.027 & 14.352 & 14.351 & 5440.0 & 13.0  & -2.6 &  ...  & 16BC008 \\
236.4846 & 10.6903 & 14.29 & 13.84 & 0.044 &  ...  & 15.367 & 14.579 & 13.842 & 14.084 &  ...   &  ...  & -3.3 &  ...  & 16AC031 \\
236.7139 & 09.6084 & 14.44 & 13.07 & 0.036 &  ...  & 15.650 & 14.533 & 14.185 & 13.541 & 4851.0 &  ...  & -3.7 &  ...  & 16BC008 \\
237.8344 & 10.5901 & 14.63 & 13.70 & 0.045 &  ...  & 15.840 & 14.743 & 14.309 & 14.059 & 5402.0 & 13.4  & -2.5 &  ...  & 16AC096 \\
240.0340 & 13.8279 & 14.57 & 13.55 & 0.044 &  ...  & 16.250 & 14.791 & 14.179 & 13.914 &  ...   &  ...  & -2.5 &  ...  & 16AC096 \\
245.1095 & 08.8947 & 14.18 & 13.58 & 0.056 &  ...  & 15.395 & 14.459 & 13.681 & 13.824 & 5567.0 &  ...  & -3.1 &  ...  & 16BC008 \\
250.8789 & 12.1101 & 14.14 & 12.69 & 0.046 &  ...  & 15.662 & 14.273 & 13.808 & 13.154 &  ...   &  ...  & -2.5 &  ...  & 16AC031 \\
252.4199 & 12.6477 & 14.43 & 13.44 & 0.045 &  ...  & 15.498 & 14.487 & 14.151 & 13.822 &  ...   &  ...  & -2.4 &  ...  & 16AC031 \\
252.4908 & 15.2984 & 14.37 & 13.14 & 0.059 &  ...  & 15.424 & 14.353 & 14.077 & 13.548 &  ...   &  ...  & -2.6 &  ...  & 16AC096 \\
254.0653 & 14.2694 & 13.39 & 12.99 & 0.060 &  ...  & 14.523 & 13.657 & 12.885 & 13.187 &  ...   &  ...  & -2.5 &  ...  & 16AC031 \\
254.7759 & 13.8207 & 13.67 & 13.16 & 0.064 &  ...  & 14.848 & 14.026 & 13.068 & 13.360 &  ...   &  ...  & -3.2 &  ...  & 16AC096 \\
\hline
\end{tabular} 
\end{table*} 

\begin{table*}
\caption{Gaia DR2 parallaxes, and the derived distances (D), temperatures (T), and surface gravities (log~g) 
from the Bayesian inference method (see Section~\ref{sect:Bayesian}, assuming [Fe/H]$_P$).  Corresponding uncertainties 
are listed as dpar, dD, dT, and dlg, respectively.  Metallicities are from our "Quick Six" analysis (see Section~\ref{sect:quick6}) 
as the individual ion abundances, the weighted average [Fe/H]$_{\rm Q6}$, the standard deviation $\sigma$Fe$_{\rm Q6}$, 
and total number of lines used.
For four targets, a second distance (D2 in Com) satisifies the Bayesian inference analyais but does not significantly
affect the stellar parameters.  For four other targets, a dwarf or giant solution has equal probability, and we examine 
both solutions here independently.
Stars no longer in the {\it Pristine} catalogue have been excluded since they are not metal-poor targets, with only one
exception (RA=213.7879, DEC=+08.4232, noted as **).  
}
\label{tab:newfs}
\begin{tabular}{llrlrlrlrlrrrrll} 
\hline
RA$_{\rm SDSS}$ & DEC$_{\rm SDSS}$ & par & dpar & D & dD & T & dT & log(g) & dlg & RV  & 
  \ion{Fe}{I} & \ion{Fe}{II} & [Fe/H]$_{\rm Q6}$ & $\sigma$Fe$_{\rm Q6}$ & N,Com \\
\hline
\multicolumn{12}{l}{High probability for [Fe/H]$_P \le-2.5$:} \\
180.2206 & +09.5683 & 0.09 & 0.05 & 12.08 & 0.47 & 5070.6 & 20.9 & 1.86 & 0.05 & 23.0   & 4.54 & 4.31 & -3.04 & 0.18 & 6 \\
181.2243 & +07.4160 & 0.47 & 0.05 &  2.28  & 0.07 & 6454.9 & 99.9 & 3.81 & 0.06 & -147.0 & 4.46 & 4.18 & -3.16 & 0.16 & 5 \\
181.3464 & +11.6698 & 1.23 & 0.03 &  0.60  & 0.01 & 6208.3 & 17.1 & 4.57 & 0.01 & 12.0   & 7.55 & 6.86 & -0.22 & 0.15 & 5 \\
181.4395 & +01.6294 & 0.08 & 0.05 & 16.76 & 0.29 & 4934.9 & 8.2 & 1.41 & 0.02 & 206.0 & 4.83 & 4.57 & -2.76 & 0.17 & 6 \\ 
181.6953 & +13.8076 & 0.77 & 0.03 &  2.74 & 0.11 & 5708.0 & 69.0 & 3.24 & 0.04 & 79.0 & 6.23 & 6.31 & -1.24 & 0.18 & 5 \\
182.5364 & +00.9431 & 0.45 & 0.06 &  2.21 & 0.12 & 6370.1 & 202.0 & 3.81 & 0.11 & 222.0 & 5.96 & 5.83 & -1.59 & 0.1 & 5 \\
183.6849 & +04.8620 & 0.95 & 0.28 &  1.06 & 0.05 & 6491.0 & 42.0 & 4.44 & 0.03 & 40.0 & 4.38 & 4.3 & -3.16 & 0.17 & 4, D2 \\
185.4110 & +07.4777 & 1.08 & 0.04 &  0.8 & 0.02 & 6304.0 & 26.0 & 4.53 & 0.01 & 176.0 & 5.93 & 5.48 & -1.75 & 0.28 & 5 \\
187.9785 & +08.7294 & 0.74 & 0.04 &  5.84 & 0.59 & 5418.0 & 42.6 & 2.66 & 0.11 & -55.0 & 6.68 & 6.58 & -0.86 & 0.2 & 5 \\
189.9449 & +11.5535 & 0.61 & 0.03 &  1.69 & 0.03 & 6491.0 & 133.0 & 3.83 & 0.04 & 32.0 & 4.78 & 4.5 & -2.82 & 0.12 & 6 \\
190.6313 & +08.5137 & 0.58 & 0.05 &  3.27 & 0.21 & 5591.6 & 35.4 & 3.21 & 0.07 & -64.0 & 6.72 & 6.63 & -0.82 & 0.25 & 5 \\
193.1159 & +08.0557 & 1.13 & 0.03 &  0.79 & 0.02 & 6359.3 & 22.6 & 4.49 & 0.01 & 43.0 & 5.6 & 5.6 & -1.9 & 0.34 & 6 \\
193.5533 & +11.5037 & 1.42 & 0.03 &  0.56 & 0.01 & 6078.8 & 17.3 & 4.6 & 0.01 & 16.0 & 6.84 & 6.87 & -0.65 & 0.15 & 5 \\
193.8390 & +11.4150 & -0.02 & 0.04 & 19.12 & 0.4 & 4764.0 & 32.0 & 1.22 & 0.03 & 4.0 & 4.73 & 4.54 & -2.83 & 0.12 & 6 \\
196.3755 & +08.5138 & 0.07 & 0.03  & 11.86 & 0.44 & 5011.6 & 20.0 & 1.73 & 0.04 & -72.0 & 4.77 & 4.77 & -2.73 & 0.15 & 6 \\
196.4117 & +14.3176 & 1.35 & 0.03 & 0.55 & 0.01 & 5802.4 & 16.3 & 4.67 & 0.01 & -50.0 & 6.3 & 6.64 & -1.06 & 0.14 & 5 \\
196.5453 & +12.1211 & 0.74 & 0.03 & 1.28 & 0.05 & 6700.0 & 17.0 & 4.25 & 0.03 & 55.0 & 5.15 & 5.15 & -2.35 & 0.11 & 5 \\
198.5486 & +11.4124 & 0.64 & 0.03 & 1.63 & 0.03 & 6493.6 & 46.5 & 3.83 & 0.03 & 77.0 & 5.06 & 4.8 & -2.55 & 0.2 & 5 \\
200.0999 & +13.7229 & 0.01 & 0.03 & 20.72 & 0.45 & 4775.0 & 33.0 & 1.24 & 0.03 & 191.0 & 5.07 & 4.93 & -2.48 & 0.23 & 5 \\
200.5298 & +08.9768 & 0.46 & 0.04 & 7.14 & 0.57 & 5363.3 & 34.7 & 2.53 & 0.08 & 98.0 & 6.35 & 5.77 & -1.38 & 0.21 & 5 \\
201.1158 & +15.4382 & 1.03 & 0.03 & 2.19 & 0.06 & 5689.0 & 66.0 & 3.37 & 0.03 & -5.0 & 5.41 & 5.55 & -2.03 & 0.13 & 5 \\
201.3732 & +08.4513 & 0.55 & 0.04 & 2.80 & 0.13 & 5733.0 & 102.0 & 3.28 & 0.06 & 43.0 & 6.63 & 6.39 & -0.97 & 0.08 & 5 \\
201.8711 & +07.1811 & 0.95 & 0.04 & 0.96 & 0.03 & 6501.0 & 27.0 & 4.44 & 0.02 & 32.0 & 4.55 & 4.49 & -2.97 & 0.09 & 4 \\
202.3435 & +13.2291 & 0.46 & 0.02 & 2.28 & 0.09 & 5749.0 & 93.0 & 3.3 & 0.05 & 102.0 & 6.26 & 6.8 & -1.02 & 0.15 & 5, D2 \\
203.2831 & +13.6326 & 0.04 & 0.03 & 13.13 & 0.3 & 5008.0 & 12.1 & 1.73 & 0.03 & -140.0 & 5.17 & 4.84 & -2.44 & 0.21 & 6, D2 \\
204.9008 & +10.5513 & 0.66 & 0.04 & 1.43 & 0.07 & 6717.9 & 17.4 & 4.22 & 0.03 & -246.0 & 4.74 & 4.62 & -2.8 & 0.19 & 5 \\
205.1342 & +13.8234 & 0.13 & 0.05 & 3.79 & 0.27 & 5462.3 & 28.8 & 2.9 & 0.07 & 127.0 & 5.68 & 6.05 & -1.63 & 0.25 & 4 \\
205.8131 & +15.3832 & 0.74 & 0.03 & 1.27 & 0.05 & 6718.0 & 15.0 & 4.23 & 0.03 & 121.0 & 5.08 & 5.0 & -2.45 & 0.16 & 6 \\
206.3487 & +09.3099 & 0.71 & 0.04 & 1.49 & 0.03 & 6572.0 & 103.0 & 3.94 & 0.03 & 159.0 & 5.86 & 5.52 & -1.78 & 0.2 & 5 \\
208.0798 & +04.4267 & 0.08 & 0.04 & 3.5 & 0.23 & 5572.0 & 66.0 & 2.97 & 0.07 & -129.0 & 4.39 & 4.78 & -2.98 & 0.2 & 6 \\
209.9364 & +15.9251 & 0.52 & 0.05 & 1.93 & 0.09 & 6664.0 & 165.0 & 3.96 & 0.06 & -92.0 & 5.14 & 4.96 & -2.42 & 0.15 & 6 \\
210.0166 & +14.6289 & 0.16 & 0.03 & 5.66 & 0.31 & 5287.7 & 25.5 & 2.47 & 0.06 & 75.0 & 4.94 & 4.69 & -2.64 & 0.11 & 6 \\
210.8632 & +08.1798 & 0.42 & 0.04 & 2.68 & 0.15 & 5592.0 & 75.0 & 3.21 & 0.06 & -12.0 & 5.39 & 5.25 & -2.16 & 0.09 & 6 \\
213.2814 & +14.8983 & 0.20 & 0.05 & 2.14 & 0.03 & 6001.4 & 54.4 & 3.55 & 0.03 & -14.0 & 6.89 & 7.26 & -0.49 & 0.1 & 6 \\
214.5556 & +07.4670 & 0.54 & 0.03 & 1.82 & 0.05 & 6482.0 & 203.0 & 3.88 & 0.05 & 47.0 & 4.93 & 4.86 & -2.59 & 0.23 & 6 \\
217.3861 & +15.1651 & 0.78 & 0.03 & 1.14 & 0.04 & 6663.0 & 19.0 & 4.3 & 0.02 & -160.0 & 6.43 & 6.1 & -1.2 & 0.13 & 5 \\
217.5786 & +14.0379 & 0.03 & 0.03 & 11.61 & 0.14 & 4967.5 & 12.2 & 1.64 & 0.03 & -16.0 & 4.87 & 4.77 & -2.66 & 0.13 & 6, D2 \\
217.6443 & +15.9634 & 0.66 & 0.03 & 1.32 & 0.06 & 6687.0 & 21.0 & 4.27 & 0.03 & -20.0 & 5.92 & 5.51 & -1.72 & 0.41 & 3 \\
218.4622 & +10.3683 & 0.08 & 0.05 & 2.57 & 0.04 & 5923.1 & 41.2 & 3.5 & 0.03 & -126.0 & 5.11 & 5.75 & -2.18 & 0.41 & 6 \\
228.4607 & +08.3553 & 0.77 & 0.04 & 1.22 & 0.05 & 6624.9 & 25.0 & 4.32 & 0.03 & 9.0 & 5.09 & 4.91 & -2.47 & 0.19 & 6 \\
228.6557 & +09.0914 & 0.70 & 0.03 & 1.31 & 0.06 & 6695.0 & 19.8 & 4.26 & 0.03 & -147.0 & 5.08 & 4.98 & -2.45 & 0.15 & 6 \\
228.8159 & +00.2222 & 0.92 & 0.06  & 0.97 & 0.05 & 6519.5 & 32.2 & 4.41 & 0.02 &   6.0 & 5.52 & 5.49 & -1.99 & 0.32 & 6, dw \\
...      & ...      & ...  & ...   & 1.72 & 0.04 & 6603.2 & 33.9 & 3.96 & 0.03 &  ...  & 5.61 & 5.32 & -1.98 & 0.31 & 6, gi  \\
233.5730 & +06.4702 & 0.10 & 0.03 & 11.51 & 0.29 & 4990.9 & 12.7 & 1.69 & 0.03 & -80.0 & 4.82 & 4.38 & -2.82 & 0.26 & 6 \\
235.1449 & +08.7464 & 0.51 & 0.03 & 2.04 & 0.05 & 6167.0 & 87.9 & 3.64 & 0.05 & -156.0 & 4.73 & 4.52 & -2.84 & 0.13 & 6 \\
236.1068 & +10.5311 & -0.01 & 0.03 & 13.22 & 0.31 & 4815.9 & 15.6 & 1.47 & 0.03 & -42.0 & 5.3 & 5.05 & -2.28 & 0.19 & 6 \\
237.8246 & +10.1427 & 0.20 & 0.03 & 6.01 & 0.43 & 5405.0 & 72.0 & 2.63 & 0.08 & -165.0 & 4.76 & 3.82 & -3.05 & 0.19 & 6 \\
240.4216 & +09.6761 & 0.11 & 0.03 & 8.33 & 0.41 & 5204.1 & 24.1 & 2.16 & 0.05 & 38.0 & 4.51 & 4.2 & -3.09 & 0.16 & 6 \\
245.5747 & +06.8844 & 0.21 & 0.04 & 5.31 & 0.4 & 5424.0 & 74.0 & 2.68 & 0.08 & -189.0 & 4.69 & 4.25 & -2.98 & 0.09 & 5 \\
245.8356 & +13.8777 & 0.15 & 0.02 & 4.85 & 0.23 & 5257.0 & 22.2 & 2.28 & 0.05 & -177.0 & 4.76 & 4.4 & -2.86 & 0.19 & 6 \\
246.5144 & +05.9826 & 1.03 & 0.03 & 2.51 & 0.06 & 5736.0 & 53.0 & 3.28 & 0.03 & -6.0 & 7.71 & 7.8 & 0.25 & 0.27 & 5 \\
246.8588 & +12.3193 & 0.19 & 0.02 & 7.01 & 0.34 & 5172.4 & 24.2 & 2.21 & 0.05 & -86.0 & 5.38 & 5.12 & -2.2 & 0.22 & 6 \\
248.4959 & +15.0776 & 0.03 & 0.02 & 9.75 & 0.36 & 5068.6 & 19.8 & 1.86 & 0.04 & -74.0 & 4.95 & 5.07 & -2.51 & 0.29 & 6 \\
250.6963 & +08.3743 & 0.02 & 0.03 & 7.82 & 0.29 & 5076.2 & 20.6 & 2.0 & 0.04 & -4.0 & 4.99 & 4.99 & -2.51 & 0.31 & 6 \\
251.4082 & +12.3657 & 0.08 & 0.04 & 14.85 & 0.43 & 4918.8 & 17.4 & 1.54 & 0.04 & -5.0 & 4.3 & 4.0 & -3.3 & 0.11 & 6 \\
252.1639 & +15.0648 & 0.08 & 0.04 & 10.47 & 0.31 & 4909.2 & 18.7 & 1.65 & 0.04 & -157.0 & 5.22 & 5.18 & -2.29 & 0.09 & 5 \\
253.8582 & +15.7240 & 0.09 & 0.03 & 9.17 & 0.37 & 5077.0 & 52.0 & 2.0 & 0.05 & -97.0 & 4.92 & 4.53 & -2.71 & 0.12 & 6 \\
254.5469 & +10.9129 & 0.41 & 0.02 & 2.77 & 0.16 & 5517.7 & 24.3 & 3.04 & 0.06 & 96.0 & 6.07 & 5.91 & -1.49 & 0.12 & 5 \\
255.2671 & +14.9712 & 0.65 & 0.02 & 1.51 & 0.03 & 6479.0 & 58.2 & 3.88 & 0.04 & 28.0 & 5.01 & 4.9 & -2.52 & 0.04 & 5 \\
255.5555 & +10.8612 & 0.11 & 0.03 & 3.98 & 0.23 & 5495.4 & 23.7 & 2.86 & 0.06 & -371.0 & 4.53 & 5.22 & -2.74 & 0.25 & 6 \\
\hline
\end{tabular}
\end{table*} 

\begin{table*}
\contcaption{Gaia DR2 parallaxes, along with distances and stellar parameters (T, logg) from the Bayesian inference 
method.}
\begin{tabular}{llrlrlrlrlrrrrll} 
\hline
RA$_{\rm SDSS}$ & DEC$_{\rm SDSS}$ & par & dpar & D & dD & T & dT & log(g) & dlg & RV  & 
  \ion{Fe}{I} & \ion{Fe}{II} & [Fe/H]$_{\rm Q6}$ & $\sigma$Fe$_{\rm Q6}$ & N,Com \\
\hline
\\
\multicolumn{12}{l}{Used the SDSS ($g-r$) solution for [Fe/H]$_P \le-2.5$ :} \\
180.3789 & +00.9470 & -0.24 & 0.53  & 2.02 & 0.11 & 5683.5 & 30.6  & 4.67 & 0.01 & -58.0   & 6.05 & 6.72 & -1.18 & 0.18 & 5, dw \\
...      & ...      & ...   &  ...  & 0.40 & 0.01 & 5565.0 & 30.6  & 3.32 & 0.01 &   ...   & 6.45 & 6.55 & -1.00 & 0.16 & 5, gi  \\
181.3699 & +11.7636 & 0.74  & 0.03  & 2.36 & 0.12 & 5494.0 & 24.8  & 3.18 & 0.06 & 81.0    & 5.96 & 5.83 & -1.59 & 0.29 & 5 \\
188.1264 & +08.7740 & 0.85  & 0.04  & 5.88 & 0.35 & 5210.5 & 29.6  & 2.30 & 0.07 & -49.0   & 5.55 & 5.70 & -1.89 & 0.30 & 5 \\
200.7620 & +09.4375 & 1.01  & 0.03  & 0.75 & 0.01 & 6237.2 & 20.3  & 4.52 & 0.01 &  34.0   & 7.10 & 6.88 & -0.49 & 0.24 & 5, dw \\
...      & ...      &  ...  & ...   & 1.95 & 0.03 & 5896.8 & 49.0  & 3.61 & 0.03 &   ...   & 7.30 & 6.76 & -0.47 & 0.26 & 5, gi  \\
229.1219 & +00.9089 & 0.51  & 0.04  & 1.84 & 0.12 & 6485.3 & 183.2 & 3.88 & 0.12 & -222.0  & 4.91 & 4.82 & -2.62 & 0.12 & 6 \\
235.7578 & +09.0000 & 0.60  & 0.04  & 1.83 & 0.07 & 6653.9 & 41.1  & 4.01 & 0.04 & -58.0   & 5.69 & 5.51 & -1.87 & 0.04 & 6 \\
237.9600 & +15.4022 & 0.70  & 0.03  & 1.41 & 0.06 & 6707.0 & 15.7  & 4.09 & 0.04 & -268.0  & 5.33 & 5.06 & -2.26 & 0.26 & 6 \\
245.4387 & +08.9954 & 0.56  & 0.03  & 1.87 & 0.04 & 6463.6 & 60.6  & 3.87 & 0.04 & -83.0   & 6.07 & 5.49 & -1.62 & 0.09 & 6 \\
248.4394 & +07.9229 & 0.95  & 0.03  & 0.96 & 0.02 & 6636.8 & 14.6  & 4.31 & 0.02 & -16.0   & 5.85 & 5.65 & -1.71 & 0.17 & 6 \\
254.5207 & +15.4969 & 1.47  & 0.02  & 0.61 & 0.01 & 6327.6 & 11.7  & 4.49 & 0.01 & 25.0    & 7.30 & 7.18 & -0.25 & 0.14 & 5 \\
\\
\multicolumn{12}{l}{Low probability for [Fe/H]$_P \le-2.5$:} \\
182.8512 & +14.1595 & 0.11 & 0.03  &  8.88 & 0.23 & 4958.7 & 10.2  & 1.75 & 0.02 & 80.0   & 5.72 & 5.95 & -1.69 & 0.17 & 5 \\
196.6015 & +15.6768 & 1.40 & 0.05  &  0.55 & 0.01 & 5922.2 & 24.6  & 4.62 & 0.01 & -104.0 & 6.64 & 6.66 & -0.85 & 0.14 & 5 \\
197.9861 & +12.3578 & 0.41 & 0.10  &  2.23 & 0.12 & 6102.0 & 521.0 & 3.72 & 0.13 & 116.0  & 6.46 & 6.15 & -1.16 & 0.06 & 5 \\
209.2123 & +01.5275 & 0.56 & 0.04  &  1.95 & 0.05 & 5952.8 & 95.4  & 3.57 & 0.06 & 13.0   & 5.81 & 5.73 & -1.72 & 0.19 & 6 \\
209.7175 & +10.8612 & 1.23 & 0.03  &  0.78 & 0.02 & 6358.0 & 20.0  & 4.50 & 0.01 & -137.0 & 5.70 & 5.41 & -1.90 & 0.14 & 6 \\
210.7504 & +12.7744 & 0.01 & 0.04  & 13.82 & 0.28 & 4651.9 & 15.4  & 1.35 & 0.03 & 41.0   & 5.33 & 5.82 & -2.01 & 0.09 & 6 \\
211.2766 & +10.3280 & 0.79 & 0.03  &  1.04 & 0.03 & 6527.9 & 22.1  & 4.40 & 0.02 &  45.0  & 6.47 & 6.30 & -1.09 & 0.15 & 6, dw \\
...      & ...      & ...  & ...   &  1.99 & 0.02 & 6370.1 & 39.5  & 3.81 & 0.02 &  ...   & 6.48 & 6.08 & -1.22 & 0.14 & 6, gi  \\
212.5834 & +10.5365 & 1.37 & 0.04  &  0.66 & 0.01 & 6222.0 & 22.5  & 4.55 & 0.01 & -127.0 & 5.68 & 5.55 & -1.86 & 0.16 & 6 \\
215.6121 & +15.0163 & 0.21 & 0.04  &  3.99 & 0.22 & 5267.4 & 25.2  & 2.62 & 0.06 & -120.0 & 6.53 & 5.92 & -1.12 & 0.10 & 4 \\
216.1245 & +10.2135 & 0.34 & 0.04  &  4.04 & 0.30 & 5412.1 & 31.0  & 2.78 & 0.08 & 107.0  & 5.26 & 5.03 & -2.31 & 0.12 & 6 \\
223.5273 & +11.1353 & 0.00 & 0.04  &  10.6 & 0.25 & 4757.4 & 16.4  & 1.55 & 0.03 & 125.0  & 5.41 & 5.33 & -2.12 & 0.21 & 6 \\
227.2885 & +01.3598 & 0.97 & 0.05  &  1.79 & 0.03 & 5950.4 & 63.6  & 3.64 & 0.04 & 6.0    & 7.13 & 6.82 & -0.47 & 0.25 & 6 \\
229.0409 & +10.3020 & 1.95 & 0.04  &  0.39 & 0.00 & 5400.0 & 16.6  & 4.71 & 0.01 & -10.0  & 6.42 & 6.62 & -0.98 & 0.15 & 4 \\
232.6956 & +08.3392 & 0.34 & 0.04  &  2.96 & 0.16 & 5640.6 & 40.2  & 3.30 & 0.06 & 38.0   & 5.31 & 5.01 & -2.29 & 0.24 & 6 \\
233.9312 & +09.5596 & 0.38 & 0.04  &  2.80 & 0.14 & 5645.4 & 38.5  & 3.30 & 0.06 & -99.0  & 5.71 & 5.10 & -2.00 & 0.09 & 6 \\
234.4398 & +12.6286 & 1.31 & 0.04  &  0.53 & 0.01 & 5867.1 & 22.0  & 4.63 & 0.01 & -66.0  & 6.99 & 6.96 & -0.53 & 0.09 & 5 \\
235.9710 & +09.1864 & 0.72 & 0.03  &  1.54 & 0.04 & 6581.5 & 33.9  & 4.02 & 0.03 & -32.0  & 5.91 & 5.81 & -1.62 & 0.18 & 6 \\
236.0769 & +04.0496 & 0.78 & 0.04  &  1.77 & 0.04 & 5753.4 & 60.2  & 3.51 & 0.04 &  0.0   & 7.81 & ...  &  0.31 & 0.08 & 2 \\
241.1186 & +09.4156 & 0.55 & 0.03  &  1.76 & 0.06 & 6299.3 & 122.6 & 3.77 & 0.07 & -52.0  & 5.65 & 5.44 & -1.92 & 0.14 & 6 \\
\\
\multicolumn{12}{l}{No longer in the {\it Pristine} catalogue:} \\
213.7879 & +08.4232 & 0.31 & 0.05 & 7.85 & 0.48 & 5289.0 & 29.0 & 2.27 & 0.07 & -107.0 & 4.88 & 4.44 & -2.76 & 0.18 & 6, ** \\
\hline
\end{tabular}
\end{table*} 

\begin{table*}
\caption{Iron-group and heavy element abundances in the new 28 metal-poor stars, and our analysis of HD122563.  
[Fe/H] is calculated as the weighted mean of \ion{Fe}{I} and \ion{Fe}{II}, and $\delta$([Fe/H]) 
= $\sigma$([Fe/H])/(N\ion{Fe}{I} + N\ion{Fe}{II}))$^{\rm 1/2}$ .
} 
\label{tab:iron}
\begin{tabular}{llllllll} 
\hline
RA$_{\rm SDSS}$ & [Fe/H] $\pm\delta$ &  log(\ion{Fe}{I}) $\pm\sigma$ (N) & log(\ion{Fe}{II}) $\pm\sigma$ (N) & [Cr/Fe] $\pm\sigma$ (N) & [Ni/Fe] $\pm\sigma$ (N) & [Y/Fe] $\pm\sigma$ (N) &  [Ba/Fe] $\pm\sigma$ (N) \\ 
\hline
180.2206 & $-2.92 \pm0.03$ & 4.60 $\pm0.18$ (49) & 4.45 $\pm 0.27$ (4) & $-0.21 \pm0.13$ (2) &  $-0.05 \pm0.18$ (1)  & $<+0.49$            &  $-1.02 \pm0.18$ (1)  \\ 
181.2243 & $-3.11 \pm0.07$ & 4.50 $\pm0.17$ ( 4) & 4.17 $\pm 0.14$ (2) & ...                 &   ...                 & $<+1.85$            &  $<+1.37$             \\ 
181.4395 & $-2.52 \pm0.02$ & 4.99 $\pm0.18$ (86) & 4.80 $\pm 0.26$ (5) & $-0.37 \pm0.11$ (3) &  $+0.03 \pm0.13$ (2)  & $<-0.12$            &  $-1.13 \pm0.24$ (2)  \\ 
183.6849 & $-3.16 \pm0.07$ & 4.38 $\pm0.24$ ( 2) & 4.30 $\pm 0.01$ (2) & ...                 &   ...                 & $<+2.33$            &  $<+1.88$             \\ 
189.9449 & $-2.78 \pm0.04$ & 4.77 $\pm0.12$ ( 8) & 4.50 $\pm 0.21$ (2) & ...                 &   ...                 & $<+1.70$            &  $<+0.93$             \\ 
193.8390 & $-2.80 \pm0.02$ & 4.71 $\pm0.21$ (81) & 4.49 $\pm 0.13$ (5) & $-0.49 \pm0.12$ (3) &  $-0.24 \pm0.21$ (1)  & $<+0.02$            &  $-1.58 \pm0.21$ (1)  \\ 
196.3755 & $-2.80 \pm0.02$ & 4.70 $\pm0.19$ (61) & 4.72 $\pm 0.16$ (3) & $-0.17 \pm0.14$ (2) &  $-0.21 \pm0.19$ (1)  & $<+0.3 $            &  $<-0.56$             \\ 
198.5486 & $-2.47 \pm0.05$ & 5.08 $\pm0.16$ ( 9) & 4.80 $\pm 0.29$ (2) & $+0.04 \pm0.11$ (2) &  $<+0.45$             & $<+1.23$            &  $<+0.39$             \\ 
201.8711 & $-2.93 \pm0.11$ & 4.57 $\pm0.15$ ( 2) & $<4.71$             & ...                 &   ...                 & $<+2.10$            &  $<+1.65$             \\ 
203.2831 & $-2.74 \pm0.02$ & 4.77 $\pm0.19$ (59) & 4.55 $\pm 0.15$ (4) & $-0.27 \pm0.11$ (3) &  $-0.19 \pm0.19$ (1)  & $<+0.01$            &  $-0.76 \pm0.23$ (3)  \\ 
204.9008 & $-2.73 \pm0.08$ & 4.84 $\pm0.18$ ( 3) & 4.66 $\pm 0.26$ (2) & ...                 &   ...                 & $<+1.89$            &  $<+1.40$             \\ 
208.0798 & $-2.94 \pm0.03$ & 4.53 $\pm0.14$ (20) & 4.83 $\pm 0.26$ (2) & $-0.35 \pm0.10$ (2) &  $<+0.16$             & $<+1.15$            &  $<+0.57$             \\ 
210.0166 & $-2.59 \pm0.02$ & 4.92 $\pm0.18$ (64) & 4.75 $\pm 0.13$ (4) & $-0.20 \pm0.13$ (2) &  $-0.17 \pm0.18$ (1)  & $<+0.51$            &  $+0.74 \pm0.20$ (4)  \\ 
213.7879 & $-2.59 \pm0.02$ & 4.93 $\pm0.18$ (61) & 4.55 $\pm 0.17$ (3) & $-0.09 \pm0.11$ (3) &  $+0.11 \pm0.18$ (1)  & $<+0.43$            &  $-0.36 \pm0.14$ (2)  \\ 
214.5556 & $-2.51 \pm0.05$ & 5.01 $\pm0.15$ (10) & 4.88 $\pm 0.35$ (2) & $<+0.07$            &  $<+0.69$             & $+1.48 \pm0.11$ (2) &  $+1.77 \pm0.22$ (4)  \\ 
217.5786 & $-2.61 \pm0.02$ & 4.88 $\pm0.18$ (79) & 4.94 $\pm 0.18$ (6) & $-0.11 \pm0.16$ (4) &  $-0.04 \pm0.18$ (1)  & $+0.16 \pm0.13$ (2) &  $+0.02 \pm0.18$ (3)  \\ 
229.1219 & $-2.52 \pm0.04$ & 5.02 $\pm0.12$ ( 8) & 4.84 $\pm 0.17$ (2) & $+0.16 \pm0.12$ (1) &   ...                 & $<+1.59$            &  $<+0.44$             \\ 
233.5730 & $-2.75 \pm0.02$ & 4.77 $\pm0.20$ (75) & 4.42 $\pm 0.28$ (3) & $-0.25 \pm0.12$ (3) &   ...                 & $<+0.23$            &  $-0.15 \pm0.32$ (3)  \\ 
235.1449 & $-2.69 \pm0.04$ & 4.85 $\pm0.15$ (16) & 4.52 $\pm 0.13$ (2) & $<-0.16$            &   ...                 & $<+1.41$            &  $+0.13 \pm0.15$ (1)  \\ 
237.8246 & $-3.29 \pm0.04$ & 4.28 $\pm0.15$ (13) & 3.73 $\pm 0.04$ (2) & $<-0.32$            &  $<+0.50$             & $<+1.29$            &  $+0.61 \pm0.34$ (2)  \\ 
240.4216 & $-2.95 \pm0.03$ & 4.56 $\pm0.20$ (41) & 4.20 $\pm 0.22$ (2) & $-0.35 \pm0.14$ (2) &  $-0.19 \pm0.20$ (1)  & $<+0.64$            &  $+0.53 \pm0.29$ (4)  \\ 
245.5747 & $-3.14 \pm0.04$ & 4.37 $\pm0.20$ (18) & 4.25 $\pm 0.09$ (2) & $<-0.30$            &  $<+0.38$             & $<+1.14$            &  $<+0.52$             \\ 
245.8356 & $-2.78 \pm0.03$ & 4.73 $\pm0.21$ (52) & 4.60 $\pm 0.25$ (4) & $-0.41 \pm0.16$ (5) &  $+0.18 \pm0.21$ (1)  & $+0.73 \pm0.21$ (1) &  $-0.51 \pm0.12$ (3)  \\ 
248.4959 & $-2.59 \pm0.02$ & 4.91 $\pm0.17$ (74) & 4.90 $\pm 0.31$ (5) & $-0.26 \pm0.10$ (3) &  $-0.04 \pm0.12$ (2)  & $<+0.16$            &  $-0.18 \pm0.10$ (3)  \\ 
250.6963 & $-2.55 \pm0.03$ & 4.95 $\pm0.20$ (62) & 4.95 $\pm 0.28$ (4) & $-0.43 \pm0.12$ (3) &  $+0.07 \pm0.14$ (2)  & $<+0.18$            &  $-0.03 \pm0.16$ (3)  \\ 
251.4082 & $-3.27 \pm0.03$ & 4.23 $\pm0.18$ (31) & 4.19 $\pm 0.33$ (3) & $<-0.17$            &   ...                 & $<+0.65$            &  $-0.73 \pm0.18$ (1)  \\ 
253.8582 & $-2.72 \pm0.02$ & 4.80 $\pm0.21$ (65) & 4.54 $\pm 0.07$ (4) & $-0.15 \pm0.14$ (2) &  $-0.13 \pm0.21$ (1)  & $<+0.35$            &  $-0.51 \pm0.20$ (3)  \\ 
255.5555 & $-2.83 \pm0.03$ & 4.64 $\pm0.18$ (22) & 4.99 $\pm 0.10$ (2) & $<-0.08$            &  $<-0.08$             & $<+0.97$            &  $+0.35 \pm0.10$ (3)  \\ 
HD122563 & $-2.76 \pm0.01$ & 4.73 $\pm0.15$ (98) & 4.86 $\pm 0.16$ (5) & $-0.39 \pm0.08$ (3) &  $+0.05 \pm0.10$ (5)  & $-0.15 \pm0.11$ (2) &  $-0.77 \pm0.08$ (3)  \\  
\hline
\end{tabular}
\end{table*} 

\begin{table*}
\caption{Light element abundances in the new 28 metal-poor stars (and our analysis of HD122563).  
When the number of lines N$_X$$<$4 for species X, then $\sigma$(X) = $\sigma$\ion{Fe}{I} /$\sqrt{N_X}$. }
\label{tab:alpha}
\begin{tabular}{lllllll} 
\hline
RA$_{\rm SDSS}$ &  [Na/Fe] $\pm\sigma$ (N) & [Mg/Fe] $\pm\sigma$ (N) & [Ca/Fe] $\pm\sigma$ (N) & [Sc/Fe] $\pm\sigma$ (N) & [TiI/Fe] $\pm\sigma$ (N) & [TiII/Fe]  $\pm\sigma$ (N) \\
\hline
180.2206 & $+0.39 \pm0.13$ (2) & $+0.30 \pm0.11$ (3) & $+0.35 \pm0.10$ (4) & ...                 & $+0.22 \pm0.13$ (2) & $+0.28 \pm0.11$ (3)   \\
181.2243 & $-0.23 \pm0.17$ (1) & $+0.67 \pm0.13$ (2) & $<+0.94$            & ...                 & ...                 & $<+1.23$              \\
181.4395 & $+0.07 \pm0.13$ (2) & $+0.26 \pm0.11$ (3) & $+0.42 \pm0.19$ (9) & $-0.15 \pm0.11$ (3) & $+0.16 \pm0.19$ (9) & $+0.19 \pm0.16$ (8)   \\
183.6849 & $-0.18 \pm0.17$ (2) & $+0.13 \pm0.14$ (3) & $<+1.01$            & ...                 & ...                 & $<+1.88$              \\
189.9449 & $+0.01 \pm0.08$ (2) & $+0.17 \pm0.08$ (2) & $<+0.65$            & ...                 & ...                 & $<+0.92$              \\
193.8390 & $<+1.17$            & $+0.37 \pm0.12$ (3) & $+0.33 \pm0.14$ (8) & $+0.08 \pm0.12$ (3) & $+0.09 \pm0.12$ (8) & $+0.17 \pm0.18$ (8)   \\
196.3755 & $+0.22 \pm0.13$ (2) & $+0.29 \pm0.11$ (3) & $+0.45 \pm0.24$ (7) & ...                 & $+0.43 \pm0.25$ (4) & $+0.28 \pm0.19$ (4)   \\
198.5486 & $+0.24 \pm0.11$ (2) & $+0.25 \pm0.09$ (3) & $+0.53 \pm0.16$ (1) & $<+0.97$            & ...                 & $<+0.71$              \\
201.8711 & $-0.12 \pm0.11$ (2) & $-0.05 \pm0.11$ (2) & $<+1.16$            & ...                 & ...                 & $<+1.59$              \\
203.2831 & $+0.66 \pm0.13$ (2) & $+0.02 \pm0.13$ (2) & $+0.21 \pm0.11$ (3) & ...                 & $+0.43 \pm0.06$ (6) & $+0.50 \pm0.10$ (5)   \\
204.9008 & $-0.15 \pm0.18$ (1) & $-0.12 \pm0.13$ (2) & $<+0.96$            & ...                 & ...                 & $<+1.36$              \\
208.0798 & $+0.26 \pm0.10$ (2) & $+0.09 \pm0.08$ (3) & $+0.45 \pm0.14$ (1) & ...                 & ...                 & $<+0.78$              \\
210.0166 & $+0.06 \pm0.13$ (2) & $+0.18 \pm0.11$ (3) & $+0.36 \pm0.28$ (4) & $<+0.44$            & $+0.37 \pm0.11$ (3) & $+0.12 \pm0.13$ (2)   \\
213.7879 & $-0.12 \pm0.13$ (2) & $+0.33 \pm0.11$ (3) & $+0.44 \pm0.09$ (4) & $<+0.26$            & $+0.45 \pm0.24$ (7) & $+0.45 \pm0.13$ (2)   \\
214.5556 & $+0.02 \pm0.12$ (2) & $+0.36 \pm0.09$ (3) & $<+0.82$            & $<+1.11$            & ...                 & $<+0.96$              \\
217.5786 & $<+1.14$            & $+0.22 \pm0.11$ (3) & $+0.55 \pm0.24$ (6) & $+0.09 \pm0.13$ (2) & $+0.23 \pm0.12$ (5) & $+0.55 \pm0.16$ (10)  \\
229.1219 & $-0.13 \pm0.09$ (2) & $+0.22 \pm0.09$ (2) & ....                & ...                 & $+1.26 \pm0.12$ (1) & $+1.44 \pm0.12$ (1)   \\
233.5730 & $+0.65 \pm0.14$ (2) & $+0.24 \pm0.12$ (3) & $+0.39 \pm0.32$ (5) & $<+0.10$            & $+0.24 \pm0.15$ (4) & $+0.28 \pm0.12$ (3)   \\
235.1449 & $-0.26 \pm0.11$ (2) & $+0.09 \pm0.11$ (2) & $<+0.67$            & $<+1.16$            &  ...                & $<+0.94$              \\
237.8246 & $+0.00 \pm0.11$ (2) & $+0.04 \pm0.11$ (2) & $<+0.88$            & $<+1.12$            &  ...                & $<+0.89$              \\
240.4216 & $+0.14 \pm0.14$ (2) & $+0.14 \pm0.12$ (3) & $+0.31 \pm0.12$ (3) & $<+0.54$            & $+0.44 \pm0.14$ (2) & $+0.52 \pm0.14$ (2)   \\
245.5747 & $-0.17 \pm0.14$ (2) & $+0.14 \pm0.14$ (2) & $<+0.96$            & $<+0.99$            & ...                 & $<+0.75$              \\
245.8356 & $+0.67 \pm0.15$ (2) & $+0.29 \pm0.11$ (4) & $+0.66 \pm0.22$ (6) & $+0.30 \pm0.15$ (2) & $+0.64 \pm0.15$ (2) & $+0.46 \pm0.25$ (11)  \\
248.4959 & $+0.39 \pm0.13$ (2) & $+0.14 \pm0.10$ (3) & $+0.35 \pm0.17$ (7) & $<+0.07$            & $+0.48 \pm0.28$ (4) & $+0.36 \pm0.18$ (5)   \\
250.6963 & $<+1.71$            & $+0.11 \pm0.12$ (3) & $+0.57 \pm0.38$ (8) & $+0.03 \pm0.20$ (1) & $+0.40 \pm0.24$ (6) & $+0.39 \pm0.12$ (8)   \\
251.4082 & $<+1.61$            & $-0.20 \pm0.13$ (2) & $<+0.11$            & $<+0.52$            &  ...                & $<+0.28$              \\
253.8582 & $+0.12 \pm0.15$ (2) & $+0.23 \pm0.12$ (3) & $+0.27 \pm0.20$ (4) & $<+0.20$            & $+0.39 \pm0.24$ (4) & $+0.37 \pm0.18$ (5)   \\
255.5555 & $+0.76 \pm0.13$ (2) & $+0.29 \pm0.11$ (3) & $+0.46 \pm0.10$ (3) & $<+0.78$            &  ...                & $<+0.88$              \\
HD122563 & $+0.21 \pm0.11$ (2) & $+0.18 \pm0.09$ (3) & $+0.32 \pm0.14$ (8) & $+0.19 \pm0.11$ (2) & $+0.07 \pm0.05$ (9) & $+0.46 \pm0.08$ (9)   \\
\hline
\end{tabular}
\end{table*} 

\begin{table*}
\caption{Total systematic errors (dX) per element species (X) due to the stellar parameters (T, logg, [Fe/H])
added in quadrature, where dT and dlogg are from Table~\ref{tab:newfs} and $\delta$[Fe/H] is 
from Table~\ref{tab:iron}.
}
\label{tab:syserrs}
\begin{tabular}{lrrrrrrrrrrrr} 
\hline
RA$_{\rm SDSS}$ & dFeI & dFeII & dNa & dMg  & dCa &  dSc  &  dTiI & dTiII & dCrI &  dNiI & dYII & dBaII \\
\hline
180.2206 & 0.03 & 0.02 & 0.03 & 0.02 & 0.01 &  ... &  0.02 & 0.02 &  0.02 &  0.02 &  ... &  0.02 \\
181.2243 & 0.09 & 0.03 & 0.07 & 0.08 & ...  &  ... &   ... &  ... &   ... &   ... &  ... &  ...  \\
181.4395 & 0.01 & 0.01 & 0.01 & 0.01 & 0.01 & 0.01 &  0.01 & 0.01 &  0.01 &  0.01 &  ... &  ...  \\
183.6849 & 0.04 & 0.01 & 0.03 & 0.03 & ...  &  ... &   ... &  ... &   ... &   ... &  ... &  ...  \\
189.9449 & 0.12 & 0.02 & 0.09 & 0.09 & ...  &  ... &   ... &  ... &   ... &   ... &  ... &  ...  \\
193.8390 & 0.04 & 0.01 & 0.05 & 0.04 & 0.03 & 0.01 &  0.05 & 0.01 &  0.04 &  0.04 &  ... &  ...  \\
196.3755 & 0.02 & 0.01 & 0.02 & 0.02 & 0.01 &  ... &  0.03 & 0.01 &  0.02 &  0.02 &  ... &  ...  \\
198.5486 & 0.04 & 0.01 & 0.03 & 0.03 & 0.03 &  ... &   ... &  ... &  0.04 &  ...  &  ... &  ...  \\
201.8711 & 0.03 & ...  &  0.02 & 0.02 & ... &  ... &   ... &  ... &   ... &   ... &  ... &  ...  \\
203.2831 & 0.01 & 0.01 & 0.02 & 0.02 & 0.01 &  ... &  0.02 & 0.01 &  0.01 &  0.01 &  ... &  0.01 \\
204.9008 & 0.01 & 0.01 & 0.01 & 0.01 & ...  &  ... &   ... &  ... &   ... &   ... &  ... &  ...  \\
208.0798 & 0.07 & 0.03 & 0.07 & 0.06 & 0.05 &  ... &   ... &  ... &  0.06 &  ...  &  ... &  ...  \\
210.0166 & 0.03 & 0.02 & 0.03 & 0.03 & 0.02 &  ... &  0.03 & 0.02 &  0.03 &  0.03 &  ... &  0.03 \\
213.7879 & 0.03 & 0.03 & 0.03 & 0.04 & 0.02 &  ... &  0.03 & 0.03 &  0.03 &  0.03 &  ... &  0.03 \\
214.5556 & 0.17 & 0.03 & 0.13 & 0.14 & ...  &  ... &   ... &  ... &   ... &  ...  & 0.09 & 0.15  \\
217.5786 & 0.01 & 0.01 & 0.02 & 0.02 & 0.01 & 0.01 &  0.02 & 0.01 &  0.01 &  0.01 &  ... &  ...  \\
229.1219 & 0.16 & 0.05 & 0.12 & 0.15 & ...  &  ... &  0.15 & 0.06 &  0.15 &  ...  & 0.08 & ...   \\
233.5730 & 0.02 & 0.01 & 0.02 & 0.02 & 0.01 &  ... &  0.02 & 0.01 &  0.02 &  ...  &  ... &  0.01 \\
235.1449 & 0.08 & 0.02 & 0.06 & 0.08 & ...  &  ... &   ... &  ... &   ... &   ... &  ... &  0.06 \\
237.8246 & 0.09 & 0.03 & 0.07 & 0.07 & ...  &  ... &   ... &  ... &   ... &   ... &  ... &  0.06 \\
240.4216 & 0.03 & 0.02 & 0.02 & 0.02 & 0.02 &  ... &  0.03 & 0.02 &  0.03 &  0.03 &  ... &  0.02 \\
245.5747 & 0.08 & 0.03 & 0.07 & 0.08 & ...  &  ... &   ... &  ... &   ... &   ... &  ... &  ...  \\
245.8356 & 0.02 & 0.02 & 0.03 & 0.02 & 0.02 & 0.02 &  0.03 & 0.02 &  0.03 &  0.02 &  ... &  ...  \\
248.4959 & 0.02 & 0.01 & 0.03 & 0.02 & 0.01 &  ... &  0.02 & 0.02 &  0.02 &  0.02 &  ... &  0.02 \\
250.6963 & 0.02 & 0.01 & 0.04 & 0.02 & 0.02 & 0.02 &  0.02 & 0.02 &  0.02 &  0.02 &  ... &  ...  \\
251.4082 & 0.02 & 0.01 & 0.03 & 0.02 & ...  &  ... &   ... &  ... &   ... &   ... &  ... &  0.02 \\
253.8582 & 0.06 & 0.02 & 0.06 & 0.06 & 0.04 &  ... &  0.06 & 0.03 &  0.06 &  0.06 &  ... &  0.04 \\
255.5555 & 0.03 & 0.02 & 0.03 & 0.03 & 0.02 &  ... &   ... &  ... &   ... &   ... &  ... &  0.02 \\
HD122563 & 0.03 & 0.02 & 0.04 & 0.03 & 0.02 & 0.01 &  0.04 & 0.01 &  0.03 &  0.03 &  ... &  ...  \\
\hline
\end{tabular}
\end{table*} 

\begin{table*}
\caption{Samples of the systematic errors from the individual stellar parameters per elemental species.  These were added in quadrature
for these stars (and the rest of the sample) in Table~\ref{tab:syserrs}.   We note that if $\sigma$\ion{Fe}{I} is used instead of 
$\sigma$[Fe/H], then the errors due to metallicity remain negligible. 
} 
\label{tab:errsample}
\begin{tabular}{lrrrrrrrrrrrrr} 
\hline
RA$_{\rm SDSS}$ & Param $\pm$d & d\ion{Fe}{I} & d\ion{Fe}{II} & d\ion{Na}{I} & d\ion{Mg}{I} & d\ion{Ca}{I} & d\ion{Sc}{II} & d\ion{Ti}{I} & d\ion{Ti}{II} & d\ion{Cr}{I} & d\ion{Ni}{I} & d\ion{Y}{II} & d\ion{Ba}{II} \\
\hline
193.8390 & 4764 $\pm$32      & 0.04  & $-$0.01 &    0.05  &    0.03  & 0.03  & 0.01  & 0.05  &   0.01  & 0.04  &  0.04 & ...   & ...   \\ 
193.8390 & 1.22 $\pm$0.03    & 0.00  &   0.01  & $-$0.01  & $-$0.01  & 0.00  & 0.01  & 0.00  &   0.01  & 0.00  &  0.00 & ...   & ...   \\
193.8390 & $-2.80 \pm$0.02   & 0.00  &   0.00  &    0.00  &    0.00  & 0.00  & 0.00  & 0.00  &   0.00  & 0.00  &  0.00 & ...   & ...   \\
\\
213.7879 & 5289  $\pm$29     & 0.03  &   0.00  &    0.03  &    0.03  & 0.02  & ...   & 0.03  &   0.01  & 0.03  &  0.03 & ...   &  0.02 \\
213.7879 & 2.27  $\pm$0.07   & 0.00  &   0.03  &    0.00  & $-$0.02  & 0.00  & ...   & 0.00  &   0.02  & 0.00  &  0.00 & ...   &  0.02 \\
213.7879 & $-2.59 \pm$0.02   & 0.00  &   0.00  &    0.00  &    0.00  & 0.00  & ...   & 0.00  &   0.00  & 0.00  &  0.00 & ...   &  0.00 \\
\\
214.5556 & 6482 $\pm$203     & 0.17  &   0.02  &    0.13  &    0.14  & ...   & ...   & ...   &   ...   & ...   &  ...  &  0.09 &  0.15 \\
214.5556 & 3.88 $\pm$0.05    & 0.00  &   0.02  &    0.00  & $-$0.01  & ...   & ...   & ...   &   ...   & ...   &  ...  &  0.02 &  0.01 \\
214.5556 & $-2.50 \pm$0.05   & 0.00  &   0.00  &    0.00  &    0.00  & ...   & ...   & ...   &   ...   & ...   &  ...  &  0.00 &  0.00 \\
\hline
\end{tabular}
\end{table*}

\begin{table*}
\caption{Orbit and Action parameters for the 70 metal-poor candidates from the \Pristine\ survey.}
\setlength{\tabcolsep}{2pt}
\label{tab:orbits}
\rotatebox{90}{
\begin{tabular}{llrrrrrrrrrrrr} 
\hline
RA       &  DEC    &   pmra            & pmdec             &     U               &   V                 &  W                  &   R$_{\rm apo}$   & R$_{\rm peri}$  & ecc           & Z$_{\rm max}$      & J$_\phi$/J$_{\phi\odot}$ & J$_z$/J$_{z\odot}$ & E/E$_\odot$ \\[.2ex]
         &         &   (\pmu)          & (\pmu)            &    (\kms)           &   (\kms)            &  (\kms)             &     (kpc)         &   (kpc)         &               &    (kpc)           &                          &                    &   \\
\hline
180.2206 &  9.5683 &   1.47 $\pm$ 0.08 &  -3.68 $\pm$ 0.06 & $ 175.69 \pm$ 6.90  & $-143.29 \pm$ 5.31  & $ -26.54 \pm$ 1.96  &   23.5 $\pm$ 1.3   &  6.6 $\pm$ 0.5 & \,\, 0.57 $\pm$ 0.01 &   23.4 $\pm$ 1.3   & $-0.04 \pm$ 0.05 & $5918.45 \pm$ 289.73 & $0.477 \pm$ -0.032 \\ 
180.3789 &  0.9470 & -14.76 $\pm$ 0.92 &  -5.28 $\pm$ 0.49 & $-101.97 \pm$ 6.13  & $ -73.77 \pm$ 5.78  & $-100.36 \pm$ 2.89  &    9.4 $\pm$ 0.3   &  6.3 $\pm$ 0.2 & 0.20 $\pm$ 0.03 &    4.5 $\pm$ 0.4   & $0.67 \pm$ 0.03 & $689.95 \pm$ 82.44 & $0.893 \pm$ -0.011 \\ 
181.2243 &  7.4160 &   1.98 $\pm$ 0.08 &  -8.19 $\pm$ 0.05 & $  60.40 \pm$ 1.88  & $  -4.87 \pm$ 1.86  & $-161.93 \pm$ 0.92  &   19.0 $\pm$ 0.1   &  6.5 $\pm$ 0.1 & 0.49 $\pm$ 0.01 &    9.4 $\pm$ 0.1   & $0.87 \pm$ 0.01 & $948.17 \pm$ 13.67 & $0.58 \pm$ -0.003 \\ 
181.3464 & 11.6698 & -14.31 $\pm$ 0.06 &   2.84 $\pm$ 0.03 & $ -39.73 \pm$ 0.65  & $ -15.73 \pm$ 0.26  & $   6.31 \pm$ 0.48  &    8.8 $\pm$ 0.0   &  7.5 $\pm$ 0.1 & 0.08 $\pm$ 0.00 &    0.7 $\pm$ 0.1   & $0.87 \pm$ 0.0 & $32.49 \pm$ 0.77 & $0.881 \pm$ -0.001 \\ 
181.3699 & 11.7636 &  -9.20 $\pm$ 0.06 &  -4.69 $\pm$ 0.03 & $ -66.05 \pm$ 3.11  & $-116.49 \pm$ 4.39  & $  44.36 \pm$ 1.64  &    8.9 $\pm$ 0.1   &  3.4 $\pm$ 0.2 & 0.45 $\pm$ 0.02 &    2.9 $\pm$ 0.1   & $0.48 \pm$ 0.02 & $303.79 \pm$ 5.57 & $1.045 \pm$ -0.002 \\ 
181.4395 &  1.6294 &  -3.34 $\pm$ 0.08 &  -9.14 $\pm$ 0.04 & $ 134.17 \pm$ 3.39  & $ -768.5 \pm$ 11.62 & $ -177.4 \pm$ 6.21  & 2345.5 $\pm$ 124.2 & 18.1 $\pm$ 0.3 & 0.98 $\pm$ 0.00 & 1049.8 $\pm$ 44.4  & $-2.4 \pm$ 0.05 & $16291.42 \pm$ 799.81 & $-1.919 \pm$ -0.125 \\ 
181.6953 & 13.8075 & -16.18 $\pm$ 0.06 &  -3.44 $\pm$ 0.03 & $-164.22 \pm$ 6.23  & $ -156.8 \pm$ 5.18  & $  28.91 \pm$ 1.87  &   11.7 $\pm$ 0.3   &  2.1 $\pm$ 0.1 & 0.70 $\pm$ 0.02 &    3.7 $\pm$ 0.2   & $0.36 \pm$ 0.02 & $271.81 \pm$ 17.54 & $0.94 \pm$ -0.012 \\ 
182.5364 &  0.9431 & -43.19 $\pm$ 0.08 & -49.25 $\pm$ 0.06 & $-124.30 \pm$ 7.39  & $-702.83 \pm$ 30.92 & $-103.95 \pm$ 15.46 &  541.2 $\pm$ 437.2 &  7.9 $\pm$ 0.3 & 0.94 $\pm$ 0.04 &  164.9 $\pm$ 134.9 & $-1.77 \pm$ 0.12 & $766.27 \pm$ 167.98 & $-0.66 \pm$ -0.285 \\ 
183.6849 &  4.8619 &  -0.70 $\pm$ 0.11 & -12.87 $\pm$ 0.12 & $  31.50 \pm$ 1.67  & $ -68.32 \pm$ 2.46  & $   11.5 \pm$ 1.24  &    8.5 $\pm$ 0.1   &  4.5 $\pm$ 0.1 & 0.30 $\pm$ 0.01 &    1.2 $\pm$ 0.1   & $0.64 \pm$ 0.01 & $84.71 \pm$ 4.01 & $1.022 \pm$ -0.005 \\ 
185.4110 &  7.4777 & -70.21 $\pm$ 0.07 & -37.42 $\pm$ 0.04 & $-146.72 \pm$ 3.89  & $-305.15 \pm$ 5.95  & $  85.86 \pm$ 1.97  &   10.7 $\pm$ 0.2   &  1.5 $\pm$ 0.1 & 0.75 $\pm$ 0.01 &    4.0 $\pm$ 0.2   & $-0.27 \pm$ 0.02 & $352.7 \pm$ 24.77 & $1.002 \pm$ -0.012 \\ 
187.9785 &  8.7294 & -20.24 $\pm$ 0.10 &   2.84 $\pm$ 0.05 & $121.345 \pm$  6.39 & $ -32.49 \pm$  2.74 & $ -55.94 \pm$ 0.52  &   11.5 $\pm$ 0.2   &  6.0 $\pm$ 0.1 & 0.32 $\pm$ 0.02 &    2.9 $\pm$ 0.2   & $0.80 \pm$ 0.01 & $240.41 \pm$ 19.72 & $0.823 \pm$ 0.004 \\ 
188.1264 &  8.7740 &  -6.31 $\pm$ 0.08 &  -7.70 $\pm$ 0.05 & $ -47.33 \pm$ 2.51  & $-247.10 \pm$ 14.74 & $-126.96 \pm$ 4.56  &    9.9 $\pm$ 0.2   &  3.2 $\pm$ 0.6 & 0.51 $\pm$ 0.06 &    9.8 $\pm$ 0.2   & $-0.02 \pm$ 0.05 & $2933.74 \pm$ 261.61 & $0.96 \pm$ -0.028 \\ 
189.9449 & 11.5534 &   3.40 $\pm$ 0.06 & -55.97 $\pm$ 0.05 & $ 257.42 \pm$ 4.44  & $-358.77 \pm$ 6.12  & $ -88.47 \pm$ 2.14  &   24.8 $\pm$ 1.4   &  2.7 $\pm$ 0.1 & 0.81 $\pm$ 0.00 &    5.2 $\pm$ 0.4   & $-0.55 \pm$ 0.03 & $194.46 \pm$ 6.74 & $0.502 \pm$ -0.034 \\ 
190.6313 &  8.5137 & -13.53 $\pm$ 0.10 & -19.24 $\pm$ 0.05 & $ -34.76 \pm$ 1.66  & $-330.56 \pm$ 21.07 & $-163.11 \pm$ 6.21  &    8.7 $\pm$ 0.4   &  5.9 $\pm$ 1.2 & 0.20 $\pm$ 0.09 &    7.4 $\pm$ 0.3   & $-0.36 \pm$ 0.08 & $2128.14 \pm$ 48.96 & $0.924 \pm$ -0.057 \\ 
193.1159 &  8.0557 &   6.70 $\pm$ 0.06 & -57.85 $\pm$ 0.04 & $ 140.26 \pm$ 3.27  & $-169.78 \pm$ 3.91  & $ -30.26 \pm$ 1.81  &   10.5 $\pm$ 0.1   &  1.1 $\pm$ 0.1 & 0.81 $\pm$ 0.02 &    1.0 $\pm$ 0.1   & $0.23 \pm$ 0.02 & $44.76 \pm$ 2.42 & $1.048 \pm$ -0.005 \\ 
193.5533 & 11.5037 &   1.37 $\pm$ 0.05 &  -1.21 $\pm$ 0.03 & $   7.22 \pm$ 0.13  & $  -4.11 \pm$ 0.12  & $  14.51 \pm$ 0.48  &    9.3 $\pm$ 0.0   &  7.7 $\pm$ 0.0 & 0.10 $\pm$ 0.00 &    0.8 $\pm$ 0.1   & $0.9 \pm$ 0.0 & $40.68 \pm$ 1.03 & $0.852 \pm$ -0.0 \\ 
193.8390 & 11.4150 &   0.98 $\pm$ 0.07 &  -2.19 $\pm$ 0.04 & $ 178.73 \pm$ 4.59  & $-112.05 \pm$ 3.29  & $ -51.18 \pm$ 1.35  &   33.6 $\pm$ 2.0   & 13.8 $\pm$ 0.5 & 0.42 $\pm$ 0.01 &   33.5 $\pm$ 1.9   & $-0.1 \pm$ 0.03 & $10599.81 \pm$ 378.3 & $0.235 \pm$ -0.029 \\ 
196.3755 &  8.5138 &  -1.46 $\pm$ 0.08 &  -6.48 $\pm$ 0.06 & $ 106.90 \pm$ 4.63  & $-317.24 \pm$ 12.03 & $-179.96 \pm$ 4.05  &   23.9 $\pm$ 2.2   &  4.9 $\pm$ 0.8 & 0.66 $\pm$ 0.02 &   21.1 $\pm$ 2.1   & $-0.39 \pm$ 0.04 & $2991.74 \pm$ 412.01 & $0.49 \pm$ -0.054 \\ 
196.4117 & 14.3176 & -18.78 $\pm$ 0.06 &  -9.71 $\pm$ 0.04 & $ -35.04 \pm$ 0.48  & $ -40.59 \pm$ 0.87  & $  -51.6 \pm$ 0.49  &    8.1 $\pm$ 0.1   &  6.4 $\pm$ 0.2 & 0.12 $\pm$ 0.00 &    1.2 $\pm$ 0.1   & $0.75 \pm$ 0.0 & $86.87 \pm$ 1.97 & $0.958 \pm$ -0.002 \\ 
196.5453 & 12.1211 &  25.63 $\pm$ 0.07 & -56.26 $\pm$ 0.04 & $ 321.48 \pm$ 11.69 & $-196.02 \pm$ 6.99  & $  -44.4 \pm$ 3.69  &   31.8 $\pm$ 3.5   &  0.5 $\pm$ 0.1 & 0.97 $\pm$ 0.01 &    4.0 $\pm$ 0.7   & $0.1 \pm$ 0.03 & $96.63 \pm$ 5.86 & $0.38 \pm$ -0.061 \\ 
198.5486 & 11.4123 & -47.39 $\pm$ 0.06 & -46.31 $\pm$ 0.04 & $ -86.24 \pm$ 1.88  & $-510.48 \pm$ 8.99  & $   8.85 \pm$ 1.27  &   17.6 $\pm$ 1.6   &  7.1 $\pm$ 0.1 & 0.42 $\pm$ 0.03 &    3.2 $\pm$ 0.4   & $-1.04 \pm$ 0.03 & $174.96 \pm$ 8.22 & $0.61 \pm$ -0.041 \\ 
200.0999 & 13.7228 &  -4.09 $\pm$ 0.06 &  -1.02 $\pm$ 0.04 & $-223.20 \pm$ 6.01  & $-340.64 \pm$ 7.01  & $ 206.07 \pm$ 0.83  &   90.2 $\pm$ 10.1  &  9.9 $\pm$ 0.7 & 0.80 $\pm$ 0.01 &   89.2 $\pm$ 9.5   & $0.10 \pm$ 0.01 & $10765.9 \pm$ 780.77 & $-0.176 \pm$ -0.048 \\ 
200.5298 &  8.9768 & -20.91 $\pm$ 0.08 & -12.61 $\pm$ 0.04 & $ -69.46 \pm$ 8.35  & $-233.94 \pm$ 18.62 & $  78.19 \pm$ 1.31  &   12.9 $\pm$ 51.3  &  0.4 $\pm$ 0.2 & 0.92 $\pm$ 0.06 &    5.8 $\pm$ 19.7  & $0.00 \pm$ 0.06 & $472.31 \pm$ 14.80 & $1.17 \pm$ 0.01 \\ 
200.7620 &  9.4375 &   8.03 $\pm$ 0.06 &  -4.98 $\pm$ 0.03 & $  93.28 \pm$ 1.30  & $   1.62 \pm$ 0.24  & $   8.97 \pm$ 0.59  &   13.6 $\pm$ 0.1   &  6.0 $\pm$ 0.1 & 0.39 $\pm$ 0.00 &    3.6 $\pm$ 0.1   & $0.85 \pm$ 0.0 & $273.96 \pm$ 7.04 & $0.748 \pm$ -0.003 \\ 
201.1158 & 15.4381 &  -8.64 $\pm$ 0.06 &  -1.74 $\pm$ 0.05 & $ -62.21 \pm$ 1.65  & $ -67.21 \pm$ 1.82  & $    2.9 \pm$ 0.53  &    8.3 $\pm$ 0.0   &  4.8 $\pm$ 0.1 & 0.27 $\pm$ 0.01 &    2.3 $\pm$ 0.1   & $0.62 \pm$ 0.01 & $240.53 \pm$ 10.05 & $1.01 \pm$ -0.002 \\ 
201.3732 &  8.4513 & -17.04 $\pm$ 0.07 &  -3.15 $\pm$ 0.04 & $-144.52 \pm$ 6.99  & $-175.08 \pm$ 7.42  & $  56.68 \pm$ 0.87  &   10.1 $\pm$ 0.3   &  1.4 $\pm$ 0.2 & 0.76 $\pm$ 0.04 &    3.6 $\pm$ 0.2   & $0.24 \pm$ 0.03 & $321.08 \pm$ 11.49 & $1.049 \pm$ -0.012 \\ 
201.8711 &  7.1810 & -78.20 $\pm$ 0.08 & -13.09 $\pm$ 0.05 & $-239.77 \pm$ 7.57  & $-265.05 \pm$ 7.84  & $  58.94 \pm$ 1.00  &   15.0 $\pm$ 0.8   &  0.5 $\pm$ 0.1 & 0.94 $\pm$ 0.01 &    3.2 $\pm$ 1.0   & $-0.1 \pm$ 0.03 & $124.81 \pm$ 1.95 & $0.822 \pm$ -0.032 \\ 
202.3435 & 13.2291 &  -4.59 $\pm$ 0.05 & -26.08 $\pm$ 0.02 & $ 145.13 \pm$ 4.50  & $-264.49 \pm$ 9.62  & $  36.58 \pm$ 2.37  &   10.1 $\pm$ 0.2   &  1.0 $\pm$ 0.2 & 0.83 $\pm$ 0.04 &    6.0 $\pm$ 0.2   & $-0.14 \pm$ 0.04 & $640.0 \pm$ 15.69 & $1.035 \pm$ -0.016 \\ 
203.2831 & 13.6326 &  -0.91 $\pm$ 0.05 &  -5.35 $\pm$ 0.03 & $ 105.34 \pm$ 3.35  & $-284.55 \pm$ 6.72  & $-203.82 \pm$ 1.65  &   27.6 $\pm$ 1.3   &  2.9 $\pm$ 0.4 & 0.81 $\pm$ 0.02 &   25.5 $\pm$ 1.7   & $-0.19 \pm$ 0.02 & $2748.57 \pm$ 254.75 & $0.432 \pm$ -0.028 \\ 
204.9008 & 10.5513 &  10.65 $\pm$ 0.06 & -33.88 $\pm$ 0.05 & $ 103.14 \pm$ 8.51  & $-106.04 \pm$ 6.33  & $-310.79 \pm$ 3.75  &   31.5 $\pm$ 1.5   &  6.2 $\pm$ 0.2 & 0.67 $\pm$ 0.02 &   28.3 $\pm$ 1.6   & $0.47 \pm$ 0.03 & $3873.18 \pm$ 104.35 & $0.336 \pm$ -0.022 \\ 
205.1342 & 13.8234 &   -8.1 $\pm$ 0.07 &  -2.72 $\pm$ 0.06 & $ -46.57 \pm$ 5.60  & $-136.76 \pm$ 8.5   & $ 137.12 \pm$ 1.18  &    9.5 $\pm$ 0.3   &  3.6 $\pm$ 0.4 & 0.45 $\pm$ 0.05 &    7.2 $\pm$ 0.4   & $0.33 \pm$ 0.03 & $1413.11 \pm$ 31.15 & $0.976 \pm$ -0.003 \\ 
205.8131 & 15.3832 &  10.45 $\pm$ 0.06 & -30.98 $\pm$ 0.04 & $ 190.87 \pm$ 5.92  & $-113.35 \pm$ 4.09  & $  64.18 \pm$ 2.01  &   14.3 $\pm$ 0.3   &  2.8 $\pm$ 0.1 & 0.68 $\pm$ 0.02 &    6.3 $\pm$ 0.3   & $0.45 \pm$ 0.02 & $541.15 \pm$ 12.38 & $0.801 \pm$ -0.01 \\ 
206.3487 &  9.3099 &   1.22 $\pm$ 0.06 & -67.65 $\pm$ 0.04 & $ 324.56 \pm$ 5.30  & $-385.14 \pm$ 7.22  & $  -2.39 \pm$ 3.00  &   41.4 $\pm$ 3.5   &  2.7 $\pm$ 0.1 & 0.88 $\pm$ 0.01 &   10.2 $\pm$ 0.8   & $-0.6 \pm$ 0.03 & $246.21 \pm$ 2.21 & $0.223 \pm$ -0.045 \\ 
208.0798 &  4.4266 &   2.02 $\pm$ 0.07 & -12.87 $\pm$ 0.07 & $  81.27 \pm$ 8.37  & $-120.10 \pm$ 8.76  & $-205.75 \pm$ 5.62  &   13.8 $\pm$ 0.7   &  3.4 $\pm$ 0.4 & 0.61 $\pm$ 0.05 &   10.7 $\pm$ 0.9   & $0.34 \pm$ 0.04 & $1600.13 \pm$ 53.45 & $0.8 \pm$ -0.02 \\ 
209.9364 & 15.9251 &   1.78 $\pm$ 0.08 & -13.73 $\pm$ 0.07 & $  56.47 \pm$ 3.89  & $ -87.63 \pm$ 3.89  & $ -116.8 \pm$ 1.42  &    9.3 $\pm$ 0.1   &  4.2 $\pm$ 0.2 & 0.37 $\pm$ 0.02 &    4.3 $\pm$ 0.1   & $0.53 \pm$ 0.02 & $578.82 \pm$ 17.2 & $0.976 \pm$ -0.002 \\ 
210.0166 & 14.6289 &  -9.65 $\pm$ 0.05 & -10.23 $\pm$ 0.05 & $  -0.72 \pm$ 1.52  & $-377.07 \pm$ 19.55 & $   76.2 \pm$ 0.60  &    8.8 $\pm$ 0.5   &  5.1 $\pm$ 0.9 & 0.27 $\pm$ 0.06 &    6.4 $\pm$ 0.4   & $-0.43 \pm$ 0.05 & $1421.89 \pm$ 143.6 & $0.959 \pm$ -0.057 \\ 
210.8632 &  8.1797 &  -5.11 $\pm$ 0.06 & -13.39 $\pm$ 0.05 & $  41.58 \pm$ 2.50  & $-170.72 \pm$ 9.11  & $ -49.28 \pm$ 2.09  &    7.8 $\pm$ 0.1   &  1.4 $\pm$ 0.3 & 0.69 $\pm$ 0.05 &    2.8 $\pm$ 0.1   & $0.20 \pm$ 0.04 & $300.71 \pm$ 22.72 & $1.214 \pm$ -0.004 \\ 
213.2814 & 14.8983 &  -0.27 $\pm$ 0.08 & -13.33 $\pm$ 0.07 & $  74.25 \pm$ 1.18  & $-105.65 \pm$ 1.51  & $ -42.69 \pm$ 0.64  &    8.6 $\pm$ 0.1   &  3.0 $\pm$ 0.1 & 0.48 $\pm$ 0.01 &    2.3 $\pm$ 0.1   & $0.45 \pm$ 0.01 & $220.76 \pm$ 4.28 & $1.084 \pm$ -0.002 \\ 
213.7879 &  8.4232 &   5.67 $\pm$ 0.07 & -15.94 $\pm$ 0.07 & $ 132.56 \pm$ 27.58 & $-114.70 \pm$ 18.31 & $-194.86 \pm$ 15.19 &   15.8 $\pm$ 2.8   &  3.0 $\pm$ 0.7 & 0.67 $\pm$ 0.11 &   10.8 $\pm$ 2.4   & $0.38 \pm$ 0.08 & $1174.7 \pm$ 64.97 & $0.738 \pm$ 0.077 \\ 
214.5556 &  7.4670 & -43.60 $\pm$ 0.05 & -28.18 $\pm$ 0.05 & $-102.93 \pm$ 3.36  & $-430.78 \pm$ 11.45 & $  81.29 \pm$ 1.16  &   11.2 $\pm$ 0.8   &  5.1 $\pm$ 0.2 & 0.37 $\pm$ 0.01 &    3.5 $\pm$ 0.2   & $-0.70 \pm$ 0.04 & $340.16 \pm$ 6.67 & $0.864 \pm$ -0.042 \\ 
\hline
\end{tabular}
}
\end{table*}

\begin{table*}
\contcaption{Orbit and Action parameters for the 70 metal-poor candidates from the \Pristine\ survey.}
\setlength{\tabcolsep}{2pt}
\rotatebox{90}{
\begin{tabular}{llrrrrrrrrrrrr} 
\hline
RA       &  DEC    &   pmra            & pmdec             &     U               &   V                 &  W                  &   R$_{\rm apo}$   & R$_{\rm peri}$  & ecc           & Z$_{\rm max}$      & J$_\phi$/J$_{\phi\odot}$ & J$_z$/J$_{z\odot}$ & E/E$_\odot$ \\[.2ex]
         &         &   (\pmu)          & (\pmu)            &    (\kms)           &   (\kms)            &  (\kms)             &     (kpc)         &   (kpc)         &               &    (kpc)           &                          &                    &   \\
\hline
217.3861 & 15.1650 &   6.02 $\pm$ 0.06 & -12.35 $\pm$ 0.06 & $  -5.72 \pm$ 2.14  & $ -41.35 \pm$ 1.0   & $-171.37 \pm$ 1.03  &   10.9 $\pm$ 0.1   & \,\, 7.5 $\pm$ 0.1 & \,\, 0.19 $\pm$ 0.00 &    6.4 $\pm$ 0.1   & $0.71 \pm$ 0.01 & $1150.58 \pm$ 17.37 & $0.793 \pm$ -0.002 \\ 
217.5786 & 14.0379 &  -1.42 $\pm$ 0.05 &  -7.46 $\pm$ 0.04 & $ 191.61 \pm$ 2.64  & $-361.55 \pm$ 4.47  & $ -86.38 \pm$ 1.13  &   22.5 $\pm$ 1.0   &  7.5 $\pm$ 0.2 & 0.50 $\pm$ 0.01 &   22.4 $\pm$ 1.0   & $-0.11 \pm$ 0.0 & $6103.43 \pm$ 214.15 & $0.483 \pm$ -0.024 \\ 
217.6443 & 15.9633 &  -1.04 $\pm$ 0.06 & -16.84 $\pm$ 0.06 & $  53.01 \pm$ 2.68  & $ -85.20 \pm$ 3.63  & $ -38.39 \pm$ 1.00  &    8.2 $\pm$ 0.0   &  3.7 $\pm$ 0.1 & 0.38 $\pm$ 0.02 &    1.5 $\pm$ 0.1   & $0.55 \pm$ 0.02 & $112.75 \pm$ 7.09 & $1.076 \pm$ -0.006 \\ 
218.4622 & 10.3683 &  -2.14 $\pm$ 0.07 &  -8.22 $\pm$ 0.07 & $ -20.40 \pm$ 0.78  & $ -95.99 \pm$ 1.48  & $-130.24 \pm$ 0.58  &    7.4 $\pm$ 0.0   &  4.9 $\pm$ 0.1 & 0.20 $\pm$ 0.01 &    4.6 $\pm$ 0.1   & $0.46 \pm$ 0.01 & $968.5 \pm$ 12.36 & $1.038 \pm$ -0.003 \\ 
228.4607 &  8.3553 &  -3.84 $\pm$ 0.07 & -59.49 $\pm$ 0.09 & $ 192.77 \pm$ 7.38  & $-267.75 \pm$ 10.58 & $ -100.5 \pm$ 4.25  &   12.9 $\pm$ 0.7   &  0.9 $\pm$ 0.1 & 0.87 $\pm$ 0.01 &    4.3 $\pm$ 0.2   & $-0.11 \pm$ 0.04 & $261.09 \pm$ 20.8 & $0.909 \pm$ -0.036 \\ 
228.6557 &  9.0914 & -15.11 $\pm$ 0.05 & -33.13 $\pm$ 0.06 & $ -18.70 \pm$ 3.11  & $-232.36 \pm$ 9.37  & $-135.68 \pm$ 0.98  &    7.5 $\pm$ 0.0   &  0.6 $\pm$ 0.2 & 0.86 $\pm$ 0.05 &    7.2 $\pm$ 0.3   & $-0.0 \pm$ 0.04 & $1206.47 \pm$ 29.75 & $1.181 \pm$ -0.002 \\ 
228.8159 &  0.2222 &  -6.67 $\pm$ 0.09 & -59.12 $\pm$ 0.11 & $ 120.03 \pm$ 4.60  & $-223.37 \pm$ 8.85  & $-102.83 \pm$ 4.26  &    9.7 $\pm$ 0.4   &  0.4 $\pm$ 0.2 & 0.92 $\pm$ 0.04 &    4.1 $\pm$ 0.3   & $0.03 \pm$ 0.03 & $349.16 \pm$ 37.54 & $1.104 \pm$ -0.014 \\ 
229.1219 &  0.9089 & -31.52 $\pm$ 0.06 & -33.97 $\pm$ 0.05 & $-151.99 \pm$ 0.39  & $-409.54 \pm$ 24.74 & $-147.89 \pm$ 0.89  &   12.9 $\pm$ 1.6   &  5.2 $\pm$ 0.4 & 0.43 $\pm$ 0.02 &    7.8 $\pm$ 0.5   & $-0.59 \pm$ 0.08 & $1122.98 \pm$ 59.34 & $0.79 \pm$ -0.064 \\ 
233.5730 &  6.4702 &  -3.73 $\pm$ 0.04 &  -6.44 $\pm$ 0.05 & $  47.28 \pm$ 2.63  & $-404.88 \pm$ 9.71  & $ -71.97 \pm$ 0.93  &   11.6 $\pm$ 0.8   &  5.1 $\pm$ 0.5 & 0.40 $\pm$ 0.02 &   11.6 $\pm$ 0.8   & $0.03 \pm$ 0.02 & $4111.92 \pm$ 239.13 & $0.83 \pm$ -0.044 \\ 
235.1449 &  8.7463 &  16.68 $\pm$ 0.05 & -25.74 $\pm$ 0.05 & $ 116.75 \pm$ 5.29  & $-104.67 \pm$ 1.79  & $-296.21 \pm$ 4.42  &   25.6 $\pm$ 1.3   &  5.2 $\pm$ 0.1 & 0.66 $\pm$ 0.02 &   22.0 $\pm$ 1.3   & $0.45 \pm$ 0.01 & $2973.52 \pm$ 51.91 & $0.45 \pm$ -0.026 \\ 
235.7578 &  9.0000 & -29.80 $\pm$ 0.06 &   6.08 $\pm$ 0.06 & $-185.50 \pm$ 5.42  & $-144.08 \pm$ 4.88  & $ 133.52 \pm$ 6.46  &   12.6 $\pm$ 0.5   &  1.7 $\pm$ 0.1 & 0.76 $\pm$ 0.02 &    6.8 $\pm$ 0.7   & $0.26 \pm$ 0.02 & $663.13 \pm$ 65.3 & $0.89 \pm$ -0.022 \\ 
236.1068 & 10.5311 &  -3.87 $\pm$ 0.04 &  -4.77 $\pm$ 0.04 & $  49.11 \pm$ 2.00  & $-383.69 \pm$ 8.65  & $  18.86 \pm$ 1.48  &   10.2 $\pm$ 0.3   &  7.3 $\pm$ 0.9 & 0.17 $\pm$ 0.04 &   10.1 $\pm$ 0.3   & $0.14 \pm$ 0.02 & $4149.48 \pm$ 228.01 & $0.813 \pm$ -0.038 \\ 
237.8246 & 10.1426 & -15.19 $\pm$ 0.05 & -12.74 $\pm$ 0.04 & $ -70.00 \pm$ 2.78  & $-583.25 \pm$ 36.27 & $  34.02 \pm$ 10.0  &   34.6 $\pm$ 18.1  &  5.8 $\pm$ 0.1 & 0.67 $\pm$ 0.12 &   24.7 $\pm$ 15.2  & $-0.73 \pm$ 0.03 & $1998.84 \pm$ 520.0 & $0.348 \pm$ -0.226 \\ 
237.9600 & 15.4022 & -11.75 $\pm$ 0.04 & -23.56 $\pm$ 0.03 & $ -91.49 \pm$ 2.99  & $-242.87 \pm$ 6.53  & $-188.22 \pm$ 0.46  &    8.3 $\pm$ 0.1   &  4.0 $\pm$ 0.1 & 0.36 $\pm$ 0.01 &    8.3 $\pm$ 0.2   & $-0.06 \pm$ 0.02 & $2986.96 \pm$ 133.31 & $1.012 \pm$ -0.006 \\ 
240.4216 &  9.6761 &   1.71 $\pm$ 0.03 & -11.53 $\pm$ 0.02 & $ 329.53 \pm$ 14.23 & $-270.18 \pm$ 13.16 & $-177.61 \pm$ 9.54  &   47.9 $\pm$ 13.2  &  3.6 $\pm$ 0.4 & 0.85 $\pm$ 0.02 &   41.5 $\pm$ 11.1  & $0.34 \pm$ 0.03 & $2968.97 \pm$ 414.55 & $0.157 \pm$ -0.131 \\ 
245.4387 &  8.9954 & -25.09 $\pm$ 0.04 & -23.90 $\pm$ 0.03 & $  -2.07 \pm$ 1.29  & $-316.79 \pm$ 6.09  & $  29.47 \pm$ 1.70  &    6.8 $\pm$ 0.1   &  1.7 $\pm$ 0.1 & 0.61 $\pm$ 0.03 &    1.5 $\pm$ 0.1   & $-0.28 \pm$ 0.02 & $143.66 \pm$ 8.72 & $1.292 \pm$ -0.007 \\ 
245.5747 &  6.8844 &  -0.49 $\pm$ 0.04 &  11.07 $\pm$ 0.03 & $-307.45 \pm$ 11.56 & $ 137.48 \pm$ 13.46 & $   5.32 \pm$ 8.20  &  131.2 $\pm$ 56.6  &  3.2 $\pm$ 0.3 & 0.95 $\pm$ 0.02 &   88.0 $\pm$ 33.4  & $0.51 \pm$ 0.05 & $1883.49 \pm$ 411.94 & $-0.285 \pm$ -0.146 \\ 
245.8356 & 13.8777 &   0.98 $\pm$ 0.03 &  -2.39 $\pm$ 0.02 & $ -76.79 \pm$ 2.01  & $ -89.61 \pm$ 1.09  & $-144.63 \pm$ 1.55  &    6.9 $\pm$ 0.1   &  4.9 $\pm$ 0.1 & 0.17 $\pm$ 0.02 &    6.0 $\pm$ 0.1   & $0.27 \pm$ 0.01 & $1865.8 \pm$ 97.43 & $1.055 \pm$ -0.002 \\ 
246.5144 &  5.9826 &   2.42 $\pm$ 0.03 &   1.45 $\pm$ 0.02 & $  -5.81 \pm$ 0.40  & $  28.86 \pm$ 0.75  & $ -17.21 \pm$ 0.46  &   10.1 $\pm$ 0.1   &  6.2 $\pm$ 0.0 & 0.24 $\pm$ 0.01 &    2.2 $\pm$ 0.1   & $0.79 \pm$ 0.0 & $189.37 \pm$ 9.41 & $0.873 \pm$ -0.002 \\ 
246.8588 & 12.3193 &  -8.20 $\pm$ 0.03 &  -5.43 $\pm$ 0.02 & $ -23.27 \pm$ 1.77  & $-326.57 \pm$ 13.64 & $  83.74 \pm$ 6.31  &    6.7 $\pm$ 0.7   &  2.7 $\pm$ 0.6 & 0.43 $\pm$ 0.11 &    5.9 $\pm$ 0.6   & $-0.16 \pm$ 0.01 & $1489.34 \pm$ 239.52 & $1.226 \pm$ -0.042 \\ 
248.4394 &  7.9229 &  -7.82 $\pm$ 0.05 & -20.68 $\pm$ 0.04 & $  34.18 \pm$ 1.02  & $ -94.12 \pm$ 1.82  & $ -18.85 \pm$ 0.35  &    7.5 $\pm$ 0.1   &  3.3 $\pm$ 0.1 & 0.39 $\pm$ 0.01 &    0.6 $\pm$ 0.1   & $0.51 \pm$ 0.01 & $29.83 \pm$ 1.04 & $1.155 \pm$ -0.004 \\ 
248.4959 & 15.0776 &  -2.87 $\pm$ 0.03 & -11.51 $\pm$ 0.03 & $ 277.89 \pm$ 11.70 & $-464.93 \pm$ 15.46 & $-112.07 \pm$ 2.48  &   56.2 $\pm$ 16.2  &  4.4 $\pm$ 0.5 & 0.85 $\pm$ 0.02 &   47.1 $\pm$ 12.6  & $0.43 \pm$ 0.06 & $3269.42 \pm$ 292.39 & $0.077 \pm$ -0.134 \\ 
250.6963 &  8.3743 &  -6.84 $\pm$ 0.04 &  -1.89 $\pm$ 0.04 & $ -21.03 \pm$ 0.87  & $-205.95 \pm$ 7.35  & $ 162.04 \pm$ 5.92  &    7.5 $\pm$ 0.3   &  0.9 $\pm$ 0.2 & 0.80 $\pm$ 0.03 &    7.4 $\pm$ 0.4   & $0.01 \pm$ 0.01 & $1634.18 \pm$ 172.91 & $1.176 \pm$ -0.026 \\ 
251.4082 & 12.3657 &  -7.86 $\pm$ 0.06 &  -1.27 $\pm$ 0.05 & $ -70.42 \pm$ 2.20  & $-400.15 \pm$ 11.31 & $ 386.54 \pm$ 11.06 &  201.6 $\pm$ 102.4 &  9.5 $\pm$ 0.7 & 0.90 $\pm$ 0.03 &  155.3 $\pm$ 63.8  & $0.05 \pm$ 0.03 & $11689.1 \pm$ 785.88 & $-0.444 \pm$ -0.127 \\ 
252.1639 & 15.0648 &   1.49 $\pm$ 0.03 &  -5.48 $\pm$ 0.03 & $  94.04 \pm$ 5.86  & $-203.86 \pm$ 3.85  & $-231.61 \pm$ 4.19  &   15.2 $\pm$ 1.1   &  4.5 $\pm$ 0.2 & 0.54 $\pm$ 0.01 &   14.1 $\pm$ 1.0   & $0.26 \pm$ 0.02 & $2869.78 \pm$ 110.49 & $0.721 \pm$ -0.04 \\ 
253.8582 & 15.7240 &  -2.22 $\pm$ 0.03 &  -5.72 $\pm$ 0.03 & $  88.00 \pm$ 6.04  & $-263.85 \pm$ 8.44  & $ -57.11 \pm$ 0.63  &    7.9 $\pm$ 0.6   &  1.9 $\pm$ 0.2 & 0.62 $\pm$ 0.04 &    5.7 $\pm$ 0.5   & $0.19 \pm$ 0.02 & $968.04 \pm$ 56.69 & $1.165 \pm$ -0.031 \\ 
254.5207 & 15.4969 &  14.12 $\pm$ 0.02 &  -9.85 $\pm$ 0.03 & $  44.56 \pm$ 0.56  & $  18.18 \pm$ 0.26  & $ -28.06 \pm$ 0.72  &   11.7 $\pm$ 0.1   &  7.2 $\pm$ 0.1 & 0.24 $\pm$ 0.00 &    0.7 $\pm$ 0.1   & $0.95 \pm$ 0.0 & $20.13 \pm$ 0.98 & $0.785 \pm$ -0.001 \\ 
254.5469 & 10.9129 & -21.80 $\pm$ 0.04 & -28.68 $\pm$ 0.04 & $ 260.45 \pm$ 10.30 & $-383.71 \pm$ 23.23 & $ 133.82 \pm$ 4.69  &   25.0 $\pm$ 5.2   &  2.5 $\pm$ 0.3 & 0.82 $\pm$ 0.01 &   20.4 $\pm$ 4.0   & $-0.28 \pm$ 0.05 & $1683.88 \pm$ 96.41 & $0.501 \pm$ -0.11 \\ 
255.2671 & 14.9711 &  -8.81 $\pm$ 0.04 & -40.50 $\pm$ 0.04 & $ 208.89 \pm$ 3.70  & $-210.05 \pm$ 4.37  & $ -32.63 \pm$ 0.96  &   12.1 $\pm$ 0.2   &  0.7 $\pm$ 0.1 & 0.88 $\pm$ 0.02 &    1.0 $\pm$ 0.1   & $0.16 \pm$ 0.01 & $51.85 \pm$ 1.75 & $0.96 \pm$ -0.01 \\ 
255.5555 & 10.8612 &  -5.06 $\pm$ 0.04 & -15.51 $\pm$ 0.04 & $-108.68 \pm$ 9.34  & $-418.27 \pm$ 13.88 & $-213.57 \pm$ 1.85  &   12.8 $\pm$ 1.0   &  5.5 $\pm$ 0.2 & 0.40 $\pm$ 0.04 &    8.9 $\pm$ 0.4   & $-0.55 \pm$ 0.02 & $1544.81 \pm$ 66.8 & $0.781 \pm$ -0.028 \\ 
\hline
\end{tabular}
}
\end{table*}

\bsp	
\label{lastpage}
\end{document}